\documentclass[a4paper,fleqn,usenatbib]{mnras}
\usepackage{newtxtext,newtxmath}
\usepackage[T1]{fontenc}
\usepackage{ae,aecompl}
\usepackage{graphicx}	
\usepackage{amsmath}	
\usepackage{verbatim}
\usepackage{relsize}

\def\bhar{$\rm \overline{BHAR}$}
\def\mstar{$M_\star$}
\def\mbh{$M_{\rm BH}$}
\def\mbulge{$M_{\rm bulge}$}
\def\sigmaone{$\Sigma_{1}$}

\def\sigmagas{$\Sigma_{\rm gas}$}
\def\sigmainf{$\sigma_{\rm inf}$}

\def\re{$r_{\rm e}$}
\def\sersic{$\rm S{\acute e}rsic$}
\def\lx{$L_{\rm X}$}
\def\xray{\hbox{X-ray}}

\title[]{Revealing the relation between black-hole growth and host-galaxy compactness among star-forming galaxies}

\author[Ni et al.]{Q. Ni,$^{1,2}$\thanks{E-mail: qxn1@psu.edu}
W. N. Brandt,$^{1,2,3}$
G. Yang,$^{4,5}$\thanks{E-mail: gyang206265@gmail.com}
J. Leja,$^{1,6}$
C.-T. J. Chen,$^{7}$
B. Luo,$^{8,9,10}$
\newauthor
J. Matharu,$^{4,5}$
M. Sun,$^{11}$
F. Vito,$^{12,13}$
Y. Q. Xue,$^{14,15}$
and K. Zhang$^{16}$
\\
$^{1}$Department of Astronomy and Astrophysics, 525 Davey Lab, The Pennsylvania State University, University Park, PA 16802, USA\\
$^{2}$Institute for Gravitation and the Cosmos, The Pennsylvania State University, University Park, PA 16802, USA\\
$^{3}$Department of Physics, 104 Davey Laboratory, The Pennsylvania State University, University Park, PA 16802, USA\\
$^{4}$Department of Physics and Astronomy, Texas A\&M University, College Station, TX 77843-4242, USA\\
$^{5}$George P.\ and Cynthia Woods Mitchell Institute for Fundamental Physics and Astronomy, Texas A\&M University, College Station, TX 77843-4242, USA\\
$^{6}$Center for Astrophysics $|$ Harvard \& Smithsonian, 60 Garden St.\ Cambridge, MA 02138, USA\\
$^{7}$Marshall Space Flight Center, Huntsville, AL 35811, USA\\
$^{8}$School of Astronomy and Space Science, Nanjing University, Nanjing 210093, China\\
$^{9}$Key Laboratory of Modern Astronomy and Astrophysics (Nanjing University), Ministry of Education, Nanjing 210093, China\\
$^{10}$Collaborative Innovation Center of Modern Astronomy and Space Exploration, Nanjing, 210093, China\\
$^{11}$Department of Astronomy, Xiamen University, Xiamen, Fujian 361005, China\\
$^{12}$Instituto de Astrof{\'{\i}}sica and Centro de Astroingenier{\'{\i}}a, Facultad de F{\'{i}}sica, Pontificia Universidad Cat{\'{o}}lica de Chile, Casilla 306, Santiago 22, Chile\\
$^{13}$Chinese Academy of Sciences South America Center for Astronomy, National Astronomical Observatories, CAS, Beijing 100012, China\\
$^{14}$CAS Key Laboratory for Research in Galaxies and Cosmology, Department of Astronomy, University of Science and Technology of China, Hefei 230026, China\\
$^{15}$School of Astronomy and Space Science, University of Science and Technology of China, Hefei 230026, China\\
$^{16}$Department of Astronomy, University of California, Berkeley, CA 94720-3411, USA
}

\date{Accepted XXX. Received YYY; in original form ZZZ}

\pubyear{2020}

\begin{document}
\label{firstpage}
\pagerange{\pageref{firstpage}--\pageref{lastpage}}
\maketitle

\begin{abstract}
Recent studies show that a universal relation between black-hole (BH) growth and stellar mass (\mstar) or star formation rate (SFR) is an oversimplification of BH-galaxy co-evolution, and that morphological and structural properties of host galaxies must also be considered. 
Particularly, a possible connection between BH growth and host-galaxy compactness was identified among star-forming (SF) galaxies. 
Utilizing $\approx 6300$ massive galaxies with $I_{\rm 814W}~<~24$ at $z$ $<$ 1.2 in the COSMOS field, we perform systematic partial-correlation analyses to investigate how sample-averaged BH accretion rate (\bhar) depends on host-galaxy compactness among SF galaxies, when controlling for morphology and \mstar\ (or SFR). The projected central surface-mass density within 1~kpc, \sigmaone, is utilized to represent host-galaxy compactness in our study. We find that the \bhar-\sigmaone\ relation is stronger than either the \bhar-\mstar\ or \bhar-SFR relation among SF galaxies, and this \bhar-\sigmaone\ relation applies to both bulge-dominated galaxies and galaxies that are not dominated by bulges. This \bhar-\sigmaone\ relation among SF galaxies suggests a link between BH growth and the central gas density of host galaxies on the kpc scale, which may further imply a common origin of the gas in the vicinity of the BH and in the central $\sim$~kpc of the galaxy. This \bhar-\sigmaone\ relation can also be interpreted as the relation between BH growth and the central velocity dispersion of host galaxies at a given gas content (i.e. gas mass fraction), indicating the role of the host-galaxy potential well in regulating accretion onto the BH.
\end{abstract}

\begin{keywords}
galaxies: active -- galaxies: evolution -- galaxies: nuclei -- X-rays: galaxies
\end{keywords}



\section{Introduction} \label{s-intro}

Correlations between black-hole (BH) mass and host-galaxy properties observed in the local universe \citep[e.g.][]{Magorrian1998,MH2003,KH2013, MM2013} have inspired investigations of so-called ``BH-galaxy co-evolution'' over the past couple decades.
As the BH accretion rate of individual objects has large long-term variability that hinders us from revealing any intrinsic link between the BH growth and its host galaxy \citep[e.g.][]{Hickox2014,Sartori2018,Yuan2018},  one effective way to investigate BH-galaxy co-evolution ``in action'' is performing large-sample studies.
With X-ray emission serving as a reliable tracer of BH accretion \citep[e.g.][]{Brandt2015}, these sample studies utilize the average BH accretion rate (\bhar) of a sample of galaxies sharing similar properties to approximate the long-term average BH growth of galaxies with these properties; i.e. they take BH growth to be ergodic.
Relations between \bhar\ and \mstar\ or SFR have been revealed \citep[e.g.][]{Mullaney2012, Chen2013, Aird2017, Aird2018,Yang2017, Yang2018a}, which are considered as observational evidence of a link between BH growth and the potential well or the gas mass of host galaxies.

However, a ``universal'' relation between BH growth and \mstar\ or SFR is likely a substantial oversimplification of BH-galaxy co-evolution.
\citet{Yang2019} found that morphology must be considered when studying BH-galaxy co-evolution: for bulge-dominated (BD) galaxies, BH growth mainly depends on SFR rather than \mstar; for galaxies not dominated by bulges (Non-BD), BH growth mainly depends on \mstar\ rather than SFR. This finding is consistent with the observational result in the local universe that BH mass (\mbh) only correlates tightly with bulge mass (\mbulge), rather than \mstar\ of the whole host galaxy \citep[e.g.][]{KH2013}.
The role of compactness (which measures the mass-to-size ratio of galaxies) has also triggered attention in recent years: \citet{Kocevski2017} found an elevated active galactic nucleus (AGN)  fraction among compact star-forming (SF) galaxies when compared with mass-matched extended SF galaxies. This finding is consistent with the predicted scenario that BH growth can be triggered by the high central gas density during a wet compaction event \citep[e.g.][]{Wellons2015,Dekel2019, Habouzit2019}.

Given all these findings, \citet{Ni2019} examined the effectiveness of compactness in predicting the amount of BH growth when controlling for various other host-galaxy properties (including morphology) using galaxies in the $\approx 0.25$~deg$^2$ CANDELS survey fields \citep{Grogin2011,Koekemoer2011}.
\citet{Ni2019} found that compactness can only effectively predict \bhar\ among SF galaxies, and the central surface-mass density within 1~kpc (\sigmaone) is more effective in predicting the amount of BH growth than the surface mass density in the central regions comprising 50\% of the galaxy stellar mass.
These results led \citet{Ni2019} to speculate that the \bhar-\sigmaone\
 relation, if confirmed, could reflect a link between BH growth and the central $\sim$~kpc gas density of host galaxies (that could be related to \sigmaone\ among SF galaxies when assuming a correlation between gas density and \mstar\ density, which is supported by recent ALMA observational results; see \citealt{Lin2019}). 
\citet{Ni2019} found evidence that the relation between \bhar\ and \sigmaone\ is not simply a secondary manifestation of the \bhar-\mstar\ relation among SF Non-BD galaxies; 
while the number of SF BD galaxies in \citet{Ni2019} was too small to confirm a significant ($> 3\sigma$) \bhar-\sigmaone\ relation (when controlling for SFR), BD galaxies with relatively high SFR values suggest the link between BH growth and \sigmaone.
If a significant \hbox{\bhar-\sigmaone} relation can be confirmed among SF BD galaxies (when controlling for SFR), it will provide a natural explanation for over-massive BH ``monsters''\footnote{BH ``monsters'' are BHs that have \mbh\ significantly larger than expected from the \mbh\ relation with bulge mass ($M_{\rm bulge}$). We note that it has also been argued that some BH monsters are not real: their BH masses seem to be unexpectedly large due to the underestimation of $M_{\rm bulge}$ when improper bulge/disk decomposition is conducted \citep[e.g.][]{Graham2016}.} in the local universe that live in compact galaxies \citep[e.g.][]{KH2013,Walsh2015,Walsh2017}. 
It is plausible that a \bhar-\sigmaone\ relation that is more ``fundamental''\footnote{Throughout this paper, when A relates with both B and C, if the relation between A and B is significant when controlling for C while the relation between A and C is not significant when controlling for B in partial-correlation analyses, we say the relation between A and B is more fundamental than the relation between A and C.} than either the \bhar-\mstar\ or \bhar-SFR relation may apply for \textit{all} SF galaxies regardless of morphology. 
If so, this would provide strong evidence for a link between BH growth and the central gas density of host galaxies, which may reveal how BHs feed from gas in the central parts of galaxies: this is especially important given that it is difficult to measure the central gas density directly for a large sample of AGNs due to current observational constraints.

In this paper, we use a large sample of galaxies and AGNs at $z < 1.2$ in the $\approx 1.4$ deg$^2$ COSMOS survey field (that has UltraVISTA and ACS coverage; \citealt{Koekemoer2007, Leauthaud2007, Laigle2016}) to probe further the relation between BH growth and host-galaxy compactness among SF galaxies.
Specifically, we will address the following questions:
Is the \hbox{\bhar-\sigmaone} relation more fundamental than the \bhar-\mstar\ relation among SF Non-BD galaxies?
Is there a significant \hbox{\bhar-\sigmaone} relation when controlling for SFR among SF BD galaxies?
Is the \bhar-\sigmaone\ relation ``universal'' among all SF galaxies? If so, what are the properties of this \bhar-\sigmaone\ relation?

This paper is structured as follows. 
In Section~\ref{s-ds}, we describe the sample construction process. 
In Section~\ref{s-ar}, we perform data analyses and present the results.
In Section~\ref{s-dis}, we interpret the analyses results and present relevant discussions. 
Section~\ref{s-con} summarizes this work and discusses future prospects.
Throughout this paper, $M_\star$\ is in units of $M_\odot$; SFR and \bhar\ are in units of $M_\odot$~yr$^{-1}$; \sigmaone\ is in units of $M_\odot / {\rm kpc}^2$.
$L_X$ indicates absorption-corrected X-ray luminosity at rest-frame 2--10 keV in units of erg s$^{-1}$. Quoted uncertainties are at the $1\sigma$\ (68\%) confidence level, unless otherwise stated. 
A cosmology with $H_0=70$~km~s$^{-1}$~Mpc$^{-1}$, $\Omega_M=0.3$, and $\Omega_{\Lambda}=0.7$ is assumed.
We consider a partial correlation to be significant if it has a \hbox{$p$-value~$<$~0.0027}, which corresponds to a significance level $>3\sigma$.
Significant results throughout the paper are marked in bold in the tables.

\section{Data and sample} \label{s-ds}

Our objects are selected from the COSMOS2015 catalog \citep{Laigle2016}. 
Only sources within both the COSMOS and UltraVISTA regions are kept, and we remove saturated objects in bad areas (\texttt{FLAG\_COSMOS = 0}, \texttt{FLAG\_HJMCC = 0}, and \texttt{FLAG\_PETER = 0}).
We further limit our selection to $I_{\rm F814W} < 24$ galaxies: $I_{\rm F814W} < 24$ is a common threshold adopted in the \textit{HST} COSMOS field for morphological classifications \citep[e.g.][]{Scarlata2007}.
We obtain spectroscopic redshifts (spec-$z$) for sources from \citet[]{Marchesi2016,Delvecchio2017,Hasinger2018}; and Salvato et al.\ in prep.
We note that $\approx 60\%$ of sources utilized in Section~\ref{s-ar} have spectroscopic redshifts.
For sources without spectroscopic redshifts, we adopt the high-quality photometric redshift (photo-$z$) measurements from \citet{Laigle2016} with $\sigma_{\Delta z/(1+z_s)} = 0.007$.

For the selected COSMOS sources, in Section~\ref{ss-mstarsfr}, we measure their \mstar\ and SFR values; in Section~\ref{ss-galfit}, we measure their structural parameters including \sersic\
 index ($n$) and effective radius (\re) that will be utilized to calculate \sigmaone; in Section~\ref{ss-morph}, we classify objects as BD/Non-BD. In Section~\ref{ss-sample}, we construct samples that will be used for the analyses in Section~\ref{s-ar}. In Section~\ref{ss-bhar}, we explain how \bhar\ utilized in Section~\ref{s-ar} is estimated.

\subsection{Stellar mass and star formation rate measurements} \label{ss-mstarsfr}
We measure \mstar\ and SFR with {\sc X-CIGALE} \citep{Yang2020}, which is a new version of {\sc CIGALE} \citep[e.g.,][]{Boquien2019} with updated AGN modules.
Photometric data in 38 bands (including 24 broad bands) from NUV to FIR \citep{Laigle2016} are utilized.
For the NUV to NIR photometry, we correct the aperture flux to total flux  following Appendix~A2 of \citet{Laigle2016}. 
For the 3 {\it Herschel}/SPIRE bands, we use photometric data reported in a super-deblended catalog described in \citet{Jin2018} which utilizes the deblending technique in \citet{Liu2018}.

For X-ray undetected galaxies, we fit them with a two-run approach: we first fit them with pure galaxy templates. We adopt a delayed exponentially declining star formation history (SFH),\footnote{The delayed SFH is chosen as \citet{Ciesla2015} found that when performing SED fitting with {\sc CIGALE} for AGN hosts, the delayed SFH model provides better estimation of \mstar\ and SFR compared with other parametric SFHs.} a Chabrier initial mass function \citep{Chabrier2003}, the extinction law from \citet{Calzetti2000}, and the dust emission template from \citet{Dale2014}, following \citet{Ciesla2015} and \citet{Yang2020}. We also add nebular emission to the SED libraries.
Details of the fitting parameters can be seen in Table~\ref{cigalep}.
Then, we add an additional AGN component presented in {\sc X-CIGALE}, \texttt{SKIRTOR} (that is established based on \citealt{Stalevski2012,Stalevski2016}), during the fitting (detailed parameters can also be found in Table~\ref{cigalep}).
One free parameter in \texttt{SKIRTOR} is the fractional contribution of AGN emission to the total IR luminosity ($\rm frac_{AGN}$), which can range from 0 to 1, and we use a step of 0.1 during the fitting.
We find that while the measurements of \mstar\ are not significantly influenced by adding an AGN component, the SFR measurements are smaller by \hbox{$\approx 0.2$--0.5} dex on average when \hbox{$\rm frac_{AGN} \geqslant 0.3$}.
When \hbox{$\rm frac_{AGN} < 0.3$}, adding an AGN component affects the SFR measurements by less than $\approx$~0.2~dex. As we group sources in log SFR bins of at least $\sim 0.5$~dex-width in our analyses (see Section~\ref{ss-sfbd}), the differences in SFR measurements caused by adding an AGN component for $\rm frac_{AGN} < 0.3$ objects are negligible in the context of this work. 
Thus, when the estimated Bayesian 1$\sigma$ lower limit of $\rm frac_{AGN}$ is $\geqslant 0.25$ ($\approx$~1\% of total objects), we adopt the Bayesian \mstar\ and SFR values from the solution with an AGN component. Otherwise, we adopt the Bayesian \mstar\ and SFR values from the solution without an AGN component.

For X-ray detected galaxies, we directly fit them with both galaxy and AGN components. 
We have also incorporated the \textit{Chandra} \xray\ flux \citep{Civano2016} into the fitting following \citet{Yang2020} (through the \xray\ module in {\sc X-CIGALE}) to constrain the AGN SED contribution, as the X-ray SED of AGN is empirically connected to the UV-to-IR SED \citep[e.g.][]{Just2007}. 
\textit{Chandra} \xray\ fluxes are adopted following the preference order of hard band (2--10 keV), full band (0.5--10 keV), and soft band (0.5--2~keV), thus minimizing the effects of X-ray obscuration. We require that the deviation from this empirical SED relation ($\Delta \alpha_{\rm OX}$) is not larger than 0.2 (which corresponds to the 2$\sigma$ scatter of the empirical relation; e.g.\ \citealt{Just2007}). 
 We note that for our X-ray detected galaxies, adding the \xray\ module or not does not significantly affect the Bayesian \mstar\ and SFR measurements: the scatter between the two sets of \mstar\ (SFR) measurements is $\approx$~0.1~(0.2) dex, with negligible systematic offsets.
We verified that the analysis results in Section~\ref{s-ar} do not change qualitatively if we add random perturbations to log~\mstar/log~SFR values of \xray\ detected galaxies with a scatter of 0.1/0.2 dex.

A comparison between our SED-based \mstar\ and SFR measurements and SED-based \mstar\ and SFR measurements with \texttt{Prospector} \citep{Leja2019b} for a subset of COSMOS galaxies is presented in Appendix~\ref{a-xcigale}, showing the general consistency between the two approaches. As our \mstar\ measurements are systematically smaller than those reported in \citet{Leja2019b} by $\approx 0.15$~dex, we correct our measurements for this systematic offset in the final adopted \mstar\ values (see Appendix~\ref{a-xcigale} for details).

As the SED fitting process is ``dominated'' by the large number of UV-to-NIR bands that may underestimate SFR in the high-SFR regime \citep[e.g.][]{Wuyts2011,Yang2017}, FIR-based SFR values are adopted when available (for $\approx$4\%/26\% of objects in the SF BD/SF Non-BD samples defined in Section~\ref{ss-sample}).
When an object is detected with S/N $>$ 5 in a \textit{Herschel} band \citep{Lutz2011,Oliver2012,Laigle2016,Jin2018}, we derive its total IR luminosity from the FIR flux in this band utilizing the SF galaxy template in \citet{Kirkpatrick2012}.
Then, a weighted total IR luminosity is calculated from all available \textit{Herschel} bands with the FIR flux error serving as the weight.
The total IR luminosity is then converted to SFR following Equation~1 in \citet{Ni2019}, assuming that most UV photons are absorbed by the dust.
We have also compared our SED-based SFR values with these FIR-based SFR values, showing the consistency of these two methods (see Appendix~\ref{a-xcigale} for details). 
We note that FIR-based SFR measurement also has its shortcomings \citep[e.g.][]{Kennicutt1998b, Hodge2020}: as the stellar populations and dust properties vary from galaxy to galaxy, there are natural uncertainties associated with the simple universal rescaling from FIR luminosity to SFR. We verified that our results in Section~\ref{s-ar} do not change qualitatively if we solely adopt SED-based SFR values.

\subsection{Structural measurements with GALFIT} \label{ss-galfit}
\subsubsection{Image and noise cutouts} \label{sss-in}
We prepare image cutouts for the selected objects from ACS F814W COSMOS science images v2.0 \citep{Koekemoer2007} that have bad pixels and cosmic rays removed.
Following \citet{Matharu2019}, our cutouts have 15 $\times$ FLUX\_RADIUS pixels in the $x$/$y$-axis (FLUX\_RADIUS is the half-light radius measured by SExtractor in \citealt{Leauthaud2007}), with the target galaxy at the center.
The noise cutouts with same sizes are made following \citet{vdw2012} and \citet{Matharu2019}, where the noise is a quadrature combination of the Poisson noise of the image and other noises where the sky-background noise dominates. We estimate the sky-background noise as well as the background sky level with segmentation maps generated for each image cutout by SExtractor v2.19.5 \citep{BA1996}, following section~3 and table~1 of \citet{Leauthaud2007}. With the information provided by these segmentation maps, we select all pixels that do not belong to sources in the image cutout, and use these pixels to estimate the background sky level/noise, which is the mean/root-mean-square value of these background pixels.

\subsubsection{PSF generation} \label{sss-psf}

The PSF model used in this work is generated by the IDL wrapper of \texttt{TinyTim} \citep{Krist1995} introduced in \citet{Rhodes2006,Rhodes2007}, assuming a G8V star and a focus at $-3.0 \mu$m. This IDL wrapper can generate the PSF model with a pixel scale of 0.03'' to match the oversampled version of ACS COSMOS science images that have geometric distortion removed. We neglect the change of PSF both temporally and across the CCD at the level of a few percent \citep{Rhodes2007, Gabor2009}. We have also compared the PSF model with real stars in the COSMOS field, and we find that the differences between the encircled flux fractions at a given radius are generally small (within a few percent).

\subsubsection{GALFIT setup}
We fit our objects with a single-component \sersic\ profile in GALFIT \citep{Peng2002}:
\begin{equation} 
  I(r) = I_o \exp {\left \{ {-b_n \left[{\left(\frac{r}{r_e}\right)}^{1/n}-1\right]}\right \} },
\end{equation}

\noindent
where $n$ is the \sersic\ index, \re\ is the half-light radius, $I(r)$ represents light intensity at a radius of $r$, $I_o$ is the light intensity at \re, and $b_n$ is coupled to $n$ to make half of the total flux lie within \re.

Following \citet{vdw2012}, we set constraints in GALFIT to keep 0.2 $< n <$ 8, 0.5 $<$ \re\ $<$ 800 (in units of pixels), 0.0001 $< q <$ 1 ($q$ is the axis ratio).
Rather than fitting a single object, we fit all the sources in the cutout that are no more than 5 mag fainter than the central target source simultaneously, which can substantially improve the accuracy of fitting \citep[e.g.][]{Peng2002,Matharu2019}. We do not fit for the sky during the fit \citep[e.g.][]{Haussler2007,Barden2012}: we set the sky level as the background sky level  estimated in Section~\ref{sss-in}. 
For $\approx$ 87\% of objects, GALFIT reached a solution without hitting any constraints (we mark them with GALFIT\_flag $= 0$); for $\approx$ 4\% of objects, GALFIT hit the constraints (we mark them with GALFIT\_flag $= 2$); for $\approx$ 9\% of objects, GALFIT did not manage to converge.
Since fitting a large number of additional objects simultaneously may cause GALFIT to fail due to these objects, we fit the $\approx$ 9\% of objects where GALFIT did not manage to converge again without fitting neighboring objects: in this second run, we use the SExtractor segmentation map to mask all neighboring objects (masked pixels within an ellipse of 3 $\times$ Kron ellipse + 20 pixels of the central object are regarded to contain the source flux, so we unmask them in the segmentation map), and we only fit the target object at the center. 
If GALFIT reached a solution without hitting any constraints in this second run, we mark the object with GALFIT\_flag $= 1$; if GALFIT hit the constraints, we mark the object with GALFIT\_flag $= 2$; if GALFIT failed again, we mark the object with GALFIT\_flag $= 3$.
We will only use the $\approx 87$\% GALFIT\_flag $= 0$ and $\approx 8$\% GALFIT\_flag $= 1$ objects for our analyses, and our results do not change qualitatively if we limit our analyses to GALFIT\_flag $= 0$ objects only.
In Appendix~\ref{a-galfit}, we show the reliability of our results by comparing with the GIM2D measurements of $I_{\rm F814W}$~$<$~22.5 galaxies in COSMOS \citep{Sargent2007}.
We also assess the level of potential AGN contamination to host-galaxy light profiles in Appendix~\ref{a-galfit}. We find that for X-ray AGNs included in our sample (see Section~\ref{ss-sample} for the sample selection), the AGN contamination is largely negligible.

\subsection{Deep-learning-based morphology} \label{ss-morph}
We use a deep-learning-based method to classify \hbox{$I_{\rm F814W} < 24$} galaxies in COSMOS \citep{Leauthaud2007} as BD galaxies or Non-BD galaxies. 
Details of this deep-learning-based BD/Non-BD classification process are presented in Appendix~\ref{a-morph}.
Our selection of BD galaxies is broadly consistent with the selection of ``pure bulges'' in \citet[][see Appendix~\ref{a-morph} for details]{HC2015}.

\subsection{Sample construction} \label{ss-sample}

We first confine our sample to galaxies at $z < 1.2$, where the \textit{HST} F814W band can characterize the rest-frame optical emission of galaxies ($\approx 370-800$ nm), so that our morphological measurements are not strongly affected by the ``morphological k-correction''.
The relatively low redshift range probed here compared with the $z = 0.5$--3 sample in \citet{Ni2019} also generally enables more accurate more accurate morphological characterization.
Following \citet{Yang2018a} and \citet{Ni2019}, we remove broad-line (BL) AGNs \citep{Marchesi2016} from the sample (which make up $\approx 6\%$ of total X-ray detected galaxies), as the strong emission from BL AGNs prohibits us from obtaining reliable measurements of host-galaxy properties. The exclusion of BL AGNs should not affect the analysis results assuming the unified model \citep[e.g.][]{Netzer2015}. According to the unified model, BL AGNs and type 2 AGNs are purely orientation-based AGN classes: when our line-of-sight does not intercept the torus, a BL AGN is observed; otherwise, a type 2 AGN is observed.
Thus, as detailed in Section~2.4.1 of \citet{Ni2019}, excluding the contribution from BL AGNs when estimating sample-averaged BH growth only decreases \bhar\ by a similar fraction for utilized subsamples of galaxies in Section~\ref{s-ar}, so it will not influence our investigations of the dependence of BH growth on various host-galaxy properties.\footnote{We recognize that it has been suggested that host-galaxy gas could have column densities on the order of $10^{23-24}~{\rm cm}^{-2}$ for $z > 3$ compact SF galaxies \citep{DAmato2020}, indicating that the unified model may not be sufficient for explaining all obscured AGNs. However, as our study focuses on low-to-moderate-redshift galaxies, the gas content is not as high as that of high-$z$ galaxies. \citet{Buchner2017} suggest that at $z < 3$, galaxy-scale gas does not generally produce Compton-thick columns. This is consistent with our checking that when we group \xray\ AGNs in our sample into several \sigmaone\ bins, the average \xray\ hardness ratio does not significantly vary: if the galaxy-scale gas column density among compact SF galaxies in our sample is sufficiently large, we would expect harder \xray\ spectra in average among AGNs hosted by more compact SF galaxies. Thus, the unified model appears to be a reliable assumption to first order for our work.}
We also confine our sample to GALFIT\_flag = 0 or 1 objects, where reliable structural measurements are available (see Section~\ref{ss-galfit}). 
Through doing this, we also reject AGNs which cause strong contamination to the host-galaxy light profiles as we do not take objects with extremely large $n$. In this step, an additional $\approx 10\%$ of X-ray detected galaxies are removed.
We calculate \sigmaone\ values for the selected galaxies assuming a constant \mstar-to-light ratio throughout the galaxy, with \mstar\ measured in Section~\ref{ss-mstarsfr}, and $I(r)$ measured in Section~\ref{ss-galfit}:
\begin{equation} 
\Sigma_{1} = \frac{\int_{0}^{1~\mathrm{kpc}}I(r)2\pi rdr}{\int_{0}^{\infty}I(r)2\pi rdr}\frac{M_{\star}}{\pi (1~\mathrm{kpc})^{2}}.
\label{eq:sigmaone}
\end{equation}
When assuming a constant \mstar-to-light ratio throughout the galaxy, we are actually assuming a rather homogeneous stellar population constitution across the whole galaxy. As discussed in \citet{Whitaker2017} and references therein, \mstar\ profiles typically follow the rest-frame optical light profiles well, though they are more centrally concentrated in general. Thus, \sigmaone\ may be underestimated when we use Equation~\ref{eq:sigmaone} to perform the extrapolation. We have compared the \sigmaone\ values in \citet{Ni2019} (which are also measured utilizing Equation~\ref{eq:sigmaone}, but for the CANDELS fields) with the \sigmaone\ values reported in \citet{Barro2017} that are derived from spatially-resolved SED fitting with multi-band \textit{HST} light profiles. The extrapolated \sigmaone\ values of SF galaxies (which are the objects of study in this work) are systematically smaller by $\approx 0.15$ dex than the \sigmaone\ values measured in \citet{Barro2017}, with a scatter of $\approx 0.3$ dex. When we limit the comparison to X-ray detected galaxies, the offset and scatter are similar. We also note that the offset and scatter do not vary significantly with SFR or \lx\ among SF galaxies. This indicates that our assumption of a constant \mstar-to-light ratio roughly holds.

\begin{figure}
\begin{center}
\includegraphics[scale=0.6]{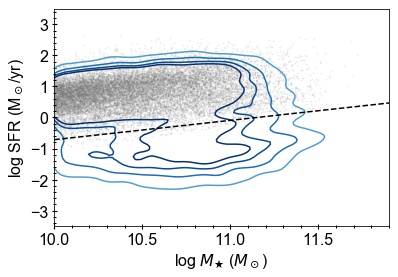}
\caption{Selected SF galaxies (gray dots) in the SFR vs. $M_\star$ plane. The contours encircle 68\%, 80\%, 90\%, and 95\% of $I_{\rm 814W} < 24$  massive (log~\mstar\ > 10) galaxies at $z < 1.2$ in the COSMOS field. The black dashed line shows the division between SF galaxies and quiescent galaxies at $z = 0.6$.}
\label{sfg}
\end{center}
\end{figure}

We use the star formation main sequence derived in \citet{Whitaker2012} at the appropriate redshift to select SF galaxies: if the SFR value of a galaxy is above the star formation main sequence or no more than 1.4~dex below the star formation main sequence, we classify this galaxy as a SF galaxy. This division roughly corresponds to galaxies lying above
the local minimum in the distribution of SFRs at a given \mstar\ (see Figure~\ref{sfg}).

\begin{figure}
\begin{center}
\includegraphics[scale=0.6]{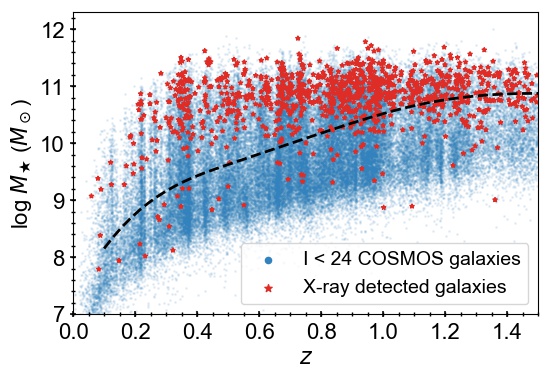}
\caption{$M_\star$ as a function of redshift. The background blue dots depict all $I_{\rm 814W} < 24$ galaxies in the COSMOS field. The red stars represent X-ray detected sources. The dashed curve indicates the $M_\star$ completeness limit as a function of redshift.}
\label{mz}
\end{center}
\end{figure}

We construct a SF Non-BD sample and a SF BD sample to study the role of \sigmaone\ in predicting BH growth when controlling for morphology and  \mstar\ (or SFR).
The SF Non-BD sample will be used in Section~\ref{ss-sfnonbd} to assess if the \bhar-\sigmaone\ relation is more fundamental than the \bhar-\mstar\ relation. 
As the relation between \bhar\ and \mstar\ or \sigmaone\ has cosmic evolution 
\citep[e.g.][]{Mullaney2012,Yang2018a, Ni2019}, we require that the SF Non-BD sample is mass-complete and has a uniform mass cut across the entire probed redshift range, so that the probed relation will not be significantly affected by the cosmic evolution.
The \mstar\ completeness curve as a function of redshift for $I_{\rm F814W} < 24$ COSMOS galaxies is shown in Figure~\ref{mz}. 
The limiting \mstar\ is derived following Section~3.2 of \citet{Ilbert2013} and Section~2.4.1 of \citet{Ni2019}. 
By selecting log \mstar\ $>$~10.2 SF Non-BD galaxies at $z<$~0.8 (log \mstar\ $=$ 10.2 is the limiting \mstar\ at $z = 0.8$), we constitute the SF Non-BD sample with a sample size of $\approx 6000$, six times that in \citet{Ni2019} in similar \mstar\ and $z$ ranges. 
We note that $\approx 78\%$ of total Non-BD galaxies in the same \mstar\ and $z$ ranges are SF Non-BD galaxies. Thus, studying the relations between BH growth and various host-galaxy properties in the SF Non-BD sample can help us investigate BH-galaxy co-evolution in the majority of the Non-BD population.

The SF BD sample will be used in Section~\ref{ss-sfbd} to test if the \bhar-\sigmaone\ relation exists when controlling for SFR. 
According to \citet{Yang2019}, \bhar\ among BD galaxies follows a linear relation with SFR in the log-log space with no obvious additional dependence on \mstar, and no evident cosmic evolution is found for this relation.
Thus, a mass-complete sample of SF BD galaxies or a sample in a narrow redshift bin is not necessary to test if the conclusion of \citet{Yang2019} holds true, or if \sigmaone\ is indeed playing an important role in predicting the amount of BH growth in this sub-population.
Therefore, we select all massive (log \mstar\ $>$ 10) SF BD galaxies at $z < 1.2$ to constitute the SF BD sample, which gives us a large sample of $\approx 1000$ galaxies, three times that in \citet{Ni2019} in the similar redshift range. 
We note that while galaxies in the SF BD sample only make up $\approx 20\%$ of total BD galaxies in the same \mstar\ and $z$ ranges, $\approx$ 76\% of the BH growth takes place within these $\approx 20\%$ of objects (we estimate the amount of BH growth as described in Section~\ref{ss-bhar}), which makes characterizing the relation between BH growth and host-galaxy properties particularly important for this subsample.

\begin{figure*}
\begin{center}
\includegraphics[scale=0.47]{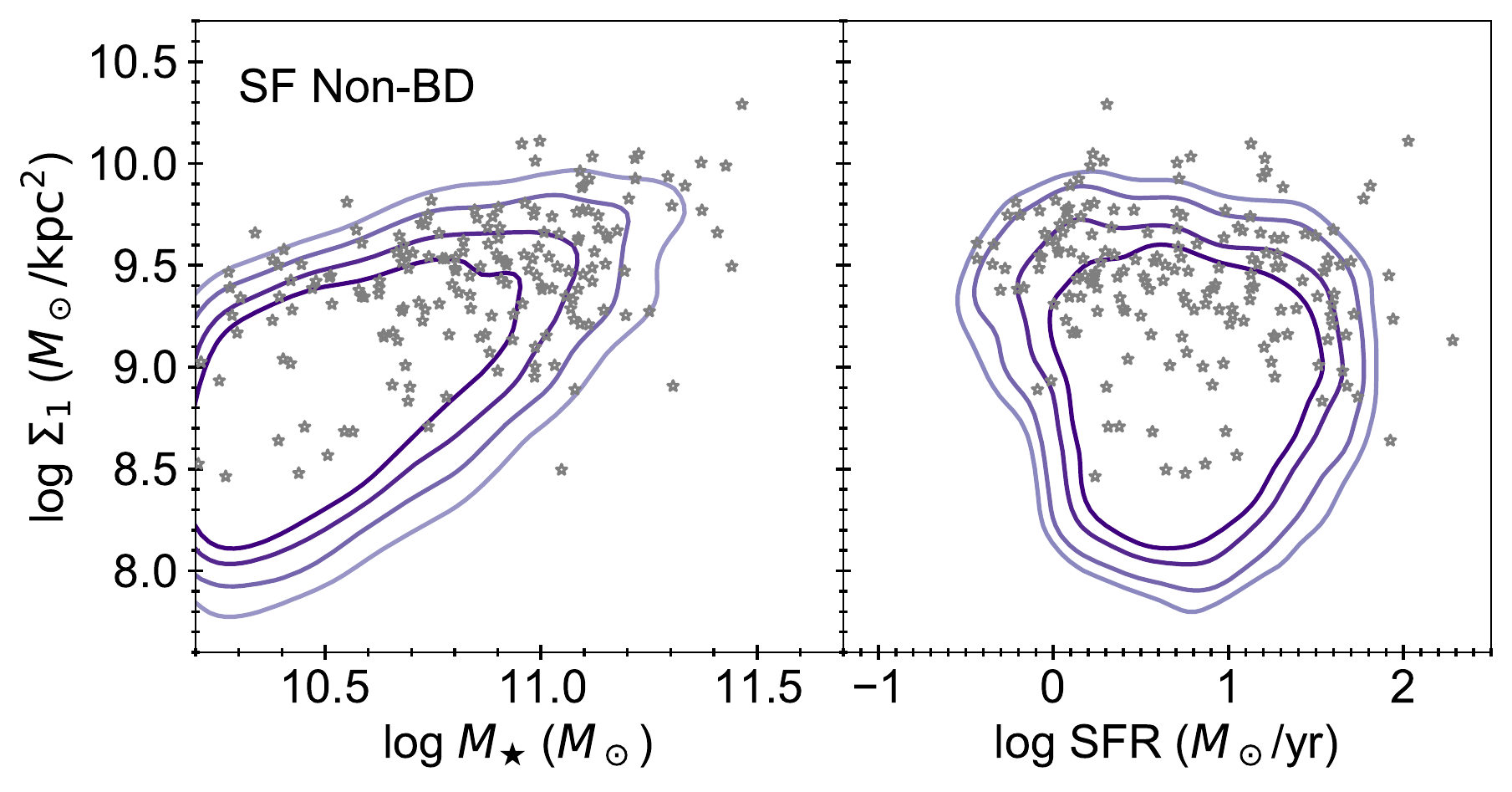}
~
\includegraphics[scale=0.47]{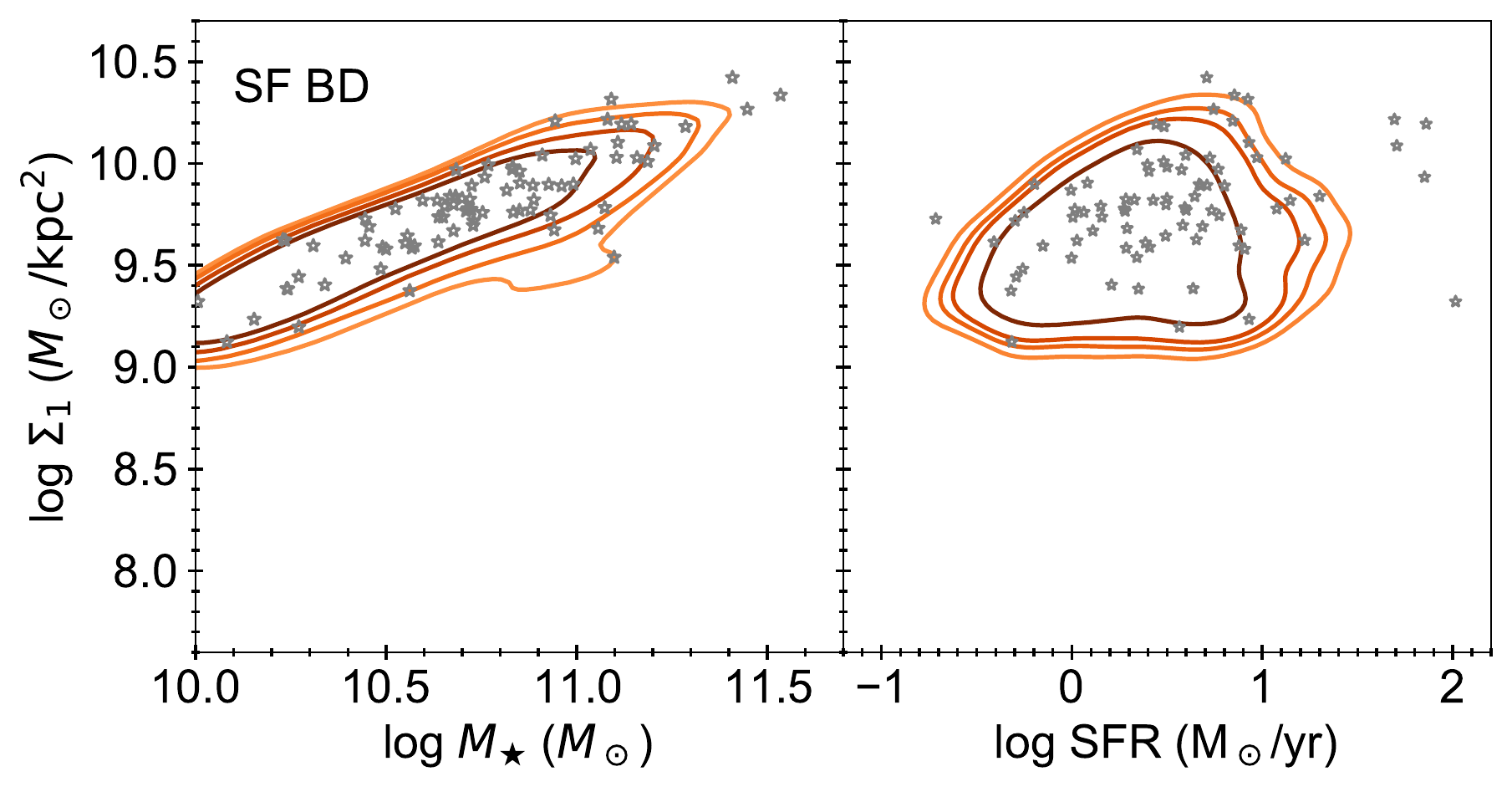}
\caption{\textit{Left panel:} \sigmaone\ vs. \mstar\ and \sigmaone\ vs. SFR for galaxies in the SF Non-BD sample. The contours encircle 68\%, 80\%,  90\%, and 95\% of galaxies. The silver stars represent X-ray detected galaxies. \textit{Right panel:} Similar to the left panel, but for galaxies in the SF BD sample.}
\label{msigma1}
\end{center}
\end{figure*}

The properties of the SF Non-BD sample and the SF BD sample are shown in Table~1.
In Figure~\ref{msigma1}, we show the \sigmaone\ vs. \mstar\ and \sigmaone\ vs. SFR distributions for the SF Non-BD sample and the SF BD sample, demonstrating the parameter space probed in this work.

We also construct a sample of SF galaxies to study the properties of the \hbox{\bhar-\sigmaone} relation regardless of morphology in Section~\ref{ss-allsf}.
This sample (we call it the ALL SF sample in short hereafter) is a mass-complete sample with a sample size of $\approx 6300$, constituted by all SF galaxies with log~\mstar~$>$~10.2 at $z <$~0.8. The properties of the ALL SF sample are also listed in Table~1.

\begin{table*}
 \begin{center}
 \caption{Summary of sample properties. (1) Name of the sample. (2) Redshift range of the sample. (3) \mstar\ range of the sample. (4) Number of galaxies in the sample. (5) Number of spec-$z$/photo-$z$ sources. (6) Number of X-ray detected galaxies.}
  \begin{tabular}{ccccccccccc}
  \hline\hline
Sample &  Redshift  &  Mass   & Number of   & Number of            & Number of \\
Name   &   Range    &  Range  &  Galaxies   & Spec-$z$/Photo-$z$   & X-ray Detections\\
(1)    &  (2)       &  (3)    &  (4)        & (5)                  & (6)    \\
\hline
SF Non-BD &  0--0.8    & log\mstar\ $>$ 10.2  & 5979 & 3823/2156 & 179 \\ \vspace{0.1 cm}
SF BD     &  0--1.2    & log\mstar\ $>$ 10    & 1020  & 421/599   & 81 \\ 
ALL SF     &  0–0.8    & log\mstar\ $>$ 10.2  & 6334  & 4041/2293    & 206\\

\hline
  \end{tabular}
  \end{center}
  \label{ts}
\end{table*}

\subsection{Sample-averaged black-hole accretion rate} \label{ss-bhar}

Following \citet{Yang2018b} and \citet{Ni2019}, we calculate \bhar\ for a given sample of galaxies sharing similar properties with contributions from both X-ray detected sources and X-ray undetected sources to cover all BH accretion, thereby estimating the \textit{long-term average BH growth} (see Section~\ref{s-intro}).

The X-ray fluxes of detected sources are adopted from the COSMOS-Legacy X-ray survey catalog \citep{Civano2016}, which is obtained from deep \textit{Chandra} observations in the field. 
We convert the X-ray fluxes (following the preference order of hard band, full band, and soft band, thus minimizing the effects of X-ray obscuration) to \lx\ assuming a power-law model with Galactic absorption and $\Gamma = 1.7$ \citep[e.g.][]{Marchesi2016b,Yang2016}. 
As discussed in \citet{Yang2018b}, the underestimation of X-ray flux due to obscuration in this scheme is small on average ($\approx 20\%$). We account for this systematic effect of obscuration by increasing the X-ray fluxes of detected sources by 20\%, following \citet{Yang2019} and \citet{Ni2019}.
The \xray\ emission of a group of X-ray undetected sources is taken into account via X-ray stacking techniques using the full-band {\it Chandra} X-ray image. Details of this stacking process can be seen in section~2.4.2 of \citet{Yang2018b}.
 
Following section~2.3 of \citet{Ni2019}, the average AGN bolometric luminosity ($\overline{L_{\rm bol}}$) for a given sample can be calculated from \lx\ of each X-ray detected source and the average X-ray luminosity of all the X-ray undetected sources ($\overline{L_{\rm X,stack}}$) obtained via stacking, assuming the \lx-dependent bolometric correction from \citet{Hopkins2007}. 
We also subtract the contributions from \xray\ binaries (XRBs) from \lx\ and $\overline{L_{\rm X,stack}}$ before applying the bolometric correction.
The XRB luminosity ($L_{\rm X,XRB}$) can be estimated through a redshift-dependent function of $M_\star$ and SFR (model 269, \citealt{Fragos2013}), which is derived utilizing observations in \citet{Lehmer2016}.\footnote{For the subsamples utilized in this work, the contribution from XRBs makes up $\approx 1$--10\% of the total X-ray emission, so that our analyses should not be affected materially by uncertainties related to the XRB modeling.} 
The equation for calculating $\overline{L_{\rm bol}}$ is

\begin{align}\label{equ:lx}
 \overline{L_{\rm bol}}  = \frac{ \bigg[{\mathlarger{\sum\limits_{n=0}^{N_{\rm det}}}} (L_{\rm X} -L_{\rm X,XRB}) k_{\rm bol}\bigg]+
 		(\overline{L_{\rm X, stack}} -  \overline{L_{\rm X, XRB}} )N_{\rm non}  \overline{k_{\rm bol}}}
		{N_{\rm det}+N_{\rm non}},
\end{align}
where $N_{\rm det}$ ($N_{\rm non}$) represents the number of X-ray detected (undetected) galaxies; $L_{\rm X,XRB}$ ($\overline{L_{\rm X, XRB}}$) is the expected XRB luminosity in each individual X-ray detected galaxy (the average XRB luminosity expected for all X-ray undetected galaxies); $k_{\rm bol}$ ($\overline{k_{\rm bol}}$) is the \lx-dependent bolometric correction applied to each individual X-ray detected galaxy (all X-ray undetected galaxies) calculated from \lx\ $-$ $L_{\rm X,XRB}$ for this object ($\overline{L_{\rm X, stack}} -  \overline{L_{\rm X, XRB}}$ of all \xray\ undetected galaxies). In this equation, X-ray detected sources contribute most of the numerator (i.e. the total $L_{\rm bol}$ of the sample); X-ray undetected sources mainly contribute to the denominator by $N_{\rm non}$ (assuming ergodic BH growth, averaging the total $L_{\rm bol}$ over the whole sample is equivalent to averaging the total $L_{\rm bol}$ over the whole duty cycle).
Then, $\overline{L_{\rm bol}}$ can be converted to \bhar\ adopting a constant radiative efficiency of 0.1:\footnote{Though it has been argued that for BHs accreting at low Eddington ratios or extremely high Eddington ratios, $\epsilon$ can be much smaller than 0.1 \citep[e.g.][]{Abramowicz2013,YN2014}, observational constraints suggest that $\epsilon \gtrsim 0.1$ holds for most of cosmic BH growth \citep[e.g.][]{Brandt2015,Shankar2020}.}

\begin{equation}\label{equ:bhar}
\begin{split}
\overline{\mathrm{BHAR}} &= \frac{(1-\epsilon)
			\overline{L_{\rm bol}}}{\epsilon c^2} \\
	 &= \frac{1.58 \overline{L_{\rm bol}}}{10^{46}\ \rm{erg~s^{-1}}}
	    M_{\sun}\ \mathrm{yr}^{-1}
\end{split}
\end{equation}
The uncertainty of \bhar\ can be obtained via bootstrapping the sample (i.e. randomly drawing the same number of objects from the sample with replacement) 1000 times. For each bootstrapped sample, \bhar\ is calculated, and the 16th and 84th percentiles of the obtained \bhar\ distribution give the estimation of the 1$\sigma$ uncertainty associated with \bhar\ of the sample.

\section{Analyses and results} \label{s-ar}

In Section~\ref{ss-sfnonbd}, we will study if the \bhar-\sigmaone\ relation is a more fundamental relation than the \bhar-\mstar\ relation among SF Non-BD galaxies. 
In Section~\ref{ss-sfbd}, we will study if the \bhar-\sigmaone\ relation also exists among SF BD galaxies.
In Section~\ref{ss-allsf}, we will first study if the \bhar-\sigmaone\ relation among  SF Non-BD galaxies is the same \bhar-\sigmaone\ relation among SF BD galaxies, and if the \bhar-\sigmaone\ relation that could apply to all SF galaxies seamlessly is a more fundamental relation than either of the \bhar-\mstar\ or \bhar-SFR relations. We will then study the properties of the \bhar-\sigmaone\ relation and its cosmic evolution.

\subsection{A \bhar-\sigmaone\ relation that is more fundamental than the \bhar-\mstar\ relation among SF Non-BD galaxies} \label{ss-sfnonbd}

\citet{Ni2019} found that the \bhar-\sigmaone\ relation among SF Non-BD galaxies is not likely to be a secondary manifestation of the \hbox{\bhar-\mstar} relation, and it is plausible that the \bhar-\sigmaone\ relation is indeed more fundamental than the \bhar-\mstar\ relation.
In this section, we test the significance of the \bhar-\sigmaone\ (\bhar-\mstar) relation when controlling for \mstar\ (\sigmaone) among galaxies in the SF Non-BD sample with partial-correlation (PCOR) analyses. 
If we find a significant \bhar-\sigmaone\ relation when controlling for \mstar\ but do not find a significant \bhar-\mstar\ relation when controlling for \sigmaone, we can conclude that the \bhar-\sigmaone\ relation is more fundamental than the \bhar-\mstar\ relation among SF Non-BD galaxies.

\begin{figure}
\begin{center}
\includegraphics[scale=0.5]{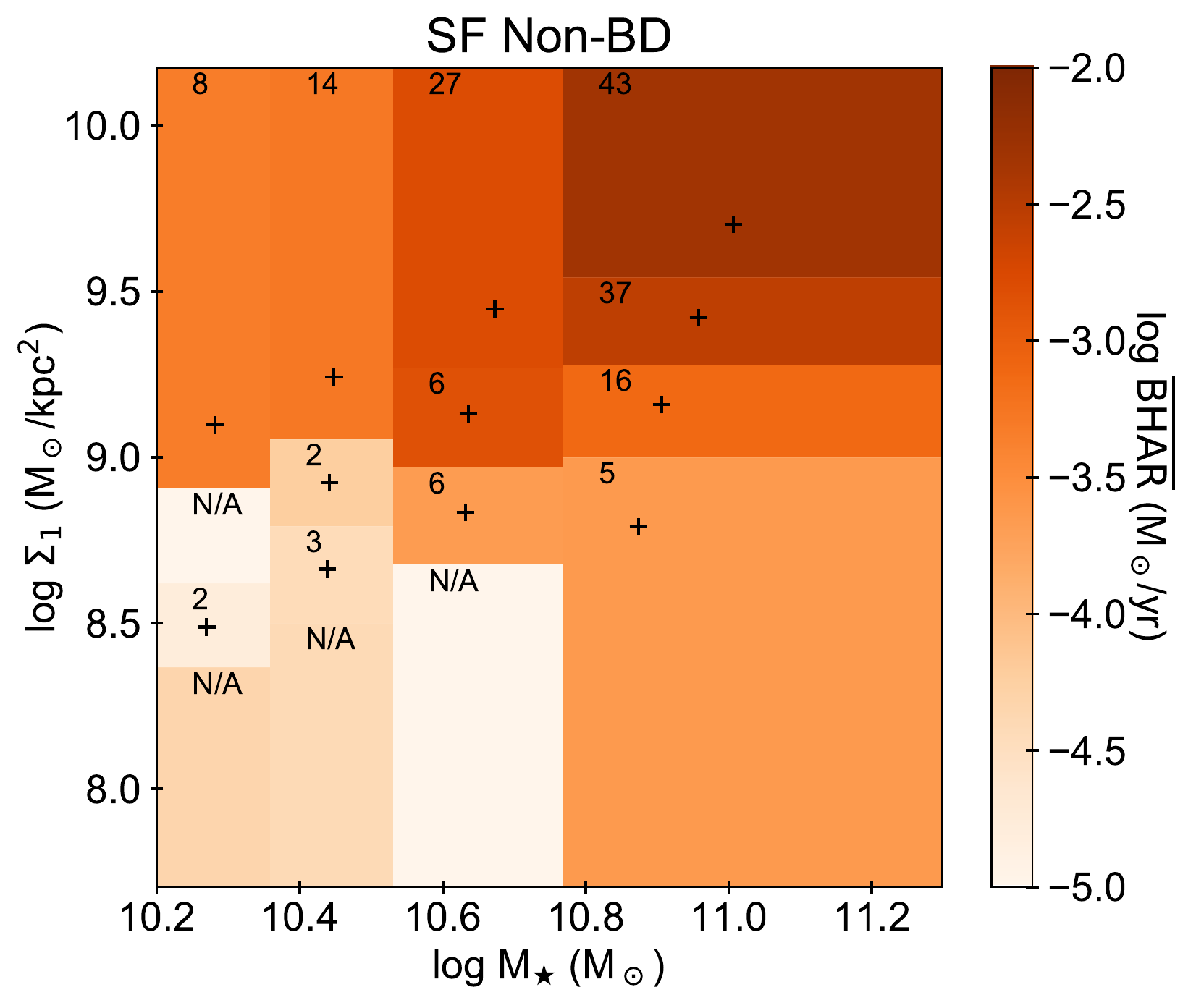}
\caption{Color-coded $\rm \overline{BHAR}$ in different bins of \mstar\ and \sigmaone\ for galaxies in the SF Non-BD sample. Each 2D bin contains $\approx$ 370 sources. The black plus sign indicates the median \mstar\ and \sigmaone\ of the sources in each bin. 
For each bin, the number of X-ray detected galaxies is listed.
For bins where \bhar\ does not have a lower limit $>0$ from bootstrapping or the number of X-ray detected galaxies is less than 2, `N/A' is shown instead.
}
\label{sfnonbd_2d}
\end{center}
\end{figure}

We bin galaxies in the SF Non-BD sample based on both \mstar\ and \sigmaone\ and calculate \bhar\ for each bin. 
The bins are chosen to include approximately the same numbers of sources ($\approx 370$; see Figure~\ref{sfnonbd_2d} for the 2D bins).
Bins where \bhar\ does not have a lower limit $>$ 0 or the number of X-ray detected galaxies is less than 2 (which will introduce large uncertainty into the estimated \bhar) will be excluded from the PCOR analyses.
We input the median log \mstar, median log \sigmaone, and log \bhar\ of valid bins to \texttt{PCOR.R} in the \texttt{R} statistical package \citep{Kim2015}, and the significance levels of the \bhar-\sigmaone\ (\bhar-\mstar) relation when controlling for \mstar\ (\sigmaone) with both the Pearson and Spearman statistics are calculated. 
The PCOR test results are summarized in Table~\ref{sfnonbd-pcor}.
The parametric Pearson statistic is used to select significant results (we note that both the \bhar-\mstar\ and \bhar-\sigmaone\ relations are roughly linear in log-log space; see \citet{Yang2019} and \citet{Ni2019} for details), and the nonparametric Spearman statistic is also listed for reference.
We can see from Table~\ref{sfnonbd-pcor} that the \hbox{\bhar-\sigmaone} relation turns out to be more fundamental than the \bhar-\mstar\ relation among SF Non-BD galaxies. Our results do not qualitatively change with different binning approaches (see Appendix~\ref{a-bin} for details).

\begin{table}
\begin{center}
\caption{$p$-values (significances) of partial correlation analyses for the SF Non-BD sample}
\label{sfnonbd-pcor}
\begin{tabular}{ccccc}
\multicolumn{3}{c}{SF Non-BD} \\ \hline \hline
Relation &   Pearson & Spearman \\\hline
\bhar-\sigmaone\            & $\boldsymbol {2 \times 10^{-5}~(3.8\sigma)}$  & $\boldsymbol {2 \times 10^{-5}~(3.7\sigma)}$ &  \\
\bhar-$M_\star$               &  $0.12~(1.6\sigma)$  & $0.03~(2.1\sigma)$ \\
 \hline\hline
\end{tabular}                                         
\end{center}
\end{table}

\subsection{The existence of a \bhar-\sigmaone\ relation among SF BD galaxies} \label{ss-sfbd}
\begin{figure}
\begin{center}
\includegraphics[scale=0.5]{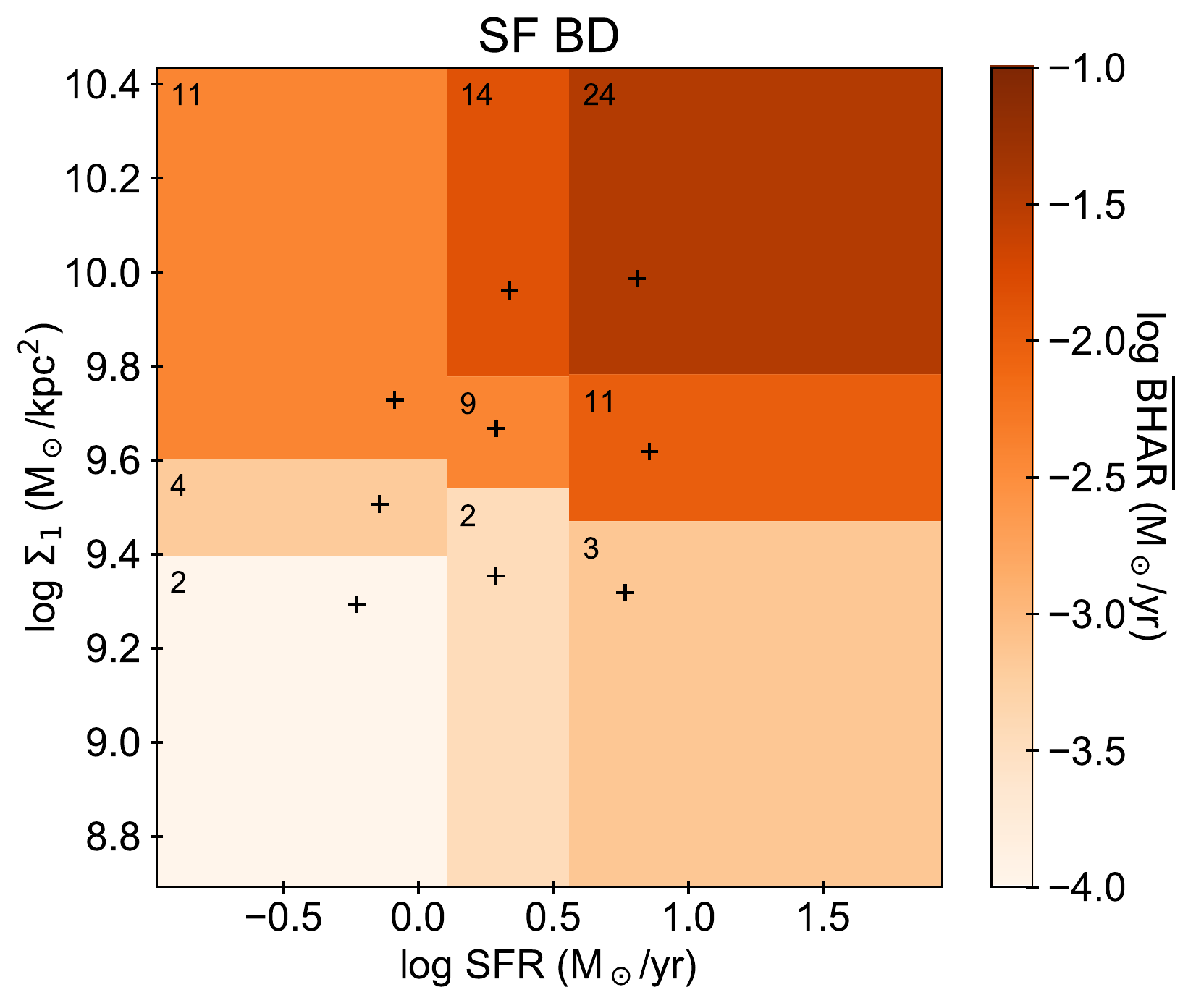}
\caption{Color-coded $\rm \overline{BHAR}$ in different bins of SFR and \sigmaone\ for galaxies in the SF BD sample. Each 2D bin contains $\approx$ 110 sources. The black plus sign indicates the median SFR and \sigmaone\ of the sources in each bin. 
For each bin, the number of X-ray detected galaxies is listed.
}
\label{sfbd_2d}
\end{center}
\end{figure}

We test the significance of the \bhar-\sigmaone\ relation when controlling for SFR among galaxies in the SF BD sample with PCOR analyses, as \citet{Yang2019} concluded that \bhar\ among BD galaxies mainly correlates with SFR.
We bin galaxies in the SF BD sample based on both SFR and \sigmaone\ and calculate \bhar\ for each bin. 
The bins are chosen to include approximately the same numbers of sources ($\approx 110$; see Figure~\ref{sfbd_2d} for the 2D bins).
We input the median log SFR, median log \sigmaone, and log \bhar\ of valid bins to \texttt{PCOR.R}, and the significance levels of the \hbox{\bhar-\sigmaone} (\hbox{\bhar-SFR}) relation when controlling for SFR (\sigmaone) with both the Pearson and Spearman statistics are calculated.
The PCOR test results are summarized in Table~\ref{sfbd-pcor}. 

From the PCOR test results, we can see that the \bhar-\sigmaone\ relation is significant when controlling for SFR, suggesting the important role of \sigmaone\ in predicting the amount of BH growth.
The \bhar-SFR relation when controlling for \sigmaone, at the same time, does not satisfy the 3$\sigma$ criterion we adopted in Secion~\ref{s-intro} for the Pearson statistic to select significant correlations (though it is marginally significant).
We note again that our results do not qualitatively change with different binning approaches (see Appendix~\ref{a-bin} for details).
We also use the PCOR analyses to assess the significance levels of the \bhar-\sigmaone\ relation when controlling for \mstar\ in a similar manner, and the results are listed in Table~\ref{sfbd-pcor}.
The \hbox{\bhar-\sigmaone} relation remains significant, demonstrating that the observed \bhar-\sigmaone\ relation in the SF BD sample is not simply a manifestation of the \bhar-\mstar\ relation.
Thus, we can conclude that the \bhar-\sigmaone\ relation exists among SF BD galaxies. We note that our findings do not challenge the existence of the \bhar-SFR relation among BD galaxies in general (see Appendix~\ref{a-allbd}).

\begin{table}
\begin{center}
\caption{$p$-values (significances) of partial correlation analyses for the SF BD sample}
\label{sfbd-pcor}
\begin{tabular}{ccccc}
\multicolumn{3}{c}{SF BD} \\ \hline \hline
Relation &   Pearson & Spearman \\\hline
\bhar-\sigmaone\            & $\boldsymbol {5\times 10^{-5}~(4.1\sigma)}$  & $\boldsymbol {3\times 10^{-4}~(3.7\sigma)}$ &  \\
\bhar-SFR              &  $4\times 10^{-3}~(2.8 \sigma)$  & $\boldsymbol {2\times 10^{-3}~(3.1\sigma)}$ \\
\hline
\bhar-\sigmaone\            & $\boldsymbol {9\times 10^{-4}~(3.3\sigma)}$   & $\boldsymbol {1\times 10^{-3}~(3.2\sigma)}$ &  \\ 
\bhar-\mstar\            &  $0.33~(1.0 \sigma)$  & $0.39~(0.9\sigma)$ \\
\hline \hline
\end{tabular}                                         
\end{center}
\end{table}

\subsection{A \bhar-\sigmaone\ relation among all SF galaxies} \label{ss-allsf}

We have confirmed the \bhar-\sigmaone\ relation in both the SF Non-BD sample (see Section~\ref{ss-sfnonbd}) and the SF BD sample (see Section~\ref{ss-sfbd}).
We will now study if the \bhar-\sigmaone\ relation among SF BD galaxies and the \bhar-\sigmaone\ relation among SF Non-BD galaxies make consistent predictions at a given \sigmaone, so that no ad hoc morphological division among SF galaxies is needed to study this relation.
As the SF BD sample and the SF Non-BD sample are selected with different \mstar\ and $z$ criteria (and only the SF Non-BD sample is a mass-complete sample), we use the 355 SF BD galaxies with log \mstar\ > 10.2 at $z < 0.8$ to perform the comparison.  
For each of these 355 galaxies, we select two galaxies from the larger SF Non-BD sample that have the closest \sigmaone\ values to it (not allowing duplications) to constitute a comparison sample.
We find that the log \bhar\ of these 355 SF BD galaxies is $-2.62^{+0.12}_{-0.16}$, and the log \bhar\ of SF Non-BD galaxies in the comparison sample is $-2.41^{+0.11}_{-0.15}$, showing the consistent predictions of the \bhar-\sigmaone\ relation among SF BD galaxies and SF Non-BD galaxies.\footnote{We do not directly derive the \bhar-\sigmaone\ relation among SF BD galaxies and SF Non-BD galaxies separately and compare, as quantifying the \bhar-\sigmaone\ relation solely among SF BD galaxies will suffer from uncertainty that is too large to conduct any meaningful comparison.}

\begin{figure}
\begin{center}
\includegraphics[scale=0.5]{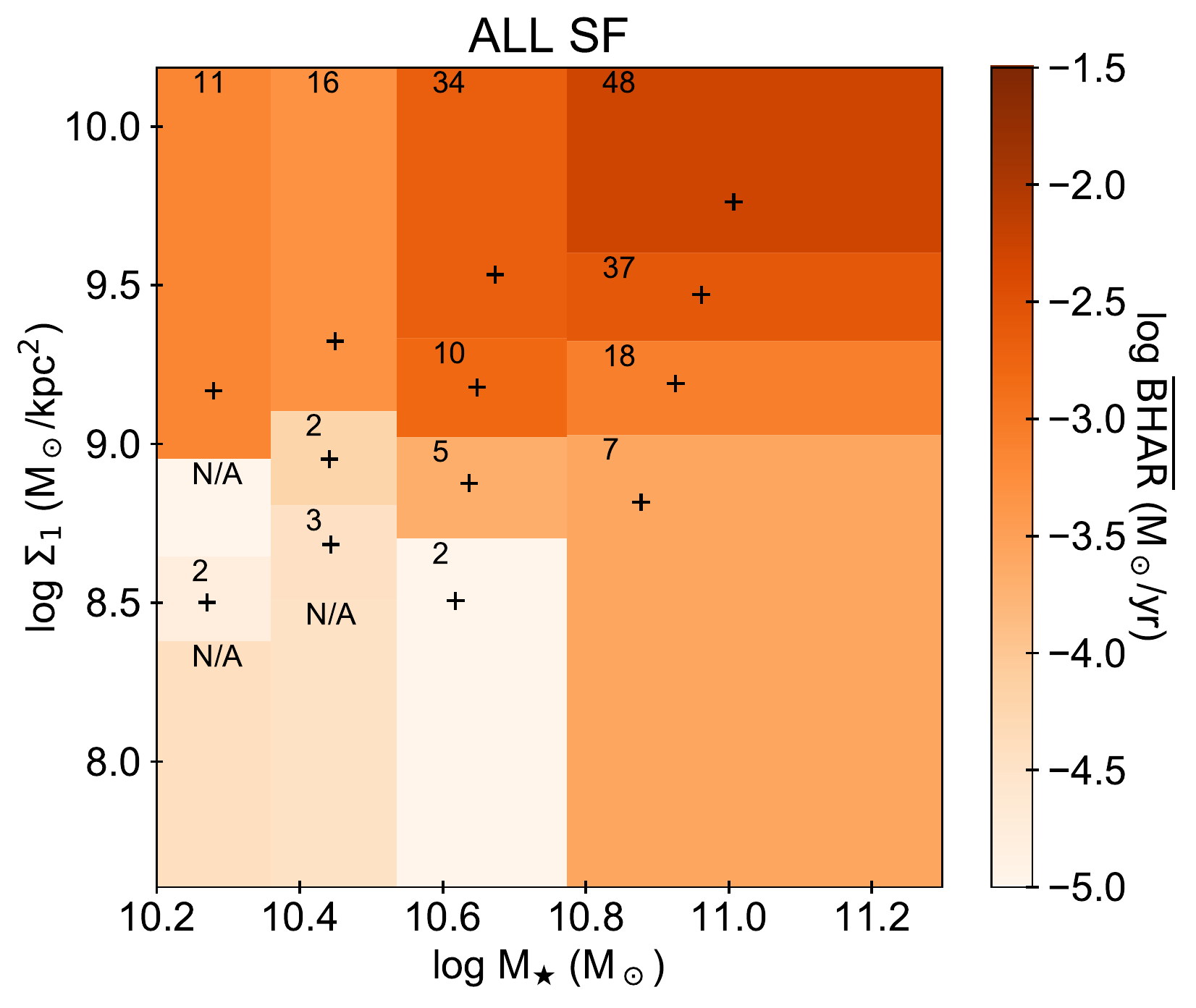}
\caption{Color-coded $\rm \overline{BHAR}$ in different bins of \mstar\ and \sigmaone\ for galaxies in the ALL SF sample. 
Each 2D bin contains $\approx$ 390 sources. 
The black plus sign indicates the median \mstar\ and \sigmaone\ of the sources in each bin. 
For each bin, the number of X-ray detected galaxies is listed.
For bins where \bhar\ does not have a lower limit $>0$ from bootstrapping or the number of X-ray detected galaxies is less than 2, `N/A' is shown instead.
}
\label{allsf_2d}
\end{center}
\end{figure}

We will now study this \bhar-\sigmaone\ relation that does not depend on morphological classes utilizing the ALL SF sample, which is constituted of all SF galaxies with log \mstar\ $>$ 10.2 at $z < 0.8$. This sample of SF galaxies is mass-complete, so the derived \bhar-\sigmaone\ relation will not be significantly affected by the cosmic evolution of this relation.
We use PCOR analyses to assess if the \bhar-\sigmaone\ relation in the ALL SF sample is still more fundamental than the \bhar-\mstar\ relation.
The 2D bins in the \sigmaone\ vs.\ \mstar\ plane that are utilized for PCOR analyses are presented in Figure~\ref{allsf_2d}.
As expected from the dominant number ($\approx 94\%$) of SF Non-BD galaxies in the sample, the \bhar-\sigmaone\ relation is significant when controlling for \mstar, and the \bhar-\mstar\ relation is not significant when controlling for \sigmaone\ (see Table~\ref{allsf}). This result does not qualitatively change with different binning approaches (see Appendix~\ref{a-bin} for details).
We also perform the PCOR analyses in a similar manner for \sigmaone\ and SFR, and it turns out that the \bhar-\sigmaone\ relation is more fundamental than the \bhar-SFR relation (see Table~\ref{allsf}).

\begin{table}
\begin{center}
\caption{$p$-values (significances) of partial correlation analyses for the ALL SF sample}
\label{allsf}
\begin{tabular}{ccccc}
\multicolumn{3}{c}{ALL SF} \\ \hline \hline
Relation &   Pearson & Spearman \\\hline
\bhar-\sigmaone\            & $\boldsymbol {2 \times 10^{-5}~(4.2\sigma)}$  & $\boldsymbol {2 \times 10^{-4}~(3.7\sigma)}$ &  \\
\bhar-$M_\star$               &  $0.28~(1.1\sigma)$  & $0.14~(1.5\sigma)$ \\ \hline
\bhar-\sigmaone\            & $\boldsymbol {6 \times 10^{-5}~(4.0\sigma)}$  & $\boldsymbol {9 \times 10^{-6}~(4.4\sigma)}$ &  \\
\bhar-SFR               &  $0.46~(0.7\sigma)$  & $0.18~(1.3\sigma)$ \\
 \hline\hline
\end{tabular}                                         
\end{center}
\end{table}

\begin{figure}
\begin{center}
\includegraphics[scale=0.45]{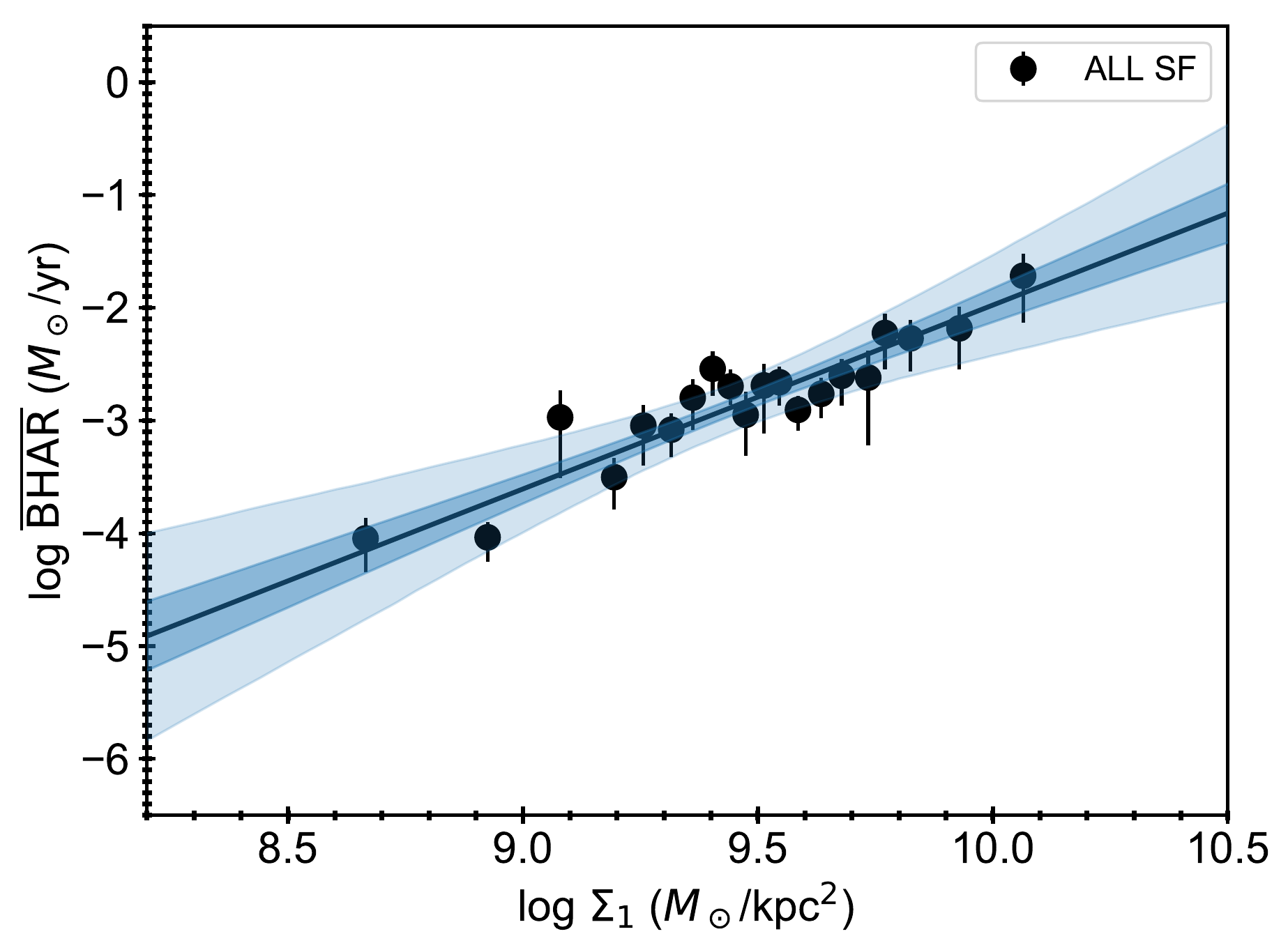}
\caption{The \bhar-\sigmaone\ relation among SF galaxies. Galaxies in the ALL SF sample are divided into bins according to their \sigmaone\ values, with $\approx$ 10 \xray\ detected galaxies in each bin. The horizontal position of each data point indicates the median \sigmaone\ of the sources in the bin; the error bars represent the 1$\sigma$ confidence interval of \bhar\ from bootstrapping. 
The black solid line and the dark/light blue shaded region represent the best-fit \bhar-\sigmaone\ relation and the 1$\sigma$/3$\sigma$ pointwise confidence intervals on the regression line.}
\label{allsf_trend}
\end{center}
\end{figure}

To study the properties of the \bhar-\sigmaone\ relation, we divide galaxies in the ALL SF sample into \sigmaone\ bins with approximately the same number of X-ray detected galaxies ($\approx 10$) per bin , and calculate \bhar\ and its 1$\sigma$ confidence interval for each bin. In Figure~\ref{allsf_trend}, we plot \bhar\ of these bins as a function of the median \sigmaone\ value of each bin.
We use the python package \texttt{emcee} \citep{emcee} to fit a log-linear model to the \bhar-\sigmaone\ relation, where the maximum-likelihood method is implemented by the Markov chain Monte Carlo Ensemble sampler. By fitting all the data points in Figure~\ref{allsf_trend}, we obtain:
\begin{equation} \label{eq-bharsigma}
 \rm{log~\overline{BHAR}} = (1.6 \pm 0.2) \times  log~\Sigma_1 + (-18.3 \pm 2.6).
 \end{equation}
The best-fit model and its 1$\sigma$/3$\sigma$ pointwise confidence intervals are also shown in Figure~\ref{allsf_trend}.

\begin{figure}
\begin{center}
\includegraphics[scale=0.45]{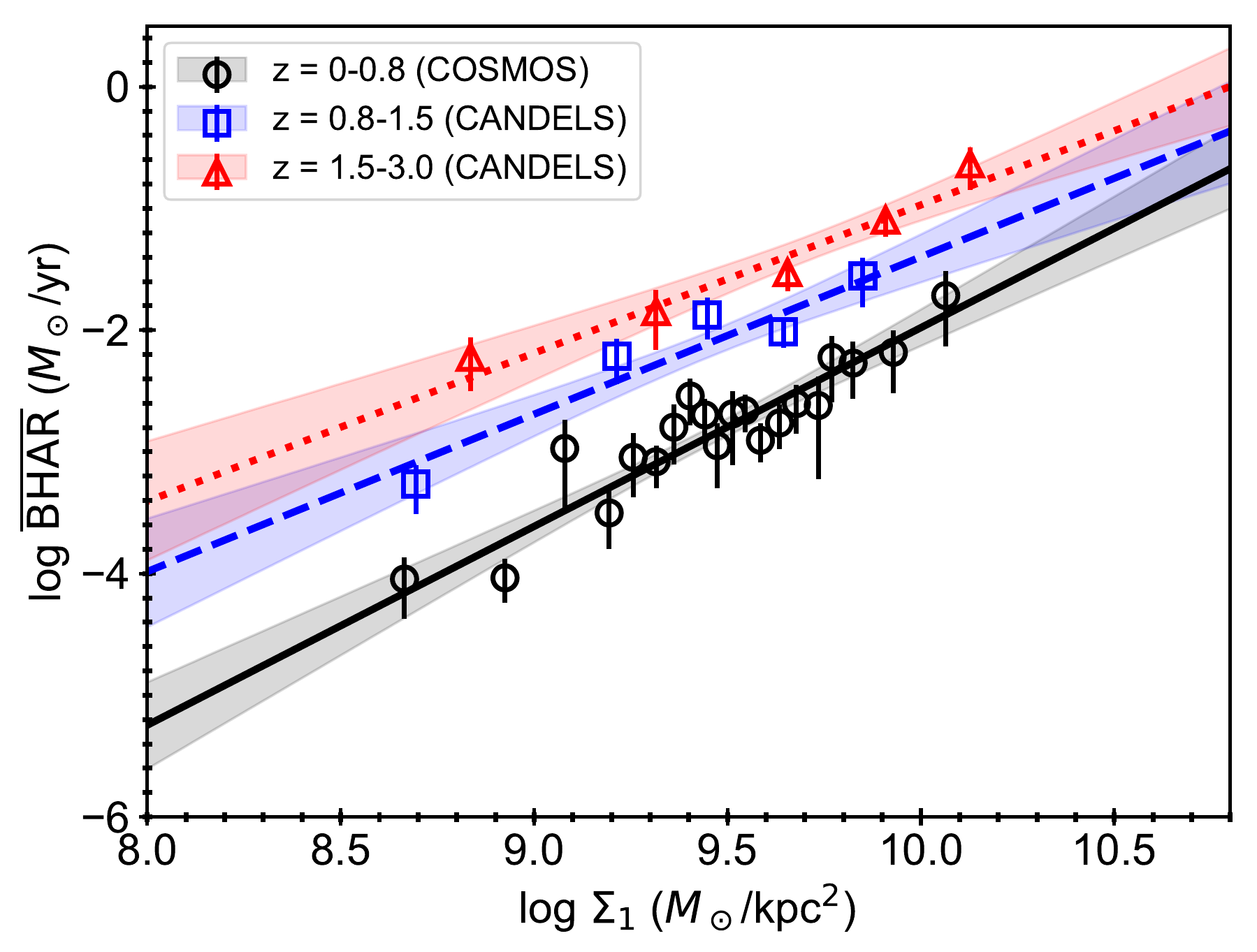}
\caption{The cosmic evolution of the \bhar-\sigmaone\ relation among SF galaxies. $z = 0$--0.8 galaxies from COSMOS/$z = 0.8$--1.5 galaxies from CANDELS/$z = 1.5$--3 galaxies from CANDELS with log \mstar\ $> 10.2$ are divided into several bins according to their \sigmaone\ values with approximately the same number of X-ray detected galaxies in each bin, represented by the black circles/blue squares/red triangles. The horizontal position of each data point indicates the median \sigmaone\ of the sources in the bin; the error bars represent the 1$\sigma$ confidence interval of \bhar\ from bootstrapping. The black solid line and the gray shaded region represent the best-fit \bhar-\sigmaone\ relation and its 1$\sigma$ pointwise confidence interval for $z = 0$--0.8 galaxies; the blue dashed line and the blue shaded region are for the $z = 0.8$--1.5 galaxies; the red dotted line and the red shaded region are for the $z = 1.5$--3 galaxies.}
\label{z_trend}
\end{center}
\end{figure}

Two subsamples of \hbox{$H_{\rm 160W} < 24.5$} SF galaxies with \hbox{log~\mstar\ $> 10.2$} in the CANDELS fields drawn from the \citet{Ni2019} sample will also be utilized to probe how the \hbox{\bhar-\sigmaone} relation evolves over the history of the Universe: one subsample is constituted of $\approx 1500$ SF galaxies at $z = 0.8$--1.5 (where \sigmaone\ values are inferred from $J_{\rm 125W}$-band light profiles), and the other subsample is constituted of $\approx 1800$ SF galaxies at $z = 1.5-$3 (where \sigmaone\ values are inferred from $H_{\rm 160W}$-band light profiles).
Though the utilized \textit{HST} bands are different, we note that the light profiles are always measured in the rest-frame optical.
\hbox{$H_{\rm 160W} < 24.5$} galaxies in the CANDELS fields are mass-complete at log~\mstar~$> 10.2$ up to $z = 3$, so these subsamples are also mass-complete samples.
For each subsample, we divide objects into \sigmaone\ bins,\footnote{As \mstar\ values utilized in \citet{Ni2019} to calculate \sigmaone\ are also measured with parametric SFHs that tend to underestimate the true \mstar\ \citep{Leja2019a}, we apply a $\approx 0.15$ dex correction to \sigmaone\ values of galaxies in the two CANDELS subsamples to maintain consistency with the \mstar\ scheme utilized in this paper.} and calculate \bhar\ and its 1$\sigma$ confidence interval for each bin. The \bhar\ values of these bins as a function of \sigmaone\ are shown in Figure~\ref{z_trend} along with the data points in Figure~\ref{allsf_trend} (which show \bhar\ as a function of \sigmaone\ in the ALL SF sample that is constituted by $z = 0$--0.8 SF galaxies in the COSMOS field).
We then use the \texttt{emcee} package to fit a log-linear model to the \bhar-\sigmaone\ relation among each subsample, as we did for the ALL SF sample.
The best-fit \bhar-\sigmaone\ relations of SF galaxies in different redshift ranges are presented together in Figure~\ref{z_trend}. 
We can see that while the slope of the best-fit log-linear model does not change significantly with redshift, for a given \sigmaone\ value, the expected \bhar\ is higher at higher redshift: \bhar\ at $z = 1.5$--3 is higher than that at $z = 0$--0.8 by $\sim$ 1~dex when controlling for \sigmaone.

\section{Discussion} \label{s-dis}
\subsection{What is implied by the apparent link between BH growth and host-galaxy compactness?} \label{ss-gas}

\subsubsection{The link between BH growth and the central gas density of host galaxies: a common origin of the gas in the vicinity of the BH and the central $\sim$~kpc?}

\begin{figure}
\begin{center}
\includegraphics[scale=0.45]{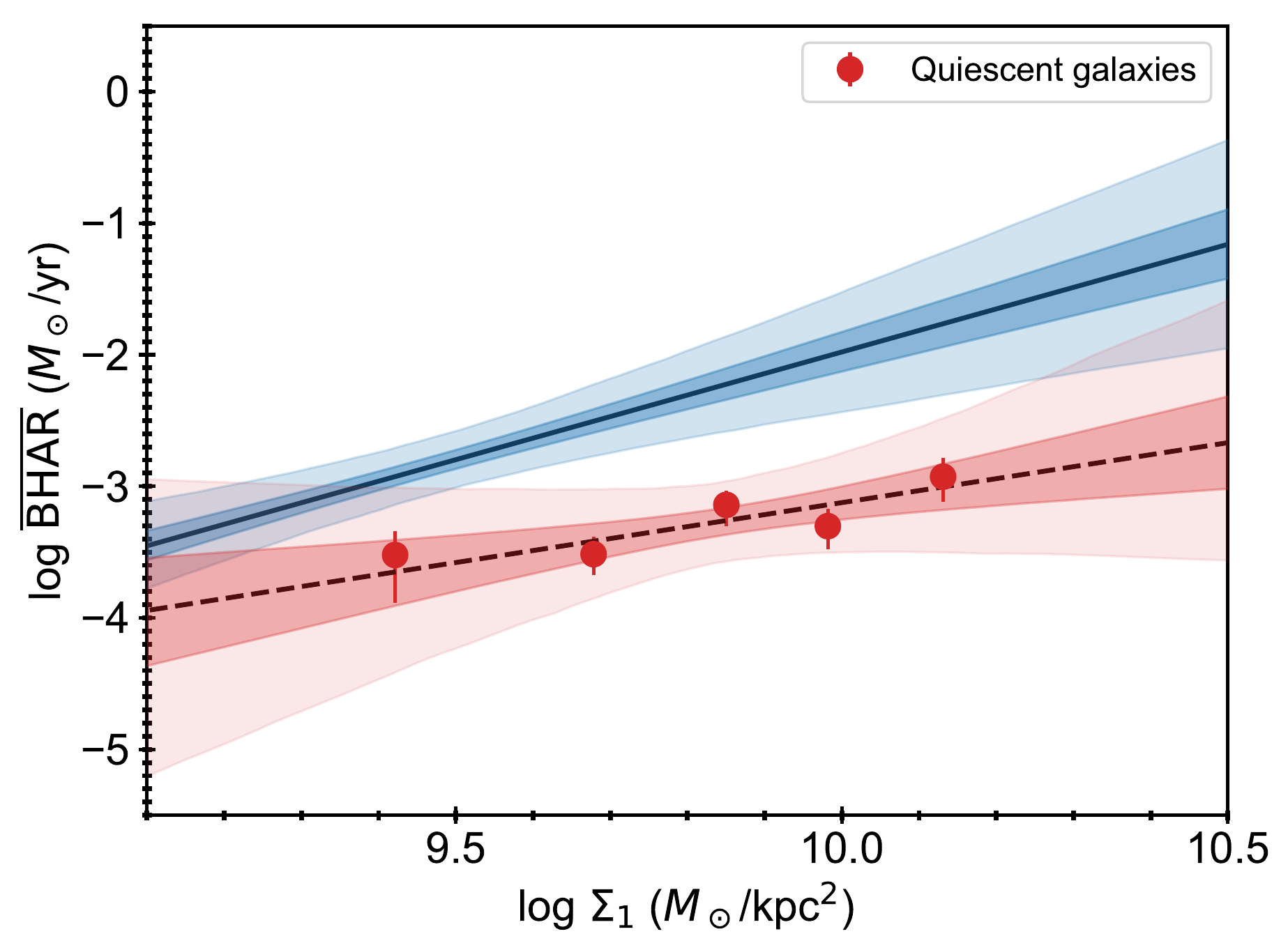}
\caption{The \bhar-\sigmaone\ relation among quiescent galaxies. 3400 quiescent galaxies with log \mstar\ $> 10.2$ at $z < 0.8$ are divided into bins according to their \sigmaone\ values, with $\approx$ 20 X-ray detected galaxies in each bin. The horizontal position of each data point indicates the median \sigmaone\ of the sources in the bin; the error bars represent the 1$\sigma$ confidence interval of \bhar\ from bootstrapping. 
The black dashed line and the dark/light red shaded region represent the best-fit \bhar-\sigmaone\ relation among quiescent galaxies and the 1$\sigma$/3$\sigma$ pointwise confidence intervals on the regression line.
The black solid line and the dark/light blue shaded region are adopted from Figure~\ref{allsf_trend}, representing the best-fit \bhar-\sigmaone\ relation among SF galaxies in the same \mstar\ and $z$ ranges and the 1$\sigma$/3$\sigma$ pointwise confidence intervals on the regression line. 
At a given \sigmaone, the \bhar\ values of quiescent galaxies are below the 3$\sigma$ lower limit of the best-fit \bhar-\sigmaone\ relation among SF galaxies, and the best-fit slope of the log~\bhar-log~\sigmaone\ relation among quiescent galaxies is much flatter compared with the best-fit slope among SF galaxies.}
\label{q_trend}
\end{center}
\end{figure}

In Section~\ref{s-ar}, we confirmed a \bhar-\sigmaone\ relation that is more fundamental than either the \bhar-\mstar\ or \bhar-SFR relation among SF galaxies, which reveals the link between long-term average BH growth and host-galaxy compactness.
This \bhar-\sigmaone\ relation is only significant among SF galaxies \citep{Ni2019}. If we plot \bhar\ as a function of \sigmaone\ for quiescent galaxies (see Figure~\ref{q_trend}), we can see that \bhar\ does not vary significantly with \sigmaone\ among quiescent galaxies, and the fitted slope ($0.9 \pm 0.5$) of the log~\bhar-log~\sigmaone\ relation among quiescent galaxies is flatter compared with the slope among SF galaxies,  being consistent with zero at a $\approx 2\sigma$ level.
This led us to speculate that the \hbox{\bhar-\sigmaone} relation reflects a link between BH growth and the central gas density (on the $\sim$~kpc scale) of host galaxies (among quiescent galaxies, \sigmaone\ cannot effectively trace the central gas density).\footnote{We note that there is still limited SF activity among the quiescent galaxies we selected (see Section~\ref{ss-sample}), so that there may still be a shallow trend between \sigmaone\ and the central gas density (though with large scatter), which could explain the observed shallow slope of the log~\bhar-log~\sigmaone\ relation in Figure~\ref{q_trend}.}
The observed cosmic evolution of the \bhar-\sigmaone\ relation in Section~\ref{ss-allsf} supports our speculation: observations show that the average (molecular) gas fraction among galaxies increases by a factor of $\sim$~10 from $z \approx 0.4$ to $z \approx 2$ \citep[e.g.][]{Schinnerer2016,Tacconi2018}, and this could well-explain our observed result that for a given \sigmaone, \bhar\ increases by a factor of $\sim$ 10 from $z = 0-$0.8 to $z = 1.5-$3.

If we can approximate \sigmaone\ as a function of gas surface density (\sigmagas) in the central $\sim$ kpc of galaxies, we will be able to convert the \bhar-\sigmaone\ relation to a \bhar-\sigmagas\ relation. 
According to the Kennicutt-Schmidt law \citep[e.g.][]{Kennicutt1998}, the SFR surface density ($\Sigma_{\rm SFR}$) is tightly linked with \sigmagas\ with a power-law index $\approx 1.4 \pm 0.15$. 
Also, observations and simulations suggest that $\Sigma_{\rm SFR}$ on the $\sim$~kpc scale correlates with \mstar\ density ($\Sigma_{M_\bigstar}$) on the same scale in SF regions \citep[e.g.][]{Cano2016,Hsieh2017,Trayford2019,Hani2020}, though the reported slope values ($\beta$) of the log~$\Sigma_{\rm SFR}$-log~$\Sigma_{M_\bigstar}$ relation vary from $\approx 0.7$ to $\approx 1$. 
Given all these findings, we can approximate \sigmaone\ as a power-law function of \sigmagas\ in the central $\sim$~1~kpc of galaxy with an index of $\approx 1.4/\beta$, suggesting that the \bhar-\sigmagas\ relation has a power-law index of $\sim 2-$4.
Further studies that utilize high-resolution ALMA observations to resolve the gas density will help to quantify directly the relation between \sigmaone\ and \sigmagas, thus making the conversion from the \bhar-\sigmaone\ relation to a \bhar-\sigmagas\ relation more reliable; with ALMA observations of a sample of 32 galaxies at $z \lesssim 0.1$, a tight relation between \sigmagas\ and $\Sigma_{M_\bigstar}$ has already been suggested in \citet{Lin2019}, and a larger sample size is needed for further quantification of this relation.
Alternatively, future accumulation of ALMA observations in combination with deep \xray\ observations will enable us to probe the \bhar-\sigmagas\ relation directly.

Assuming \sigmaone\ serves as an indicator of \sigmagas, the \hbox{\bhar-\sigmaone} relation may indicate that gas in the vicinity of the BH that will be accreted has the same origin as gas in the central $\sim$ kpc part of galaxies.
It is plausible that gas could be transported from the inner $\approx 1$ kpc of galaxies all the way to the torus and accretion disk via gravitational instabilities (see \citealt{SB2019} and references therein).
If \sigmagas\ (on kpc scales) correlates with the ambient gas density ($\rho$) of BHs (on pc to sub-kpc scales) well, the relation between BH growth and $\rho$ may be quantitatively examined.
However, while we can convert the \hbox{\bhar-\sigmaone} relation to a \bhar-\sigmagas\ relation, this does not necessarily mean the dependence of BH growth on $\rho$ can be directly inferred.

BH growth may depend on other factors that also correlate with \sigmaone.
Bondi-type accretion models \citep[e.g.][]{Bondi1952,Springel2005} predict that the amount of BH growth should be approximately proportional to 
 $M_{\rm BH}^2 \rho c_s^{-3}$ (assuming that the gas has negligible velocity relative to the BH as an initial condition; $c_s$ is the sound speed in the gas), and both \mbh\ and $c_s$ correlate with \sigmaone\ (though with considerable scatters).\footnote{While we have necessarily assumed zero angular momentum here with the Bondi-type accretion model (that has been suggested to be a reliable approximation of BH accretion), we note that in real cases, the accretion of gas with significant angular momentum onto the BH is far more complicated than the simple picture proposed here.} 
We can roughly infer the correlation between \sigmaone\ and \mbh\ from the \mbh-\mstar\ relation and the \sigmaone-\mstar\ relation among the general galaxy population.
The power-law index of the \mbh-\mstar\ relation observed in the local universe is $\approx 1.05$ (e.g.\ \citealt{MM2013,Reines2015}); we note that this relation is not very tight, with a scatter of $\approx 0.55$ dex. Also, this \mbh-\mstar\ relation does not seem to have significant cosmic evolution at $z < 2$ \citep[e.g.][]{KH2013,Sun2015,Ding2020,Suh2020}.
The power-law index of the \sigmaone-\mstar\ relation is $\approx 1.1$ in the ALL SF sample.
We thus infer that \sigmaone\ can be expressed as a power-law function of \mbh\ with an index close to one (though with a considerable scatter). 
As $c_s^2$ scales with the temperature of the medium, it should also scale with $M/R$ assuming virial equilibrium, where $M$ and $R$ are the mass and radius of the gravitationally bound system in the vicinity of the BH. 
It has been suggested that $M/R$ scales with \mstar/($K_v(n) \times$\re) (see Section~1 of \citealt{Taylor2010}), where
\begin{equation}
 K_v(n) \cong \frac{73.32}{10.465+(n-0.94)^2}+0.954.
 \end{equation}
Through fitting objects in our ALL SF sample, we found that \sigmaone\ could be expressed as a power-law function of \mstar/($K_v(n) \times$\re) with an index of $\sim 1.3$. Utilizing this conversion, $c_s^{-3}$ should be proportional to $\sim$ \sigmaone$^{-1.1}$ (as $c_s^2$ is proportional to $M/R$ when assuming virial equilibrium, and $M/R$ scales with \mstar/($K_v(n) \times$\re)).
Thus, if Bondi-type accretion models can well approximate BH growth among SF galaxies, and if \sigmagas\ correlates well with $\rho$, we should observe a \bhar-\sigmaone\ relation with an index of $\sim$ 1.4--1.6 (as \mbh$^2$ $\propto$ \sigmaone$^2$, $\rho$ $\propto$ \sigmaone$^{\beta/1.4}$, and $c_s^{-3}$ $\propto$ \sigmaone$^{-1.1}$), which is consistent with the best-fit log-linear model in Section~\ref{ss-allsf} that has a slope of $1.6\pm 0.2$.

\begin{figure*}
\begin{center}
\includegraphics[scale=0.34]{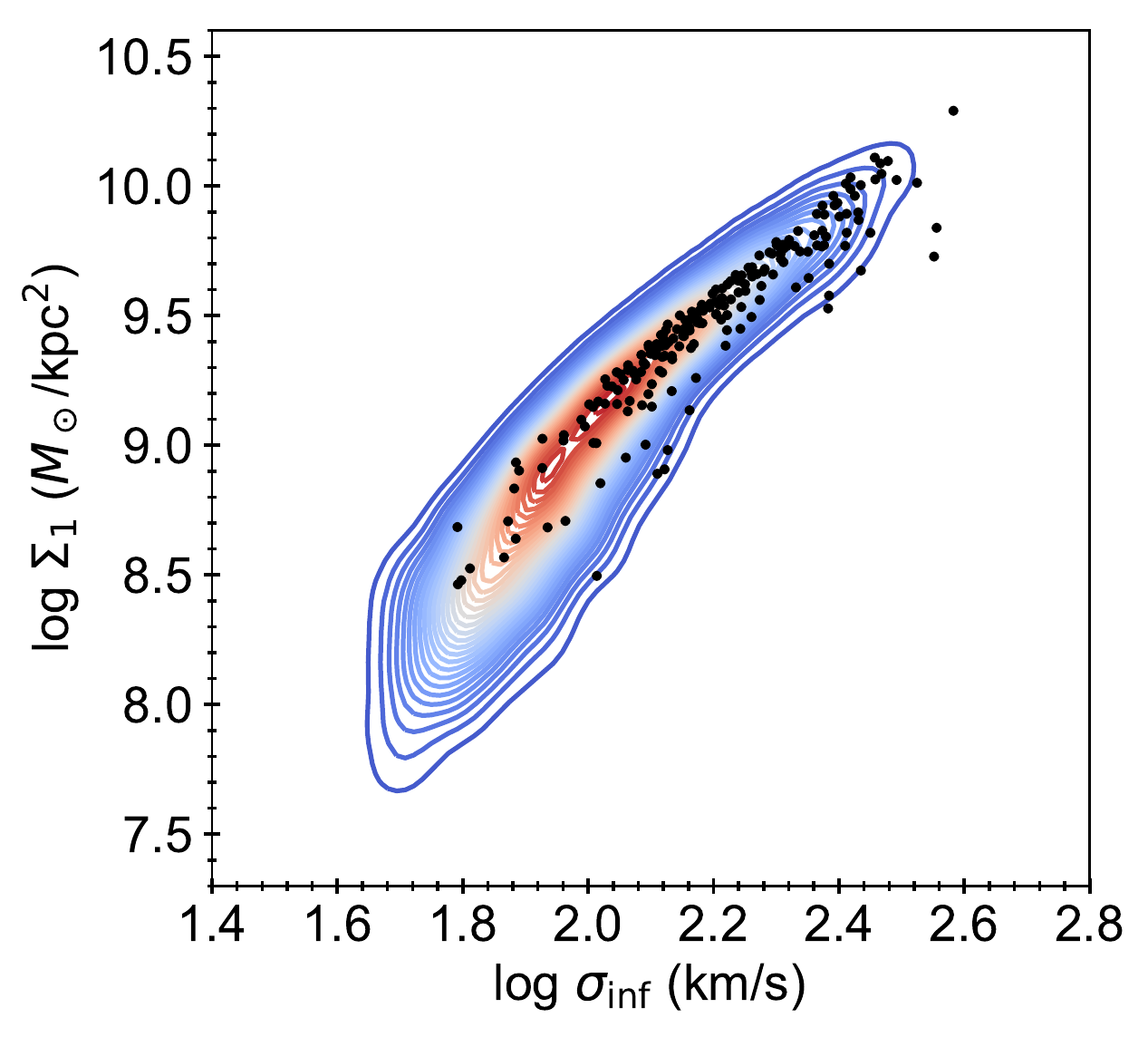}
\includegraphics[scale=0.34]{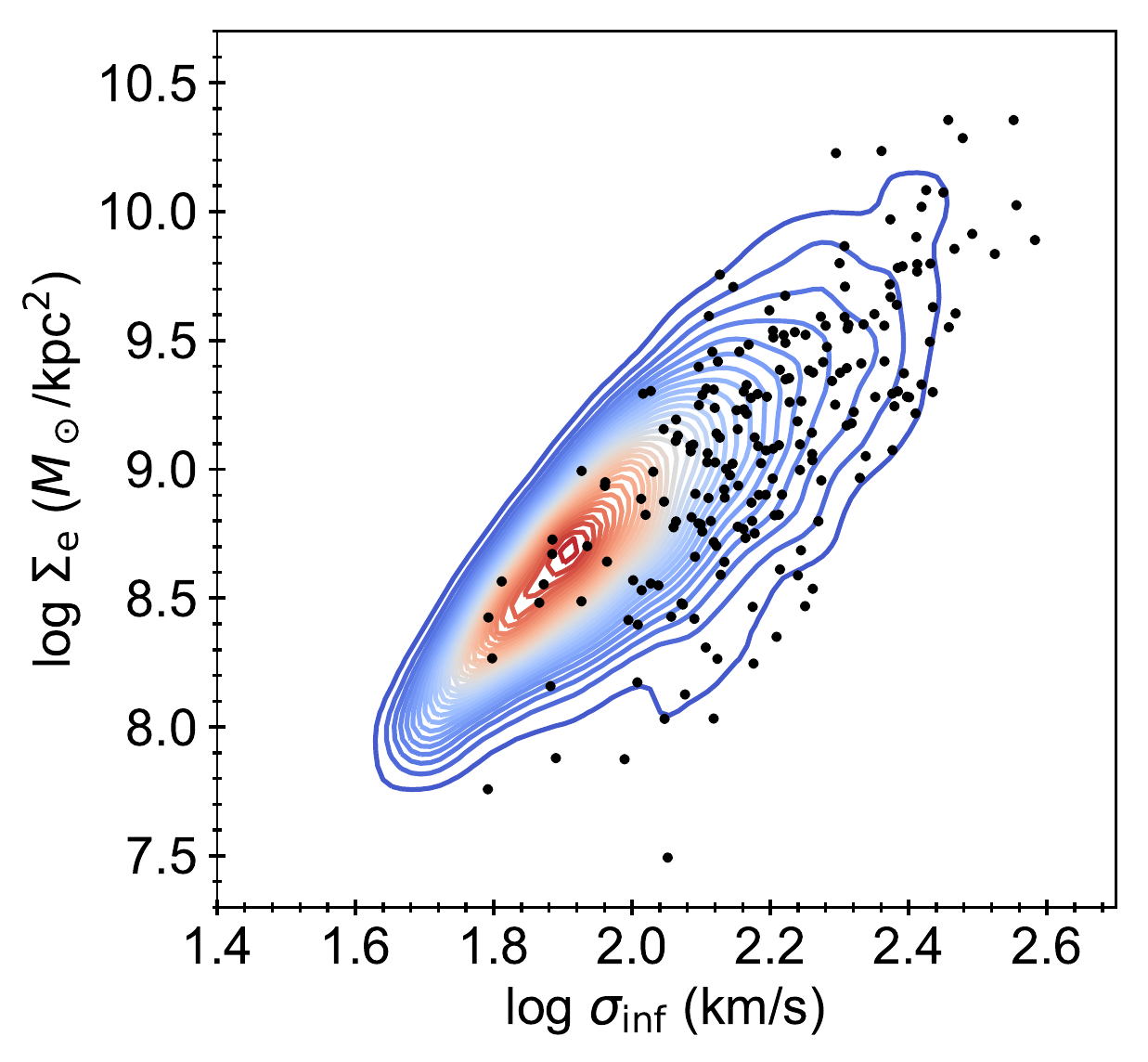}
\includegraphics[scale=0.34]{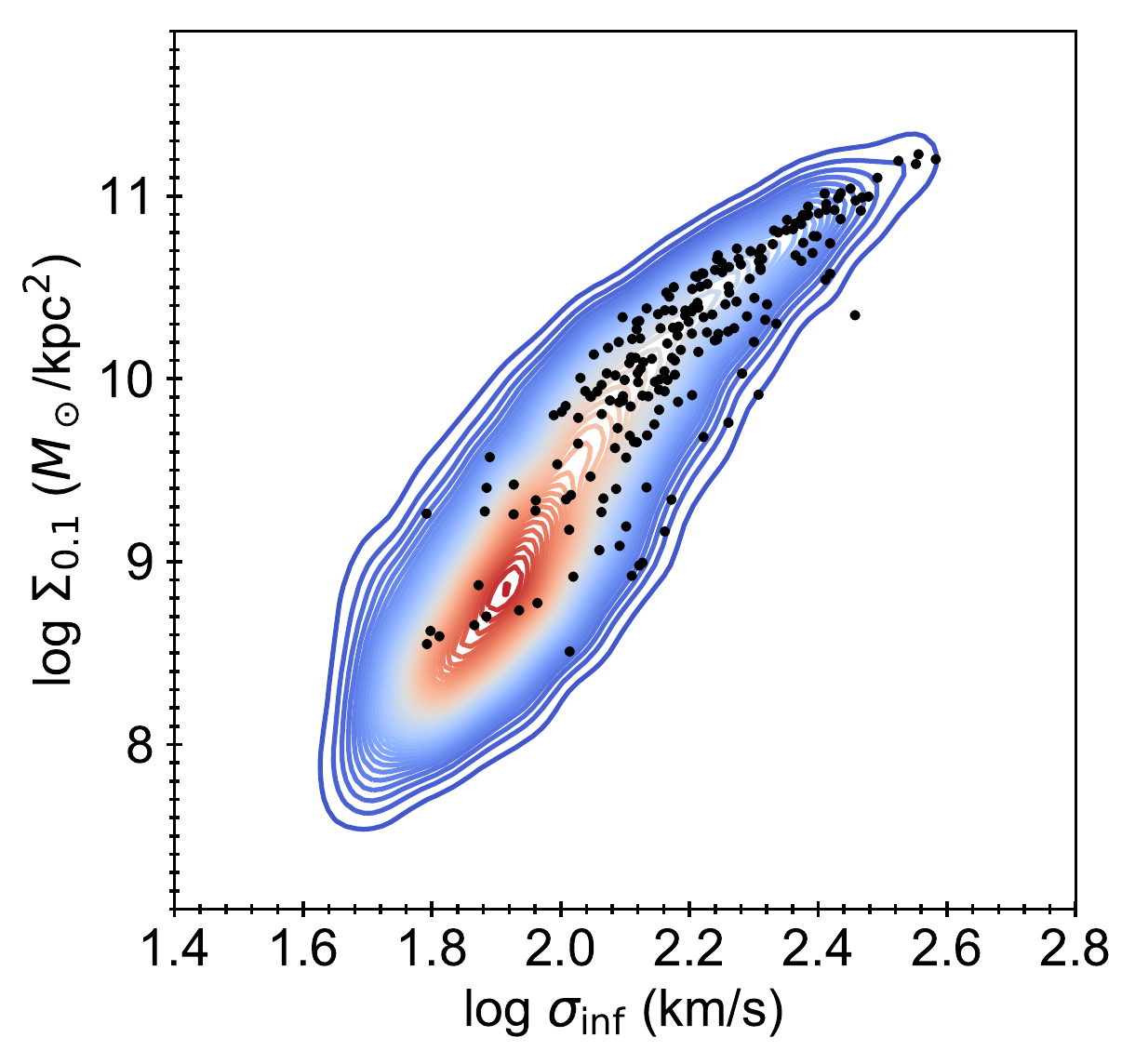}
\includegraphics[scale=0.34]{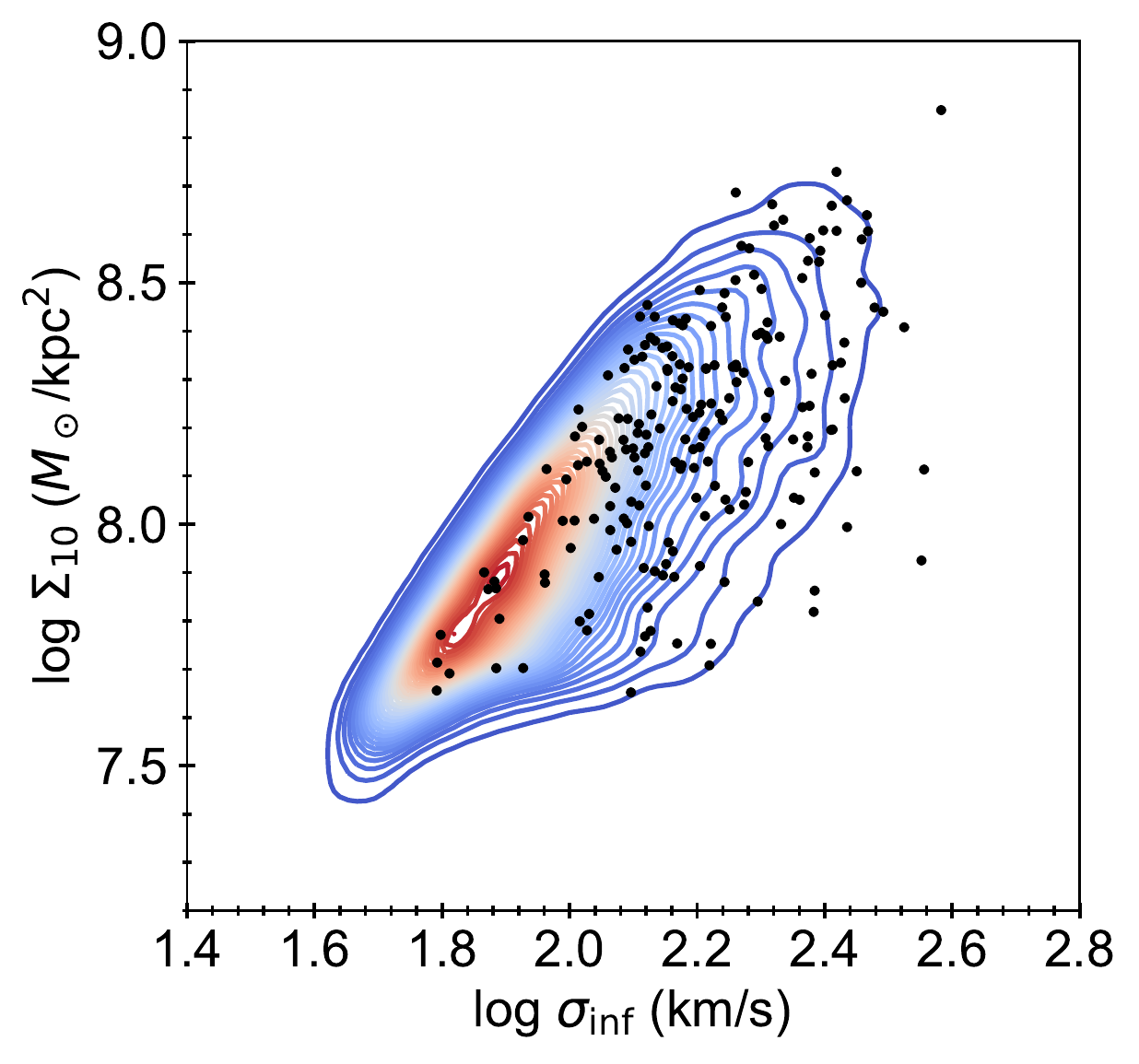}
\caption{\textit{Left to right:} 2D kernel density estimation (KDE) plot of \sigmaone, $\Sigma_{\rm e}$, $\Sigma_{0.1}$, and $\Sigma_{10}$ vs. \sigmainf\ of galaxies in the ALL SF sample. X-ray detected galaxies are represented by the black dots. Note the correlation between \sigmainf\ and \sigmaone\ is tighter than the correlations between \sigmainf\ and other compactness parameters shown.}
\label{sigmaalot}
\end{center}
\end{figure*}

\subsubsection{An indicated role of the host-galaxy potential well in feeding BHs?} 

Alternatively, the \bhar-\sigmaone\ relation may reflect a link between BH growth and the host-galaxy potential well depth at a certain gas content among SF galaxies.
We note that \sigmaone\ is tightly correlated with the inferred central velocity dispersion (\sigmainf; \citealt{Bezanson2011}) of galaxies:
\begin{equation}
\sigma_{\rm inf}=\sqrt{\frac{GM_{\star}}{K_{\star}(n)r_e}},
\end{equation}	
where
\begin{equation}
K_{\star}(n) = 0.557 \times K_v(n).
\end{equation}
At $z \sim 0$, \sigmainf\ has proven to be a good approximation of the true central velocity dispersion ($\sigma$; \citealt{Bezanson2011}), which measures the potential well depth of galaxies.\footnote{In \citet{Bezanson2011}, the comparison between \sigmainf\ and $\sigma$ is mainly performed using a large sample of SDSS galaxies at $z \sim 0$ where a good agreement is confirmed. At high redshift, only tens of objects have measurements of $\sigma$ (with large error bars). Their \sigmainf\ values are in general consistent with $\sigma$ measurements. \citet{Bezanson2011} thus assume that \sigmainf\ can also be a good approximation of $\sigma$ at high redshift. This is also the underlying assumption when we use \sigmainf\ to approximate $\sigma$ for objects in our sample.}  
In the first panel of Figure~\ref{sigmaalot}, we show galaxies in the ALL SF sample in the \sigmaone\ versus \sigmainf\ plane. We can see that most of these galaxies, and especially the X-ray detected galaxies, are ``degenerate'' in the \sigmaone\ vs. \sigmainf\ space (i.e. their \sigmaone\ values are tightly correlated with their \sigmainf\ values). 
It is possible that \sigmaone\ actually serves as a proxy for the central velocity dispersion when predicting BH growth in our study: we find that all the analysis results in Section~\ref{s-ar} do not change qualitatively if we replace \sigmaone\ with \sigmainf. If so, the effectiveness of \sigmaone\ among all possible compactness parameters could naturally be explained. \citet{Ni2019} found that \sigmaone\ is a better indicator of BH growth than the surface mass density ($\Sigma_{\rm e}$); also, if we calculate the projected central mass density within 0.1 kpc ($\Sigma_{0.1}$) or 10 kpc ($\Sigma_{10}$) by extrapolating the measured \sersic\ profiles similarly to the approach presented in Equation~\ref{eq:sigmaone}, we find that \sigmaone\ is also a better indicator compared with them. 
It is interesting and reasonable to question why it is the mass density in the central $\sim$~1 kpc part that matters most.
In the last three panels of Figure \ref{sigmaalot}, we plot $\Sigma_{\rm e}$, $\Sigma_{0.1}$, and $\Sigma_{10}$ vs. \sigmainf\ for galaxies in the ALL SF sample. None of these quantities is as tightly correlated with \sigmainf\ as \sigmaone\ (see Figure~\ref{sigmaalot}): it might be the case that the mass density in the central $\sim$~1 kpc part matters most simply because it is the best representative of the central velocity dispersion among all the compactness parameters examined (see \citealt{Fang2013} for the tight correlation between \sigmaone\ and the central velocity dispersion of SDSS galaxies).

\begin{figure}
\begin{center}
\includegraphics[scale=0.45]{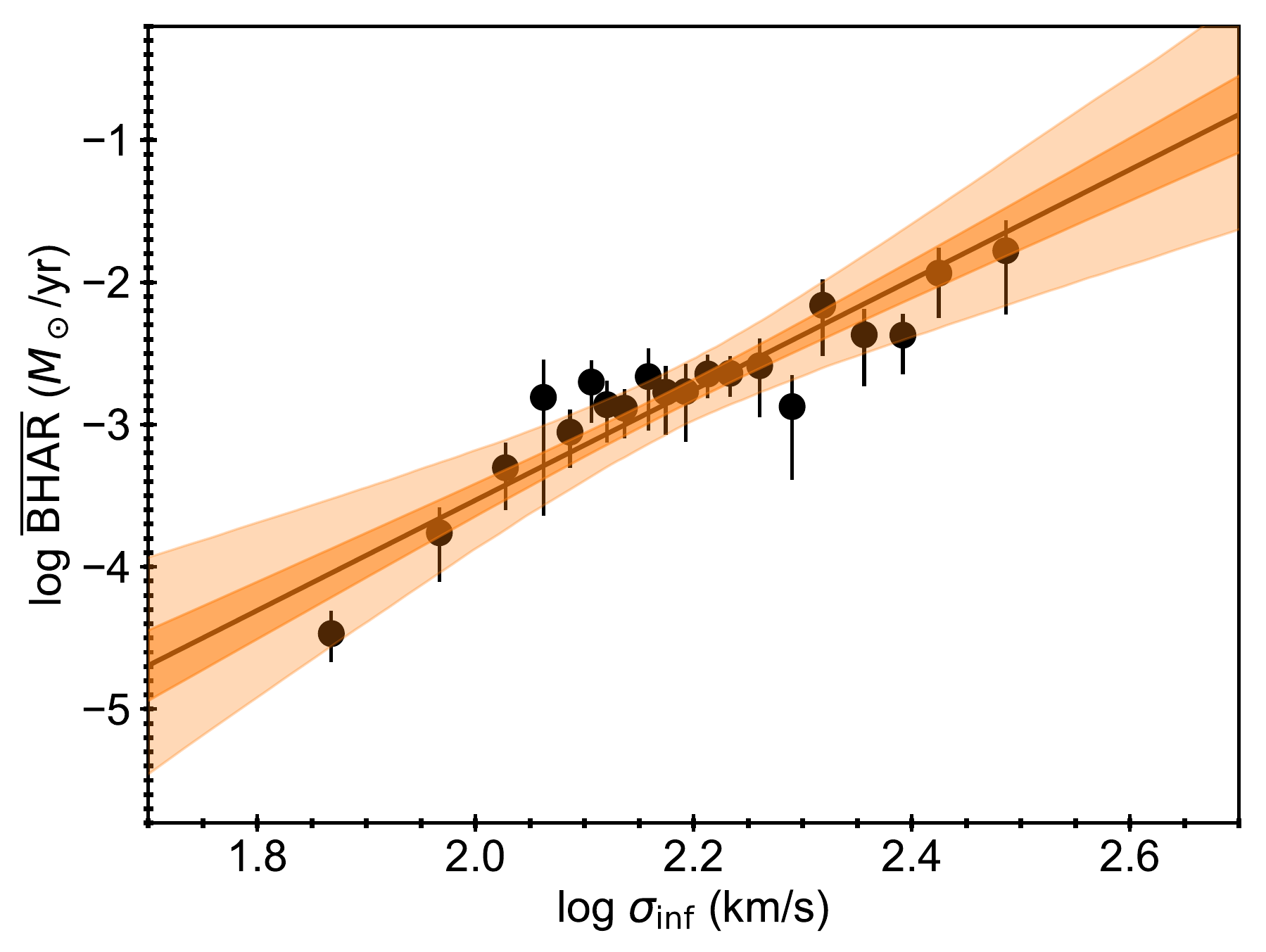}
\caption{\bhar\ as a function of \sigmainf\ among SF galaxies. Galaxies in the ALL SF sample are divided into bins according to their \sigmainf\ values, with $\approx$ 10 X-ray detected galaxies in each bin. The horizontal position of each data point indicates the median \sigmainf\ of the sources in the bin; the error bars represent the 1$\sigma$ confidence interval of \bhar\ from bootstrapping. The black solid line and the dark/light orange shaded region represent the best-fit \bhar-\sigmainf\ relation and the 1$\sigma$/3$\sigma$ pointwise confidence intervals on the regression line.}
\label{bharsigmainf}
\end{center}
\end{figure}

In Figure~\ref{bharsigmainf}, we present the \bhar-\sigmainf\ relation among SF galaxies.
This \bhar-\sigmainf\ relation suggests that the host-galaxy potential well may play a fundamental role in feeding BHs among SF galaxies where cold gas is abundant. \footnote{We note that the \bhar-\sigmainf\ relation is not necessarily ``responsible'' for producing the \mbh-$\sigma$ relation among local bulges. The \mbh-$\sigma$ relation may simply mark the turning point where both the BH and galaxy cannot be fueled efficiently \citep[e.g.][]{King2005,King2010,Murray2005}.} 
In this scenario, the link between BH growth and \sigmagas\ in the central $\sim$ 1 kpc still exists, though it actually \textit{manifests} the relation between BH growth and host-galaxy potential well depth at a given gas content.     
Fitting the data points in Figure~\ref{bharsigmainf} with \texttt{emcee}, the best-fit log-linear model of the \bhar-\sigmainf\ relation is:
\begin{equation} \label{eq-bharsigma}
 \rm{log~\overline{BHAR}} = (3.9 \pm 0.5) \times  \sigma_{\rm inf} + (-11.3 \pm 1.1).
 \end{equation}
As the 1$\sigma$ uncertainty of the slope is large, the exact form of the \bhar-\sigmainf\ relation remains unclear. 
It has been suggested that AGNs can feed efficiently from surrounding dense gas clumps, at rates close to the dynamical rate $\dot{M}_{\rm dyn}$ (assuming that the gas is initially in rough virial equilibrium; \citealt{Zubovas2019}):
\begin{equation}
\dot{M}_{\rm dyn} \propto \frac{f_{\rm g} \sigma^3}{G},
\end{equation}
where $f_{\rm g}$ is the gas fraction in the galaxy that could explain the cosmic evolution of the \bhar-\sigmaone\ (or the \bhar-\sigmainf) relation.
Among quiescent galaxies that lack gas, $\dot{M}_{\rm dyn}$ cannot be achieved, so that \bhar\ does not have strong dependence on \sigmainf\ (or \sigmaone).
The predicted slope (of 3) is within the $\sim 2\sigma$ confidence interval of the fitting result. A larger sample of galaxies/AGNs will be needed to provide further constraints on the relation that could validate or rule out this scenario, and provide more insights into the BH feeding mechanism among SF galaxies.

While the observed \bhar-\sigmaone\ relation may suggest the role of the host-galaxy potential well in feeding the BH, we note that the \mbh-$\sigma$ relation observed among local bulges may suggest the role of the host-galaxy potential well in ``shutting off'' the BH growth (AGN feedback via outflows is one possible way to achieve this ``shut-off'' process; e.g. \citealt{SR1998,King2005,Murray2005}), further indicating the connection between the host-galaxy potential well and the BH growth.

We also note that, from the side of galaxy evolution, \sigmaone\ (or $\sigma$) is linked with the color (or specific SFR) of galaxies \citep[e.g.][]{Fang2013,Whitaker2017}, and serves as a good predictor of quiescence. 
AGN activity reaches the high-point among high-\sigmaone\ SF galaxies that become quiescent later on (see Section~\ref{ss-allsf}; also see \citealt{Kocevski2017}), which indicates that there is a potential link between AGNs and the quenching of galaxies: whatever process (i.e. AGN feedback, morphological quenching, halo gas shock heating) that quenches galaxies may also slow down the BH growth (see Figure~\ref{q_trend} for the BH growth among high-\sigmaone\ quiescent galaxies).

\subsection{Potential connections between the \bhar-\sigmaone\ relation and the \mbh-\mbulge\ relation, and implications for BH ``monsters'' among local bulges}  \label{ss-monster}

In Section~\ref{ss-sfbd}, we confirmed the \bhar-\sigmaone\ relation among SF BD galaxies, and we also verified that the \bhar-\sigmaone\ relation among all SF galaxies applies to BD galaxies and Non-BD galaxies seamlessly in Section~\ref{ss-allsf}.
The \bhar-\sigmaone\ relation and the \mbh-$M_{\rm bulge}$ relation may indeed reflect the same underlying link. As discussed in Section~\ref{ss-gas}, this link may be the direct dependence of BH growth on the central $\sim$~kpc gas density of host galaxies, or the dependence of BH growth on the host-galaxy potential well depth at a given gas content (which will also manifest a link between BH growth and \sigmagas\ on the $\sim$ kpc scale).
We will show below how the \bhar-\sigmaone\ relation and the \mbh-$M_{\rm bulge}$ relation quantitatively agree.
Among ellipticals and classical bulges in the local universe, the \mbh-$M_{\rm bulge}$ relation takes the form of:
\begin{align}
M_{\rm BH} \propto M_{\rm bulge} ^{\alpha},
\end{align}
and the reported values of $\alpha$ range from $\approx 1.2$ \citep[e.g.][]{KH2013} to $\approx 1.4$ \citep[e.g.][]{Reines2015}.
If we assume that this relation also approximately holds true at higher redshift (see Figure~38c of \citealt{KH2013}) and take the (time) derivative of this formula, we obtain:
\begin{align}
dM_{\rm BH} \propto M_{\rm bulge}^{\alpha - 1} \times dM_{\rm bulge},
\end{align}
which suggests that:
\begin{align}
{\rm{\overline{BHAR}}} \propto M_{\rm bulge}^{\alpha - 1} \times {\rm SFR_{bulge}}, \label{eq-bharbulge}
\end{align} 
where SFR$_{\rm bulge}$ is the SFR of the bulge component. Through assuming that SFR$_{\rm bulge}$ is approximately proportional to $\Sigma_{\rm SFR, 1~kpc}$,
we can further express SFR$_{\rm bulge}$ as a function of \sigmaone:
SFR$_{\rm bulge}$ $\propto$ \sigmaone$^{\beta}$, where $\beta$ is the slope of the log~$\Sigma_{\rm SFR}$-log~$\Sigma_{M_\bigstar}$ relation ($\beta \sim 0.7$--1; e.g.\
\citealt{Cano2016,Hsieh2017,Trayford2019,Hani2020}).\footnote{\label{fn-sfrbulge}If we fit the SFR-\sigmaone\ relation directly among the SF BD sample, we obtain a power-law index of $\approx 0.7\pm 0.1$, consistent with the adopted $\beta$ values. The scatter of the fitted log SFR-log \sigmaone\ relation is $\approx 0.5$ dex, and a significant fraction of this scatter could be attributed to the uncertainty associated with the SFR$_{\rm bulge}$-$\Sigma_{\rm SFR, 1~kpc}$ relation, as the expected scatter associated with the log~$\Sigma_{\rm SFR, 1~kpc}$-log~\sigmaone\ relation is $\sim 0.2$--0.3 dex \citep[e.g.][]{Cano2016,Hsieh2017,Lin2019}.
While the assumption of SFR$_{\rm bulge}$ $\propto$ $\Sigma_{\rm SFR, 1~kpc}$ automatically holds true when $\Sigma_{\rm SFR}$ is uniform across the bulge, $\Sigma_{\rm SFR}$ is not uniform in real cases \citep[e.g.][]{Nelson2012}. 
Thanks to the compact sizes of bulges (on the kpc scale), we expect a considerable fraction of their star formation to be enclosed in their central 1~kpc regions. Thus, while $\Sigma_{\rm SFR}$ may be far from uniform, the assumed relation between SFR$_{\rm bulge}$ and $\Sigma_{\rm SFR, 1~kpc}$ still roughly holds, though a considerable scatter is associated with this relation, which originates from the scatter in the fraction of SFR enclosed in the central~1 kpc region among SF BD galaxies.}
We can also approximate $M_{\rm bulge}$ as a power-law function of \sigmaone: through fitting all log \mstar\ > 10.2 BD galaxies at $z < 0.8$ in our sample, we find that the power-law index is $\sim 1.6$.
We can then write the right side of equation~\ref{eq-bharbulge} as a pure function of \sigmaone:
\begin{align}
{\rm{\overline{BHAR}}} \propto \Sigma_1^ {1.6 \times (\alpha - 1)} \times \Sigma_1^{\beta}.
\end{align}
Different combinations of $\alpha$ and $\beta$ values predict the power-law index of the \bhar-\sigmaone\ relation to be $\sim$ 1.0–1.6, which is consistent with our derived index of $1.6 \pm 0.2$ in Section~\ref{ss-allsf}.
Thus, it is plausible that the observed \mbh-$M_{\rm bulge}$ relation reflects the same underlying link as the \bhar-\sigmaone\ relation.
This picture of the same underlying link for both the \bhar-\sigmaone\ and \mbh-$M_{\rm bulge}$ relations is also supported by the observed scatter of the \hbox{\mbh-$M_{\rm bulge}$} relation. The \sigmaone\ values of SF BD galaxies in our sample at a given \mstar\ have a scatter of $\sim 0.2$ dex. Considering the \bhar-\sigmaone\ relation, we would expect a $\sim$ 0.3 dex scatter for the \mbh-$M_{\rm bulge}$ relation, which is the scatter observed in Section~6.6.1 of \citet{KH2013}.

It is plausible that the \mbh-$M_{\rm bulge}$ relation cannot characterize the BH ``monsters'' (i.e. BHs found in local compact galaxies that have \mbh\ values much larger than expected from the \mbh-$M_{\rm bulge}$ relation) well simply because in cases where the bulge is so compact and the gas is so highly concentrated, the central $\sim$~kpc gas density (or the central velocity dispersion that is tightly linked with compactness) cannot be well approximated by the SFR of the whole bulge.
Our derived \bhar-\sigmaone\ relation in Section~\ref{ss-allsf}, at the same time, may manifest the underlying link better in ultra-compact SF bulges, and it has the potential to explain the local BH ``monsters''.
We will take NGC~4486B as an example, where the $\approx 6 \times 10^8 M_{\odot}$ BH is ``overmassive'' by $\approx$ 1.7 dex \citep{KH2013}.
NGC~4486B has \re\ $\approx$ 0.2 kpc and $n \approx 2.2$ \citep{Kormendy2009}.
If we compare NGC~4486B with a typical local bulge that has \re\ $\approx$ 3 kpc and $n \approx 3$,
we find that the percentage of mass concentrated in the central 1~kpc of NGC~4486B is greater than that of a typical bulge by a factor of $\approx 5$.
This means that the \sigmaone\ value of NGC~4486B is larger than the typical \sigmaone\ value at the same $M_{\rm bulge}$ by $\approx 0.7$ dex. 
If we assume this deviation approximately holds true during all BH-growth episodes of NGC~4486B, a $\approx 1.1$~dex elevation in its \mbh\ compared with typical local bulges that have similar $M_{\rm bulge}$ values will be generated according to the derived \bhar-\sigmaone\ relation (see Equation~\ref{eq-bharsigma}).
We also note that for a given \sigmaone, the expected amount of BH growth could increase by $\approx 1$ dex when the redshift rises (see Figure~\ref{z_trend}).
If NGC~4486B is a ``relic'' galaxy that had finished growing most of its \mbh\ by $z \sim 2$, an additional elevation of its \mbh\ by up to $\approx$~1~dex compared with typical local bulges that have similar $M_{\rm bulge}$ values can be expected due to the cosmic evolution of BH growth at a given \sigmaone: for these typical bulges, a significant fraction ($\sim 50\%$) of \mstar\ is likely to be assembled at $z < 2$ \citep[e.g.][]{Thomas2005,Thomas2010,DL2006}, suggesting a significant amount of BH mass assembly at $z < 2$. 
Taking all these into account, the $\approx$ 1.7 dex deviation in \mbh\ of NGC~4486B from the \mbh-$M_{\rm bulge}$ relation is understandable as BHs among SF galaxies follow the \bhar-\sigmaone\ relation.

\section{Conclusions and Future Work} \label{s-con}

Utilizing extensive multiwavelength observations in the COSMOS survey field, we have revealed and studied the dependence of BH growth on host-galaxy compactness represented by \sigmaone\ among SF galaxies. 
The main points from this paper are the following:
\begin{enumerate}
\item
We built a catalog of $I_{\rm F814W} < 24$ galaxies at $z < 1.2$ from the COSMOS survey field (Section~\ref{s-ds}).
We measured their \mstar\ and SFR values utilizing UV-to-FIR photometry (Section~\ref{ss-mstarsfr} and Appendix~\ref{a-xcigale}).
We measured their structural parameters (Section~\ref{ss-galfit} and Appendix~\ref{a-galfit}), and classify them as BD or Non-BD galaxies (Section~\ref{ss-morph} and Appendix~\ref{a-morph}) utilizing the high-resolution \textit{HST} F814W mosaics.
Drawing upon all these measurements, we compiled a sample of SF Non-BD galaxies and a sample of SF BD galaxies, as well as an ALL SF sample regardless of morphology (Section~\ref{ss-sample}). \sigmaone\ values of galaxies are calculated from \mstar\ and structural parameters. 
Deep \textit{Chandra} \hbox{X-ray} observations in the field are utilized to estimate \bhar\ for samples of galaxies (see Section~\ref{ss-bhar}).
\item
Utilizing partial-correlation analyses, we found that the \bhar-\sigmaone\ relation is more fundamental than the \bhar-\mstar\ relation among SF Non-BD galaxies (Section~\ref{ss-sfnonbd}), as we observe a significant \bhar-\sigmaone\ relation when controlling for \mstar, while we do not observe a significant \hbox{\bhar-\mstar} relation when controlling for \sigmaone.
We also found that the \bhar-\sigmaone\ relation is significant when controlling for SFR in the SF BD sample (Section~\ref{ss-sfbd}), which suggests that the \bhar-\sigmaone\ relation also exists among SF BD galaxies.
\item
We confirmed that the same \bhar-\sigmaone\ relation applies to both SF Non-BD and SF BD galaxies, and this \bhar-\sigmaone\ relation is more fundamental than either the \bhar-\mstar\ or \bhar-SFR relation among SF galaxies (Section~\ref{ss-allsf}).
Our best-fit log~\bhar-log~\sigmaone\ relation has a slope of $1.6 \pm 0.2$.
While the slope of the log~\bhar-log~\sigmaone\ relation does not exhibit significant changes with redshift, \bhar\ at a given \sigmaone\ evolves with redshift in a manner that could be well explained by the cosmic evolution of the gas content (Section~\ref{ss-allsf} and Section~\ref{ss-gas}).
The \bhar-\sigmaone\ relation among SF galaxies could suggest a link between BH growth and the central ($\sim$~kpc scale) gas density of host galaxies. 
A common origin for gas in the vicinity of the BH and in the central $\sim$ kpc part of the galaxy may be further implied by this relation. 
The \bhar-\sigmaone\ relation could also be interpreted as a relation between BH growth and the central velocity dispersion of host galaxies at a given gas content, indicating the role of the host-galaxy potential well in feeding BHs (Section~\ref{ss-gas}).
\item
The quantitatively derived \bhar-\sigmaone\ relation in Section~\ref{ss-allsf} has the potential to explain local BH ``monsters'' among compact galaxies (Section~\ref{ss-monster}).
It is plausible that both the \bhar-\sigmaone\ and \mbh-$M_{\rm bulge}$ relations manifest the same underlying link between BH growth and host galaxies discussed in Section~\ref{ss-gas}, and local BH ``monsters'' deviate from the \mbh-$M_{\rm bulge}$ relation simply because the total SFR of ultra-compact bulges cannot approximate the central $\sim$~kpc gas density (or the velocity dispersion) well. 
\end{enumerate}

In the future, deep \textit{JWST} imaging combined with deep \xray\ coverage could help to quantify the \bhar-\sigmaone\ relation among SF galaxies better with a larger sample of galaxies/AGNs that has lower limiting \mstar.
\textit{JWST} IFU observations (as well as grism observations) could measure the gas/stellar velocity dispersion of galaxies/AGNs, enabling the first characterization of the \bhar-$\sigma$ relation.
Future accumulation of ALMA pointings that have \textit{HST}-like resolution in deep X-ray survey fields could help to probe the relation between BH growth and host-galaxy central gas density directly. 
Quantifying these relations could provide insights into the feeding mechanism of BHs, and how it links with the host galaxies.

\section*{Acknowledgements}
We thank the anonymous referee for helpful feedback.
We thank Robin Ciardullo and Marc Huertas-Company for helpful discussions.
QN and WNB acknowledge support from \textit{Chandra} X-ray Center grant GO8-19076X, NASA grant 80NSSC19K0961, and the V.M. Willaman Endowment.
BL acknowledges financial support from the NSFC grants 11991053 and
11673010 and National Key R\&D Program of China grant 2016YFA0400702. 
YQX acknowledges support from NSFC-11890693, NSFC-11421303, the CAS Frontier Science Key Research Program (QYZDJ-SSW-SLH006), and K.C. Wong Education Foundation.

\section*{Data availability}
The data underlying this article were accessed from the NASA/IPAC Infrared
Science Archive (IRSA) COSMOS database (\texttt{https://irsa.ipac.caltech.edu/Missions/cosmos.html}).
The derived data generated in this research will be shared on reasonable request to the corresponding author.







\clearpage
\appendix
\section{Assessing \mstar\ and SFR measurements from {\sc X-CIGALE}} \label{a-xcigale}
\begin{figure}
\begin{center}
\includegraphics[scale=0.47]{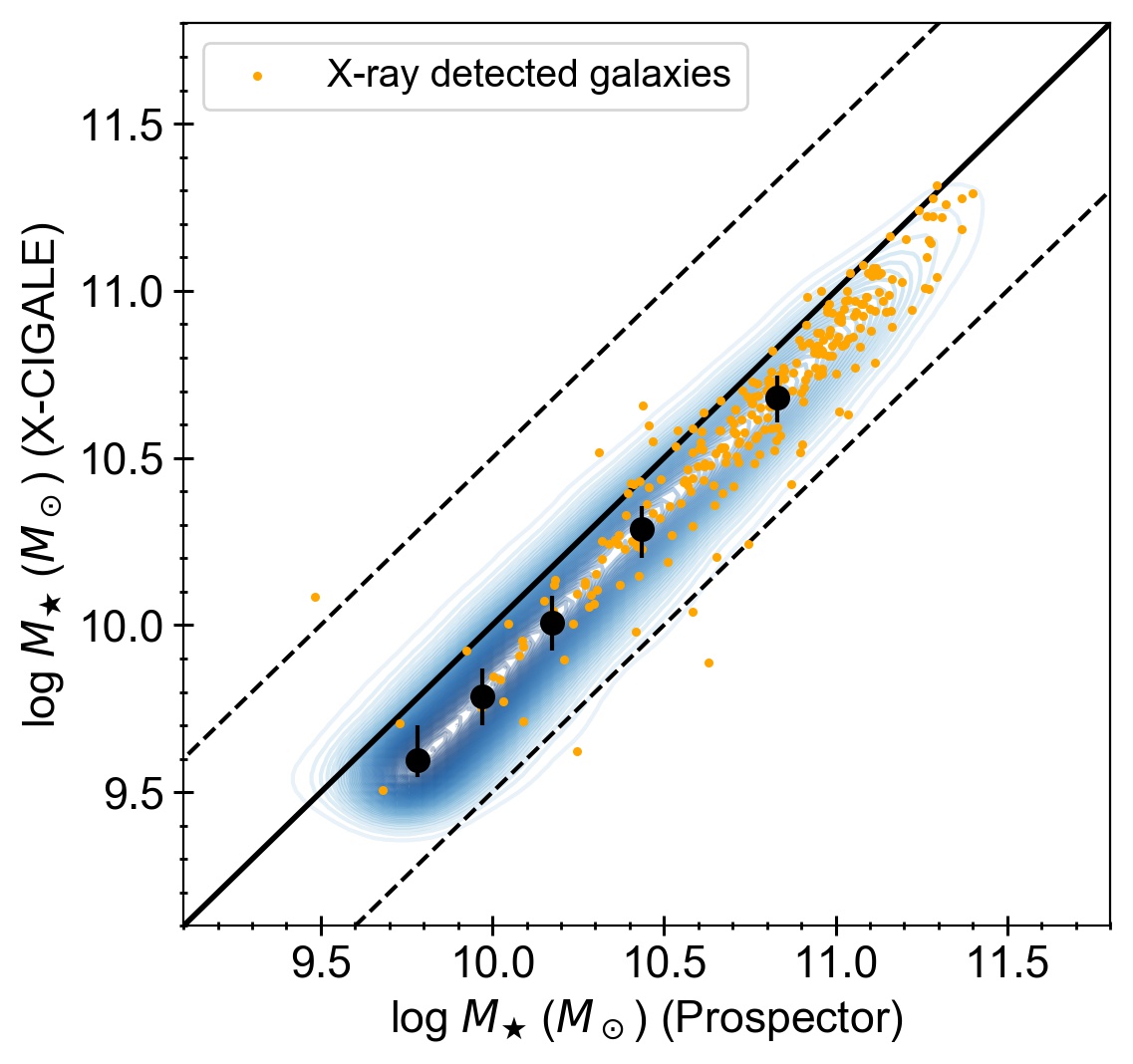}
\includegraphics[scale=0.47]{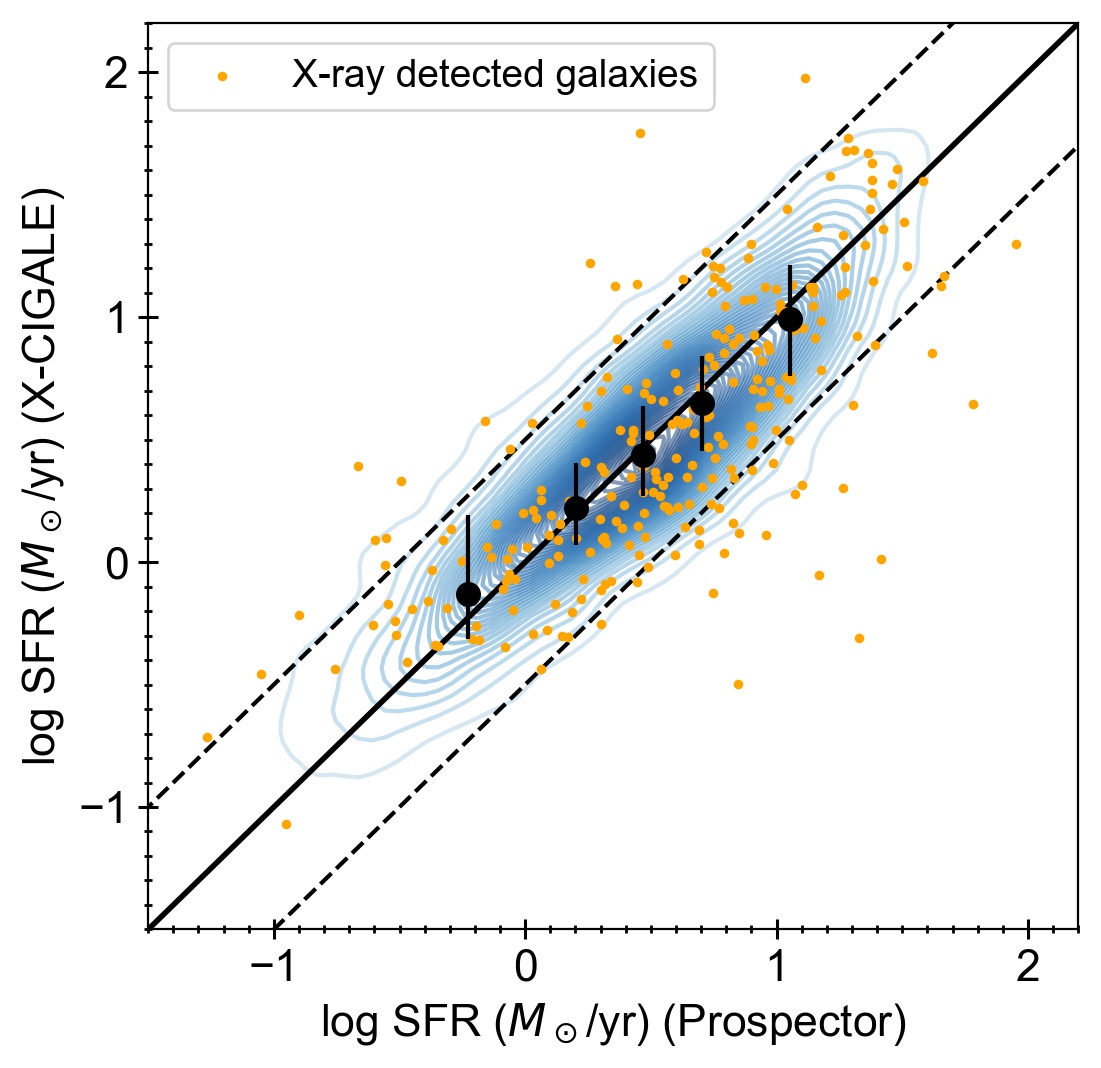}
\includegraphics[scale=0.47]{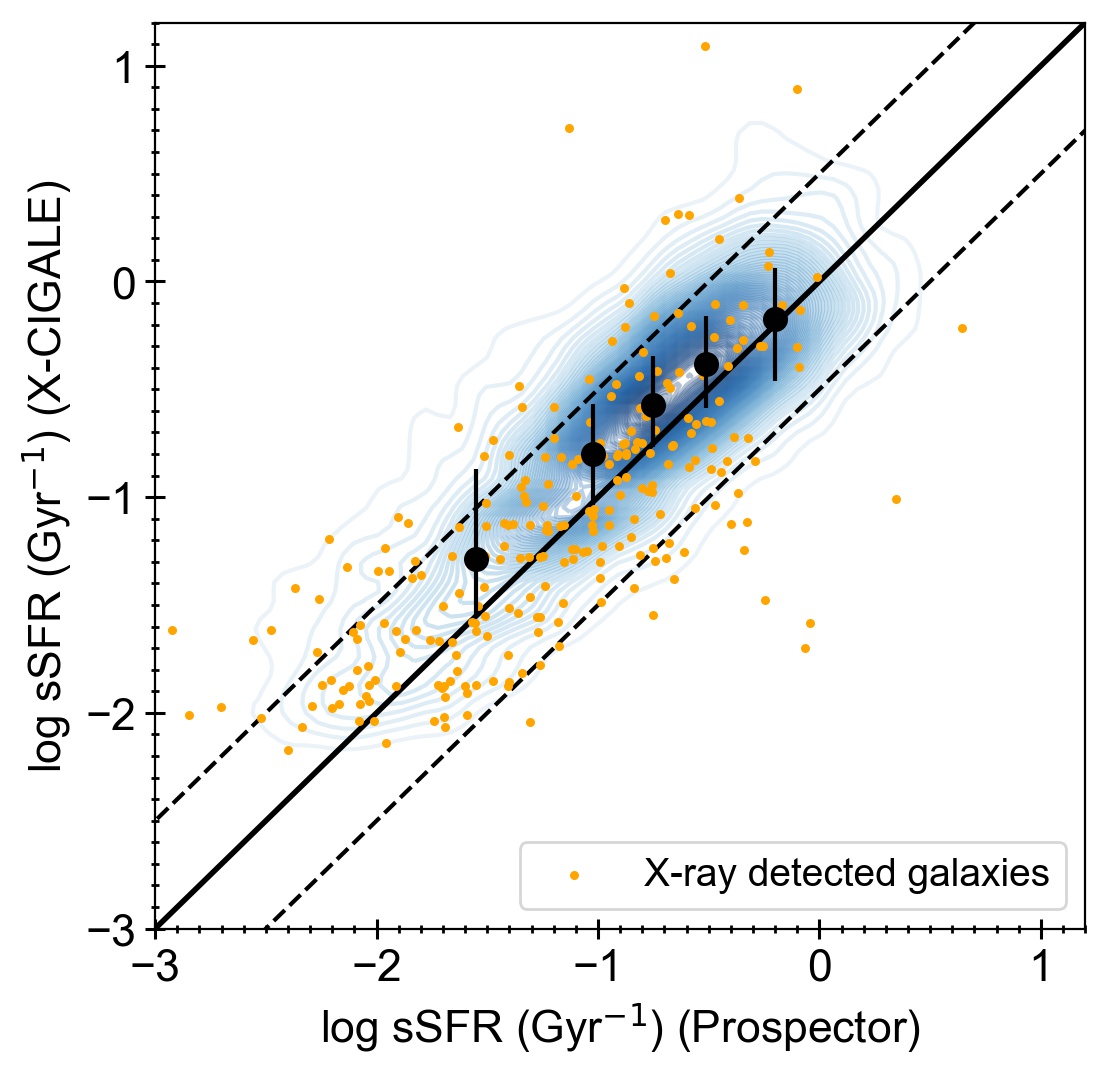}
\caption{\textit{Upper panel:} 2D KDE plot of \mstar\ measured with {\sc X-CIGALE} versus \mstar\ measured in \citet{Leja2019b} with \texttt{Prospector} in log-log space. The orange dots represent X-ray detected galaxies in our sample where an AGN component is added during the SED fitting. Black error bars represent the median {\sc X-CIGALE}-based \mstar\ value in different bins of \mstar\ measured with \texttt{Prospector} and the scatter between the two sets of measurements in each bin. The black solid line represents a 1:1 relation; the black dashed lines represent 0.5~dex offsets from the 1:1 relation. \textit{Middle panel:} Similar to the upper panel, but for the two sets of SFR measurements. \textit{Lower panel:} Similar to the upper panel, but for the two sets of sSFR measurements.}
\label{leja}
\end{center}
\end{figure}

In Table~\ref{cigalep}, we list the parameters used to construct the SED templates when fitting \mstar\ and SFR with {\sc X-CIGALE} in Section~\ref{ss-mstarsfr}.
In Figure~\ref{leja}, we show the comparison between our SED-based \mstar\ (SFR) measurements with {\sc X-CIGALE} and SED-based \mstar\ (SFR) measurements with \texttt{Prospector} in \citet{Leja2019b} for \hbox{log \mstar\ $> 9.5$} COSMOS SF galaxies at $z = 0.2$--0.8, as well as the comparison of the obtained specific SFR (sSFR; which is calculated as SFR/\mstar). In \citet{Leja2019b}, a more flexible nonparametric SFH, a more flexible dust attenuation law, and a more flexible dust-emission model are utilized, which is beyond the scope of this work due to the large amount of computational time needed.
We can see that our \mstar\ measurements are systemically smaller than those reported in \citet{Leja2019b} by $\approx 0.15$~dex.  
As reported in \citet {Leja2019a}, this offset is expected mainly due to the usage of a nonparametric SFH in \texttt{Prospector}.
We correct for this systematic offset in the final adopted \mstar\ values (by adding 0.15 dex to the obtained log \mstar\ values), though we note that as we only quantitatively study the slope of the log~\bhar-log~\sigmaone\ relation in this paper, the systematic offset in \mstar\ measurements should not affect our results.
Our SFR measurements do not show any systematic offset when compared with SFR measurements in \citet{Leja2019b}. 
The relatively small scatter of $\approx 0.1$ dex in \mstar\ and $\approx 0.2$ dex in SFR between the two sets of measurements demonstrates that though our adopted SED libraries may not be the ideal approach, they are acceptable for this analysis.
We verified that our results in Section~\ref{s-ar} are not materially affected if we add random perturbations to log~\mstar/log~SFR with a scatter of 0.1/0.2 dex.
For X-ray detected galaxies especially, the systematic offset and the scatter of \mstar\ are close to those for the general galaxy population;
there is a $\approx 0.1$~dex offset between the two sets of SFR measurements (SFR  values measured via {\sc X-CIGALE} are systematically smaller than those measured via \texttt{Prospector}) and the scatter is relatively larger ($\approx 0.35$ dex)
due to the default usage of AGN templates in our study (SED-based SFR measurements mainly depend on the UV and IR SED where the AGN component has non-negligible contributions).
We note that even with the offset and the relatively larger scatter, $\approx 81\%$ of X-ray detected objects have SFR values in the two sets of measurements agreeing within 0.5 dex.
Perturbing the log~SFR values of X-ray detected galaxies in our sample by this large scatter also does not affect the analysis results in Section~\ref{s-ar} materially.

\begin{table*}
 \begin{center}
 \caption{Utilized {\sc X-CIGALE} modules with fitting parameters. Default values are adopted for parameters not listed.}
  \begin{tabular}{ccccccccccc}
  \hline\hline
Module  & Parameters  & Values \\
\hline
Star formation history: {\it sfhdelayed} & $\tau$ (Myr) & 100, 150, 200, 250, 300, 350, 400, 500, 600, 800, \\
                                         &              &  1200, 2000, 3000, 5000, 8000  \\  
                                         & $t$ (Myr)    & 50, 100, 200, 300, 400, 500, 600, 800, 1000, 1500, 2000, \\ \vspace{0.1 cm}
                                         &              &  2500, 3000, 3500, 4000, 5000, 6000, 8000, 10000 \\

Stellar population synthesis model: {\it bc03}           & Initial mass function &   \citet{Chabrier2003} \\\vspace{0.1 cm}
                                                                                   & Metallicity &  0.02\\\vspace{0.1 cm}
Nebular emission: {\it nebular}        & -       &  - \\
Dust attenuation: {\it dustatt\_calzetti}  & $E(B-V)$ for the young population &     0.0--1.5 in a step of 0.1  \\ \vspace{0.1 cm}
                          &  $E(B-V)$ reduction factor of the old population&    0.44 \\ 
Dust emission: {\it dale2014}   & $\alpha$ in d$M_{\rm dust} \propto U^{- \alpha}{\rm d}U$ &   1.5, 2.0, 2.5 \\
AGN emission: {\it skirtor2016}  & Torus optical depth at 9.7$\mu m$ &   7 \\
                                                & Viewing angle ($^{\circ}$) &  30, 70 \\
                                               & AGN fraction in total IR luminosity ($\rm frac_{AGN}$)& 0--0.9 in a step of 0.1, 0.99\\
                                               & $E(B-V)$ of AGN polar dust & 0.1, 0.2, 0.3, 0.4, 0.5\\
\hline \hline
  \end{tabular}
  \end{center}
  \label{cigalep}
\end{table*}

\begin{figure}
\begin{center}
\includegraphics[scale=0.5]{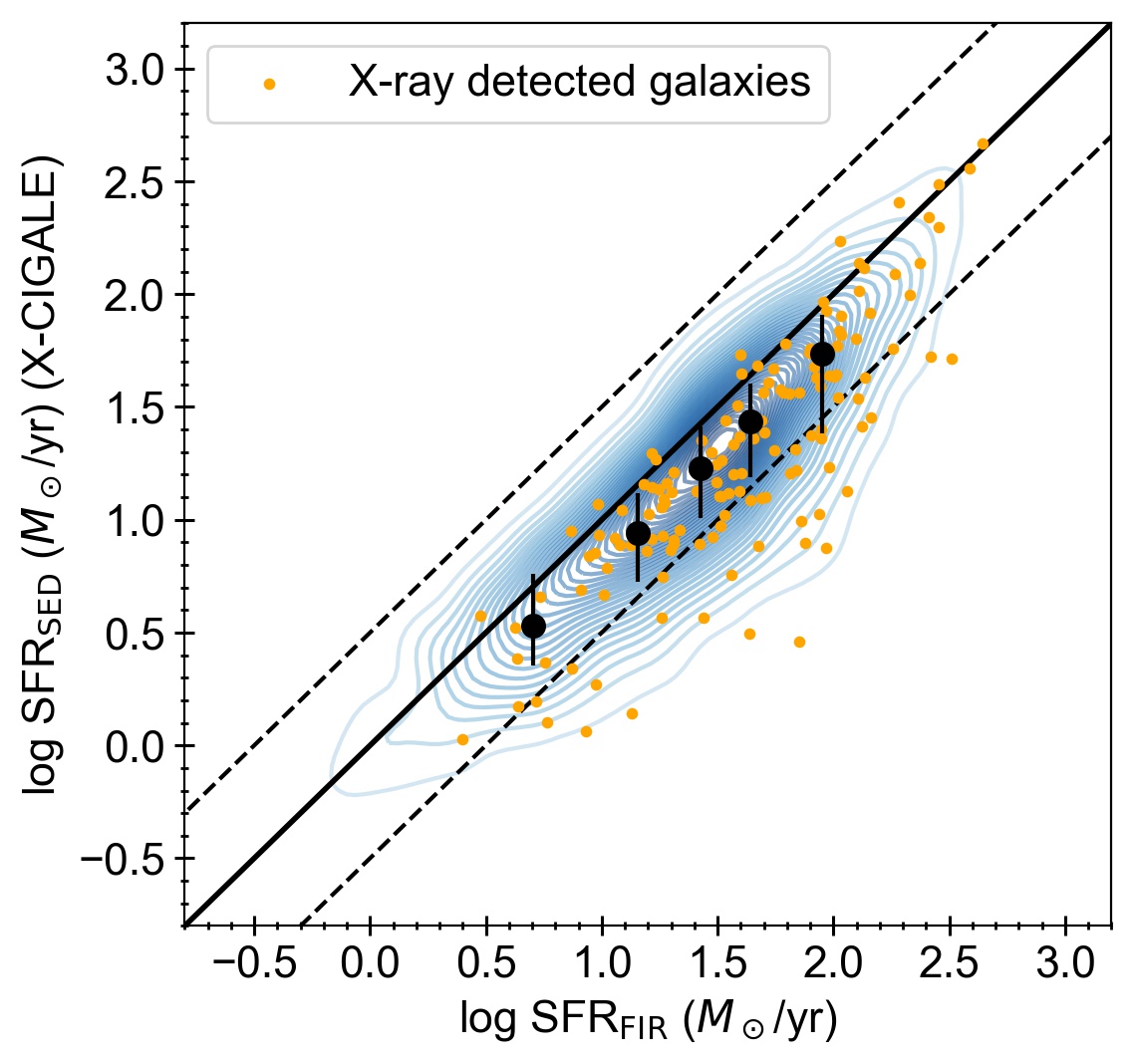}
\caption{Similar to Figure~\ref{leja}, but for SED-based SFR values (measured with {\sc X-CIGALE}) versus FIR-based SFR values for all log \mstar\ $>$ 9.5 COSMOS galaxies. 
}
\label{firsfr}
\end{center}
\end{figure}

We also compare our SED-based SFR measurements with FIR-based SFR measurements, as can be seen in Figure~\ref{firsfr}. 
We can see that the median offset between the two measurements is small ($\approx 0.19$ dex; SFR values measured via {\sc X-CIGALE} are systematically smaller than those measured from FIR luminosity). For $\approx 87$\% of the objects ($\approx 73$\% of X-ray detected objects), SFR values measured by these two methods agree within 0.5 dex. This general agreement is sufficient in the context of this work, as we group sources into log SFR bins of at least $\approx 0.5$~dex-width in our analyses (see Section~\ref{ss-sfbd}).

\section{Assessing structural measurements from GALFIT} \label{a-galfit}

\begin{figure}
\begin{center}
\includegraphics[scale=0.5]{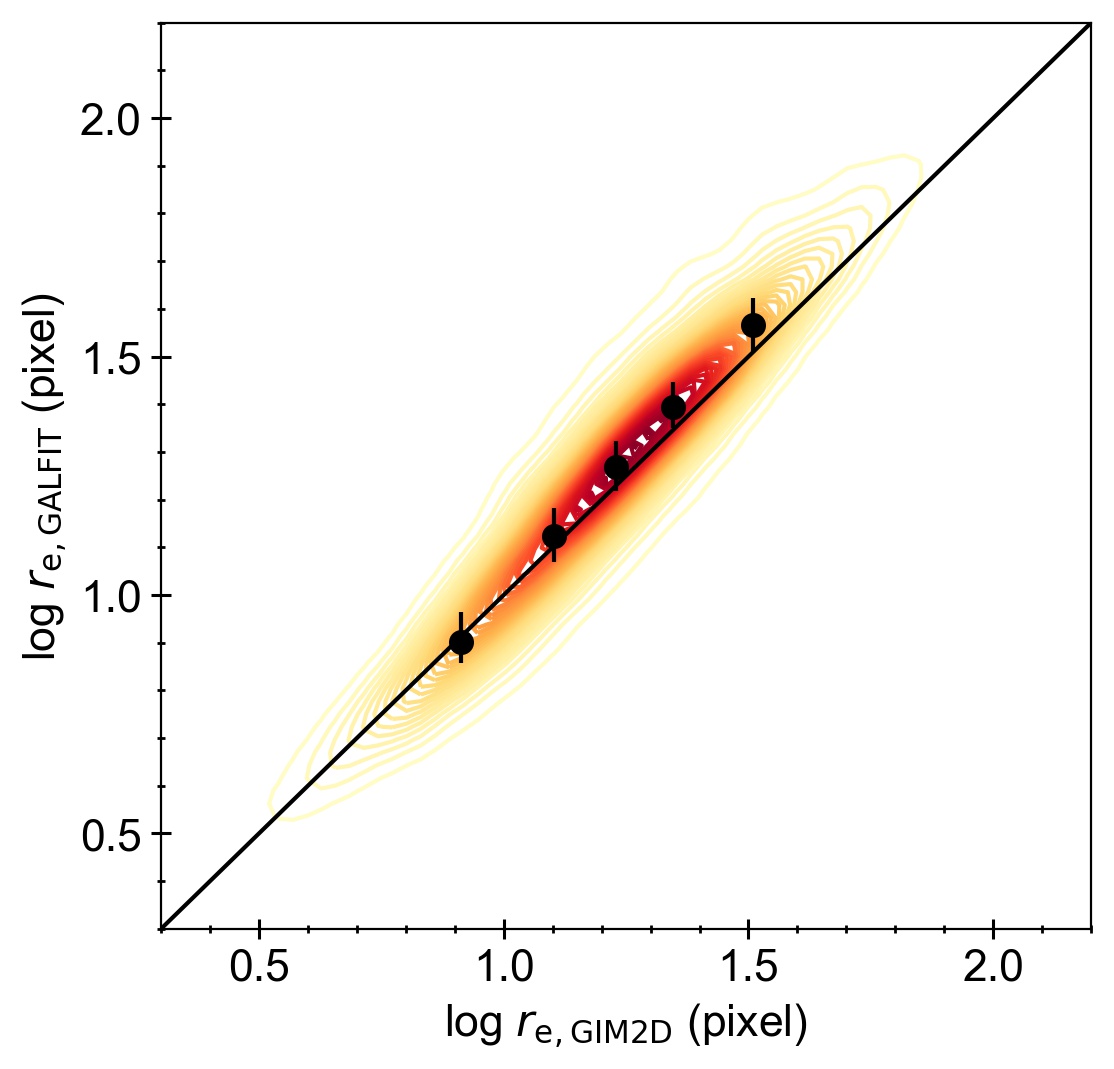}
\includegraphics[scale=0.5]{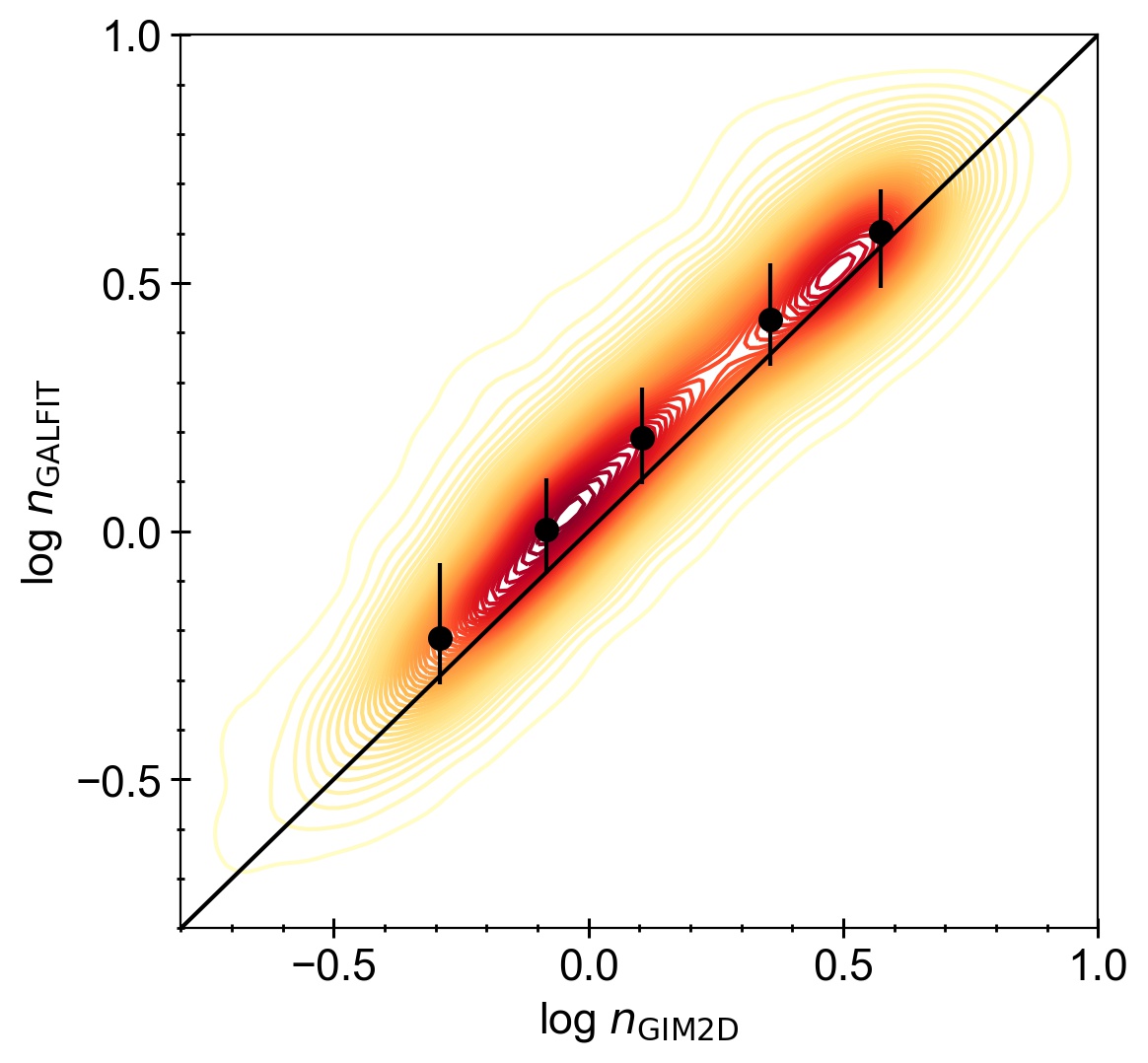}
\caption{\textit{Upper panel:} 2D KDE plot of our \re\ values measured with GALFIT versus \re\ measured with GIM2D \citep{Sargent2007} in log-log space. Black error bars represent the median GALFIT-based \re\ in different bins of GIM2D-based \re\ and the scatter between the two sets of measurements in each bin. The black solid line represents a 1:1 relation. \textit{Lower panel:} Similar to the upper panel, but for $n$ measured with GALFIT versus $n$ measured with GIM2D in \citet{Sargent2007}. }
\label{zurich}
\end{center}
\end{figure}

\citet{Sargent2007} provide GIM2D structural measurements for $I_{\rm F814W}$~$<$~22.5 objects in COSMOS.
We compare our measured \re\ values and $n$ values with those reported in \citet{Sargent2007} in Figure~\ref{zurich}. 
As can be seen from the figure, our \re\ and $n$ measurements have negligible systematic offsets when compared with \citet{Sargent2007} (our \re\ and $n$ values are slightly larger in general), and the scatter between the two sets of measurements is $\approx 0.05$ dex for \re\ and $\approx 0.1$ dex for $n$,\footnote{We note that the differences in \re\ or $n$ between the two sets of measurements do not have significant dependence on apparent magnitude.} which demonstrates that our structural measurements are consistent with \citet{Sargent2007}. 

\begin{figure}
\begin{center}
\includegraphics[scale=0.6]{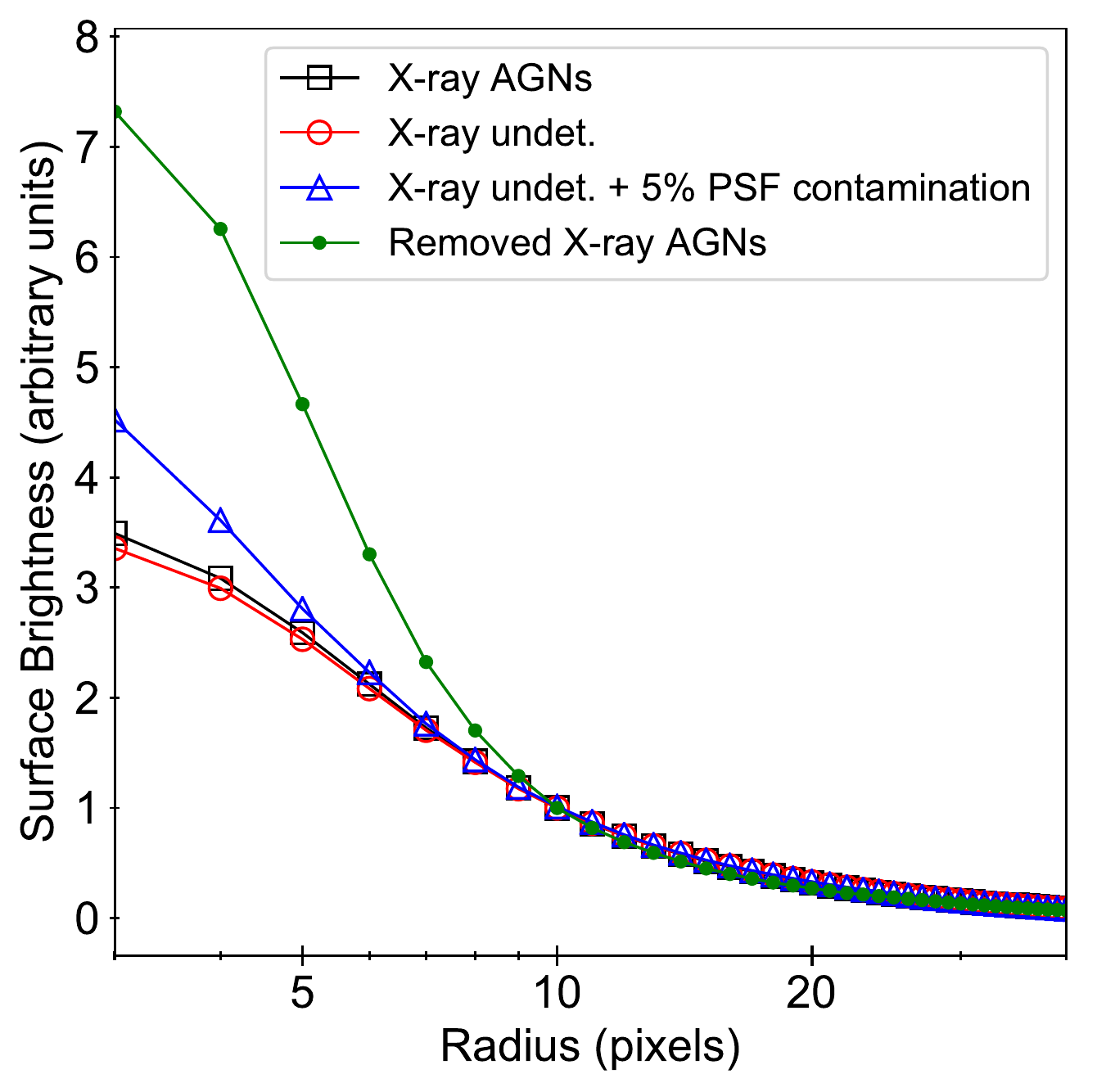}
\caption{Stacked $I_{\rm 814W}$-band surface brightness profiles of X-ray AGNs (black squares) and X-ray undetected galaxies with similar $n$ and \re\ values (red circles). The blue triangles show the modeled surface brightness profile when X-ray undetected galaxies have a $\approx 5\%$ contamination to their total integrated light from AGNs on average. The green dots show the stacked surface brightness profile of removed X-ray AGNs in Section~\ref{ss-sample}.}\label{sbcompare}
\end{center}
\end{figure}

We note that while point-like emission from AGNs has the potential to contaminate host-galaxy light profiles which may affect the reliability of structural measurements, this contamination is small in our sample (where AGNs that dominate over host galaxies are removed; see Section~\ref{ss-sample}).
We stack the $I_{\rm 814W}$-band surface-brightness profiles of $\approx 650$ X-ray AGNs (log \lx\ $>$ 42 sources) at $z < 1.2$ in our compilation. For each of these X-ray AGNs, we select one galaxy not detected in the X-ray that has the closest $n$ and \re\ values to it (without duplications). 
We then stack the $I_{\rm 814W}$-band surface-brightness profiles of these matched X-ray undetected galaxies.
The comparison between the stacked surface-brightness profiles of X-ray AGNs and X-ray undetected galaxies with similar structural parameters can be seen in Figure~\ref{sbcompare}.
We can see that the stacked surface-brightness profile of X-ray AGNs is very similar to that of X-ray undetected galaxies. There is no obvious ``excess'' in the center which would suggest nuclear contamination from AGNs (see section~3.1 of \citealt{Kocevski2017}). The median reduced $\chi^2$ of the single-component \sersic\ fits of X-ray AGNs ($\approx 1.1$) is also similar to that of X-ray undetected galaxies.
We model how the point-like emission from AGN may affect the host-galaxy surface-brightness profile: for the stacked surface-brightness profile of \xray\  undetected galaxies, we add point-like emission which accounts for $\approx 5\%$ of the total integrated light (the point-like emission is modeled utilizing the PSF generated in Section~\ref{sss-psf}).
Typically, when the PSF contamination is $\gtrsim 5$--10\%, \texttt{GALFIT} will hit the \re~=~0.5 and/or $n = 8$ constraint we set in Section~\ref{ss-galfit} for the single-component \sersic\ fitting, so that the object will not be utilized. For example, if we use \texttt{GALFIT} to fit the obtained composite light profile (the stacked light profile of matched \xray\ undetected galaxies plus a 5\% PSF contamination; see the blue curve in Figure~\ref{sbcompare}), we hit the constraints mentioned above.
In the centers of galaxies, the obtained composite surface-brightness profile clearly shows higher surface brightness than that of \xray\ AGNs (see Figure~\ref{sbcompare}).
Thus, the contamination to the host-galaxy light profile in the \textit{HST} F814W band is small for X-ray AGNs in our sample.\footnote{We note that it is unlikely for a galaxy to mimic the light profile of a galaxy with much more concentrated \sersic\ profile when there is a moderate level of AGN contamination ($\lesssim 5$--10\%), so that this mode of contamination has limited influence for the \sigmaone\ measurements. While $\approx 80\%$ of the light from a point-like source is concentrated within a radius of 5 pixels (0.15$''$) according to the COSMOS $I_{\rm 814W}$-band PSF model, a large fraction of the light from a low-to-moderate-redshift galaxy lies outside the central 5-pixel-radius region, and the \sersic-profile fitting will likely be dominated by this part of the light. For the stacked $I_{\rm 814W}$-band light profile of galaxies that host \xray\ AGNs shown in Figure~\ref{sbcompare}, $\approx 86\%$ of the light lies outside the central 5-pixel-radius region. Even for a BD galaxy with \re\ $\approx 1$ kpc at $z = 1.2$, $\approx 50\%$ of the light lies outside the central 5-pixel-radius region.}
For comparison, we also show the stacked surface-brightness profile of X-ray AGNs removed from our sample in Figure~\ref{sbcompare}, which demonstrates high levels of AGN contamination.

\section{Identifying BD galaxies in the COSMOS field} \label{a-morph}
We classify galaxies as BD/Non-BD utilizing a convolutional neural network (CNN).
This machine-learning-based approach has been widely adopted to perform morphological classification of galaxies. For example, \citet{HC2015} utilized a CNN that was trained based on the visual classification in \citet{Kartaltepe2015} to perform morphological classification for $H < 24.5$ galaxies in all the CANDELS fields, and the generated catalog has been adopted in \citet{Ni2019}.

\begin{figure}
\begin{center}
\hspace{-0.05 cm}
\includegraphics[scale=0.25,angle=0]{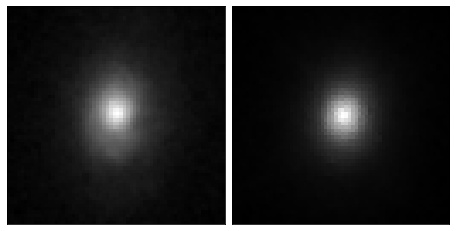}
\includegraphics[scale=0.25,angle=0]{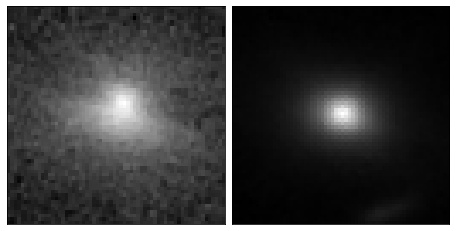}
\vspace{0.1 cm}
\newline
\includegraphics[scale=0.25,angle=0]{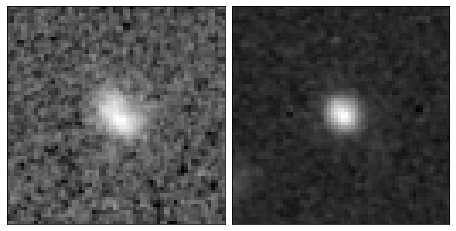}
\includegraphics[scale=0.25,angle=0]{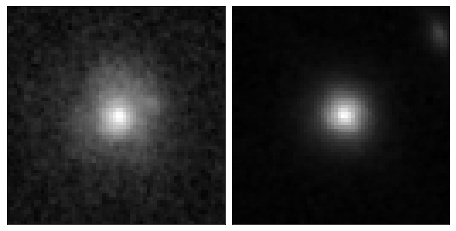}
\caption{Examples of BD galaxies classified in \citet{HC2015} with $H_{\rm 160W}$-band images, but not classified as BD galaxies in our training sample. In each subfigure, the left panel is the $I_{\rm 814W}$-band cutout of size 64 $\times$ 64 pixels, and the right panel is the $H_{\rm 160W}$-band cutout of size 64 $\times$ 64 pixels.}
\label{hvsi}
\end{center}
\end{figure}

\begin{figure}
\begin{center}
\includegraphics[scale=0.17]{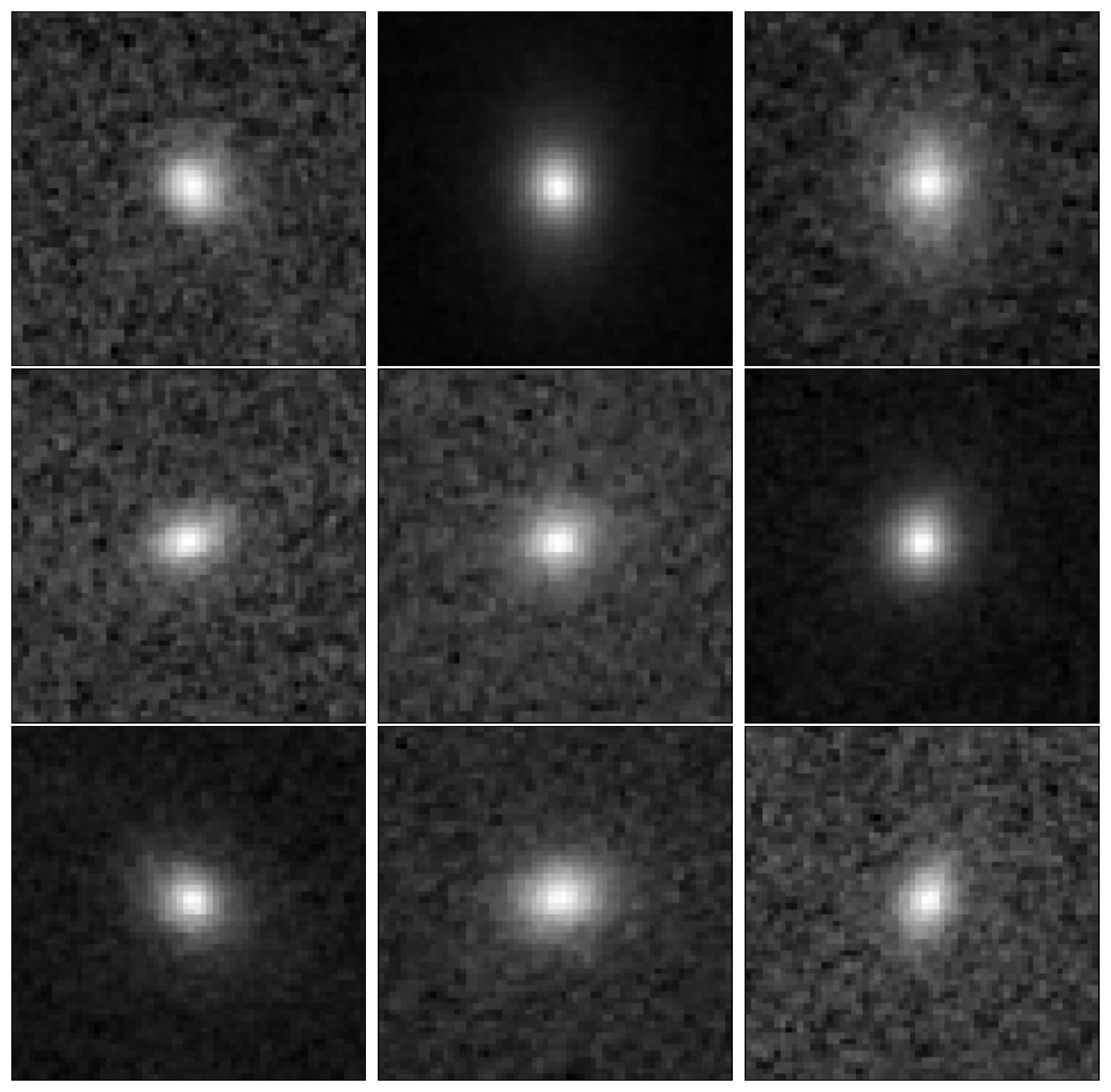}
\hspace{-0.3 cm}
\includegraphics[scale=0.17]{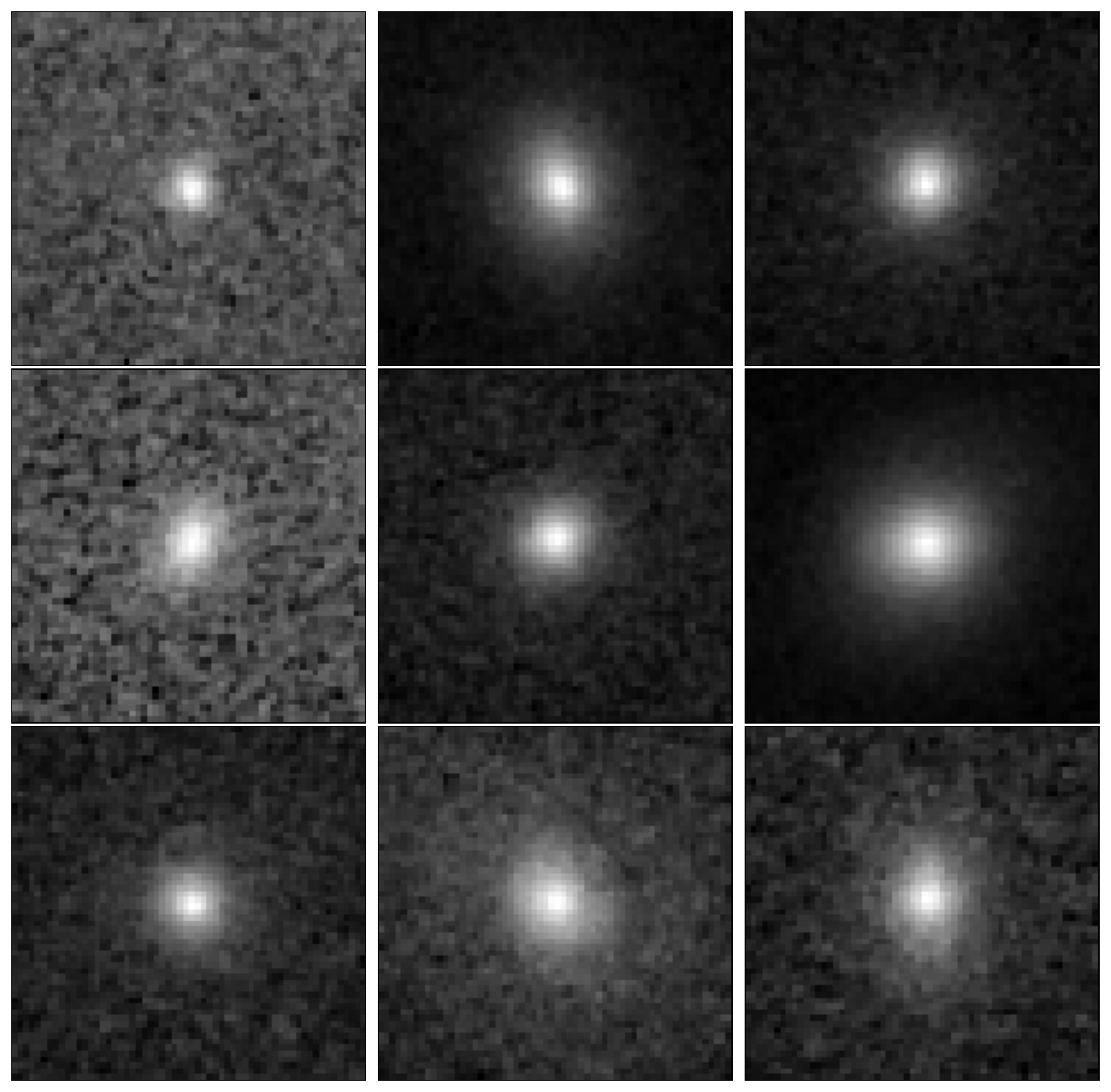}
\caption{Example $I_{\rm 814W}$-band cutouts (64 $\times$ 64 pixels) of BD galaxies visually identified in the training set.}
\label{myvs}
\end{center}
\end{figure}

To train the CNN, we select $\approx 8000$ galaxies among all $I_{\rm F814W}$ $<$ 24 galaxies ($\rm MU\_CLASS = 1$; MU\_CLASS is a star/galaxy classifier in the COSMOS ACS catalog; \citealt{Leauthaud2007}) in the COSMOS \textit{HST} field \citep{Capak2007,Leauthaud2007}, and manually assign each of them a binary label of BD (1) or Non-BD (0).
In order to make the selection of BD galaxies consistent with \citet{Kartaltepe2015} and \citet{HC2015} (here we are pointing to objects with $f_{\rm sph}$ $>$ 2/3, $f_{\rm disk}$ $<$ 2/3, and $f_{\rm irr}$ $<$ 1/10), 4015 galaxies in the training set are selected from the CANDELS-COSMOS field, where morphological classifications from \citet{HC2015} are available. 
When labeling these sources, we try to be consistent with \citet{HC2015}, and the overall agreement is $\approx$ 95\%. 
Approximately 71\%/99\% of BD/Non-BD galaxies identified in \citet{HC2015} are still labeled as BD/Non-BD galaxies. 
The reason for the relatively low level of agreement among BD galaxies can be attributed to both the morphological k-correction and the different angular resolution of CANDELS F160W images (0.06''/pixel) and COSMOS F814W images (0.03''/pixel) (see Figure~\ref{hvsi}). 
The other 4271 galaxies are randomly selected across the whole COSMOS field, and we visually classify them as consistently as possible.
Among the 8286 galaxies in total, 891 galaxies are classified as BD galaxies (see Figure~\ref{myvs}).

We split these labeled galaxies into a training set (5286 galaxies), a validation set (1500 galaxies), and a test set (1500 galaxies).\footnote{The relatively large number of objects placed in the validation/test set compared to common practice is due to the limited fraction ($\approx 10\%$) of BD galaxies: the number of BD galaxies in the validation/test set should be large enough for reasonable statistics.}
We then create cutouts for them of size 64$\times$64 pixels from ACS COSMOS science images v2.0 \citep{Koekemoer2007}, and store the normalized FITS file as \texttt{NumPy} arrays. 

Before the training, we copy the the training set nine times and add random Gaussian noise (that is small enough so that the overall galaxy morphology/structure does not have noticeable changes), which has proved to be a good approach for data augmentation \citep[e.g.][]{HC2015}.
During the training, real-time random rotations, shifts in the center position (less than 10\% of the total height and width), and zooms (between 75\% and 135\%) are also applied to the training set.

\begin{table}
 \begin{center}
\caption{Convolutional Neural Network Configuration.\label{cnn}}
  \begin{tabular}{cccc}
 \hline \hline
{Layer} & {Filter Size} & {Feature Number}  & {Output Shape} \\ \hline
Conv2D & 3$\times$3 & 32 & (64, 64, 32) \\
Conv2D & 3$\times$3 & 32 & (64, 64, 32)\\
MaxPooling2D & 2$\times$2 & - & (32, 32, 32) \\
Conv2D & 3$\times$3 & 64 & (32, 32, 64)\\
Conv2D & 3$\times$3 & 64 & (32, 32, 64)\\
MaxPooling2D & 2$\times$2 & - & (16, 16, 64)\\
Conv2D & 3$\times$3 & 128 & (16, 16, 128)\\
Conv2D & 3$\times$3 & 128 & (16, 16, 128)\\
MaxPooling2D & 2$\times$2 & - & (8, 8, 128)\\
Dense & - & 1024 & 1024 \\
Dropout(0.2) & - & - & 1024 \\
Dense & - & 1 & 1 \\
\hline \hline
 \end{tabular}
  \end{center}
\end{table}

The CNN used in this work is implemented with the \texttt{Keras} package \citep{keras}.
The architecture of the CNN can be seen in Table~\ref{cnn}.
Hyper-parameters including the network depth, filter size, and number of channels are optimized with the validation set. 
The activation function in between all the convolution and dense layers is ReLU \citep[e.g.][]{Nair2010}. 
A sigmoid function is applied to the last-layer output to compress it  into [0, 1], which can be interpreted as the probability of being BD galaxies.

We use binary cross-entropy as the loss function, and we also apply the inverse class ratio as the weight to the loss to account for the sample imbalance.
We use the ADAM optimizer \citep{Kingma2014} to minimize the loss, and set the initial learning rate to be 0.0001.
At the end of each learning epoch, we use the $F_1$ score to assess the model:
\begin{align*}
F_1 = \rm \frac{2 \times precision \times recall}{precision + recall},
\end{align*}
where
\begin{align*}
\rm precision & =  \rm \frac{True~Positive~(TP)}{True~Positive~(TP) + False~Positive~(FP)}, \\
\rm recall & =  \rm \frac{True~Positive~(TP)}{True~Positive~(TP) + False~Negative~(FN)}.
\end{align*}
The $F_1$ score is widely used to assess the quality of binary classification.
For imbalanced data sets, it is more sensitive to the true quality of classification than accuracy.
We drop the learning rate by a factor of 2 if the $F_1$ score of the validation set stops increasing for 10 epochs.
When the $F_1$ score of the validation set stops increasing for 50 epochs, we stop the training process and save the model.

We test the obtained model with the test set. 
When converting the predicted probability into a binary label, we first use the default threshold of 0.5 to classify BD/Non-BD galaxies, and we find that the number of FP is larger than the number of FN (due to the sample imbalance). 
For the purpose of this work, we require the ``contamination'' in the BD sample to be as small as possible. Thus, we use the validation set to select a higher probability threshold that can make the number of FP approximately equal to the number of FN. 
The final training results can be seen in Table~\ref{cm}. We can correctly predict $\approx$ 87.3\% of BD galaxies and 98.4\% of Non-BD galaxies in the test set, and the number of predicted BD galaxies is roughly equal to the number of true BD galaxies.

\begin{table}
 \begin{center}
\caption{Training results assessed with the test set of 1500 galaxies.\label{cm}}
  \begin{tabular}{ccccccccc} \hline  \hline 
TP   & FP  &  TN & FN & \multicolumn{3}{c}{Accuracy} & $F_1$  \\  \cline{5-7} 
     &     &     &     & BD & Non-BD & Overall  & \\ \hline 
144  &    21 & 1314 & 21 & 87.3\% & 98.4\% & 97.2\% & 0.87\\ \hline \hline
 \end{tabular}
  \end{center}
\end{table}

\begin{figure}
\begin{center}
\includegraphics[scale=0.17]{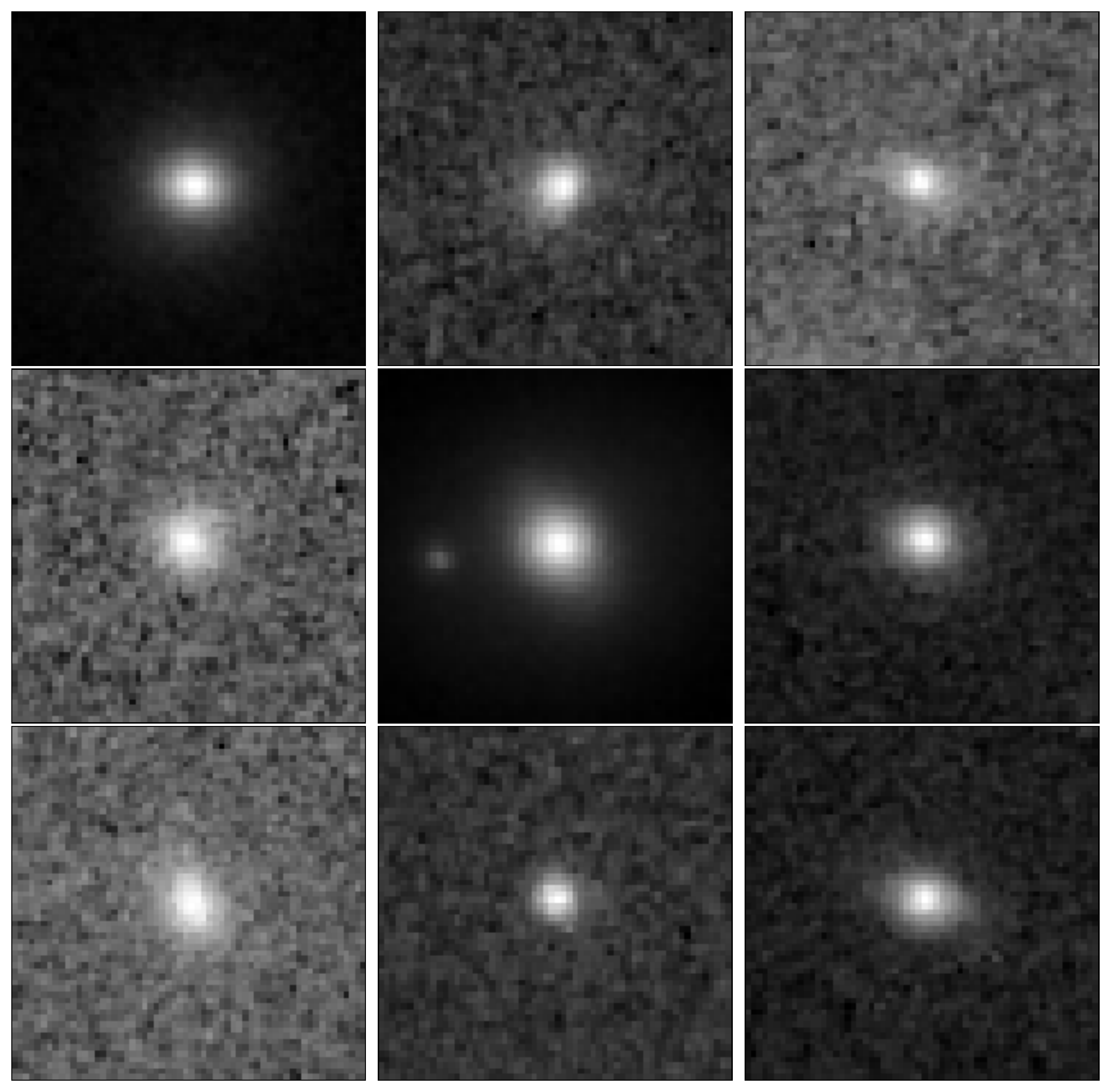}
\includegraphics[scale=0.17]{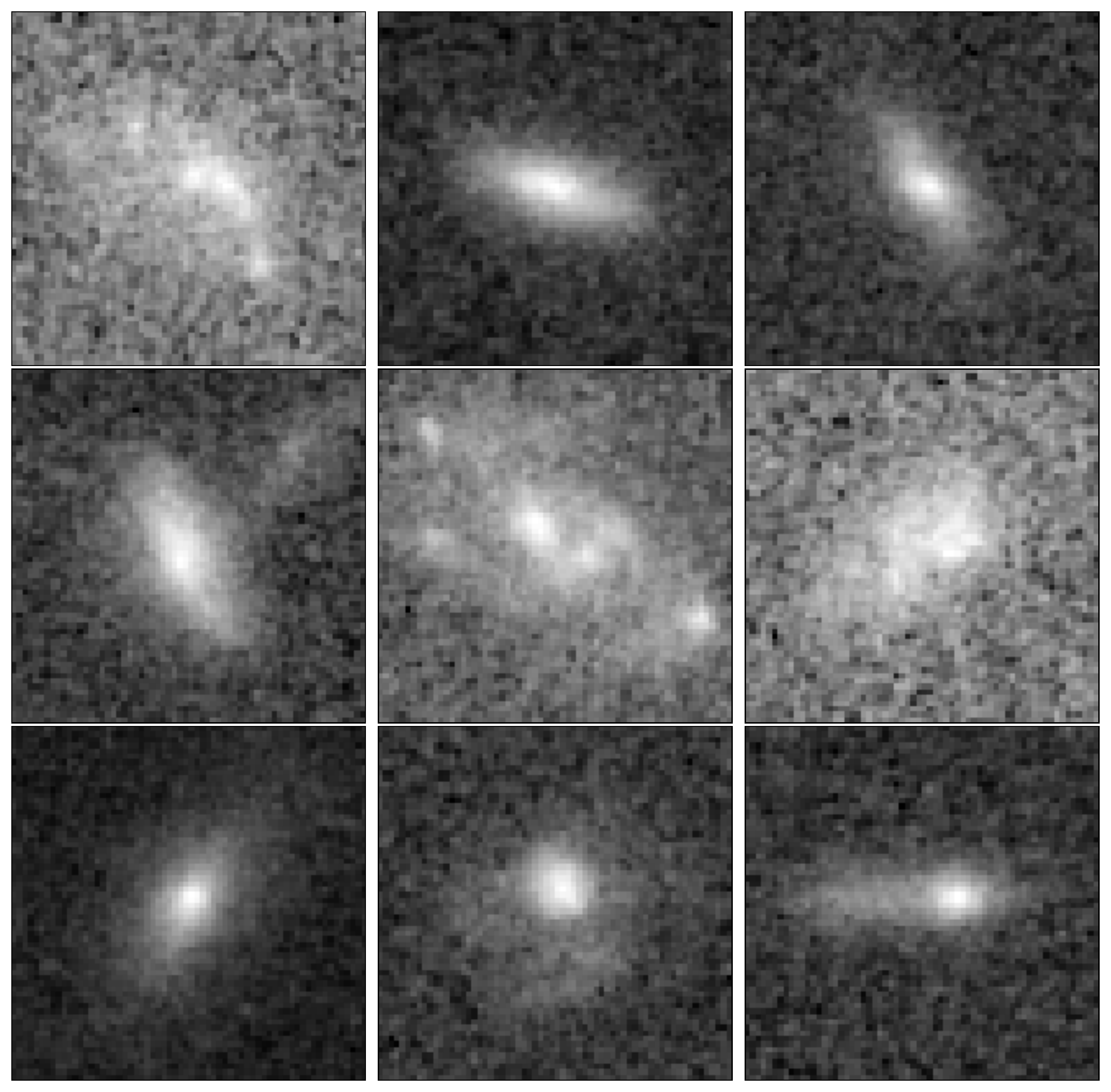}
\caption{Example $I_{\rm 814W}$-band cutouts (64 $\times$ 64 pixels) of deep-learning-predicted BD galaxies (left) and Non-BD galaxies (right) in the COSMOS field.}
\label{dlresult}
\end{center}
\end{figure}

\begin{figure}
\begin{center}
\includegraphics[scale=0.5]{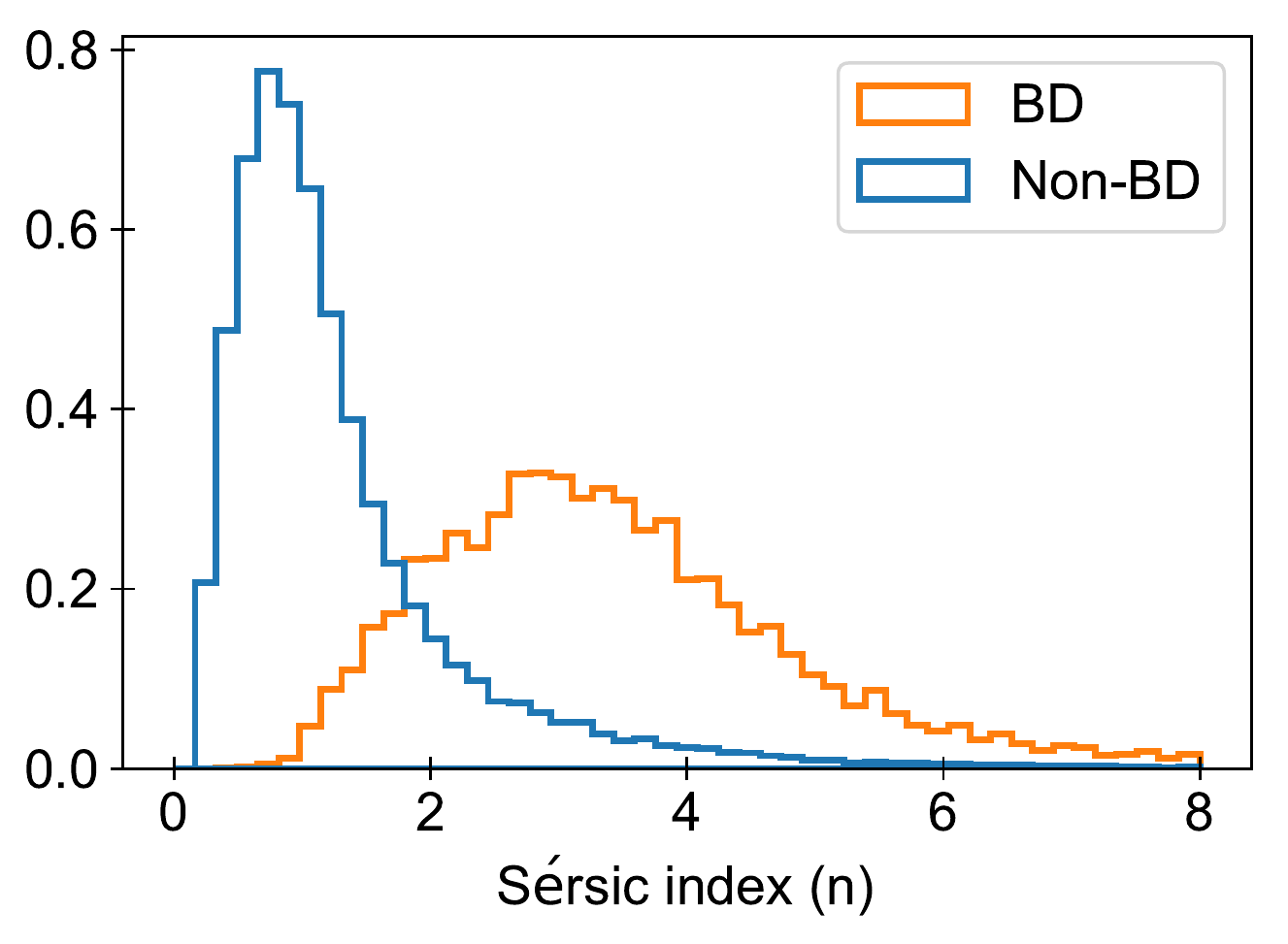}
\caption{The S$\acute{\rm e}$rsic index distributions of the predicted BD galaxies and Non-BD galaxies in the COSMOS field that demonstrate a clear distinction between the two populations.}
\label{sersic}
\end{center}
\end{figure}

With the trained CNN and the tuned probability threshold, we classify $\approx 115,000$ $I_{\rm F814W} <$ 24 galaxies in the COSMOS ACS field as BD or Non-BD.
Figure~\ref{dlresult} shows example cutouts of the predicted BD galaxies and Non-BD galaxies (the presented galaxies are randomly drawn from the sample).
In Figure~\ref{sersic} we show the distributions of $n$ among classified BD galaxies and Non-BD galaxies. The clear separation in the distribution of $n$ between the two populations demonstrates further the validity of our classification.
We also note that our classification is consistent with that of \citet{HC2015} when comparing the relative numbers of BD galaxies. At $z < 0.8$ and log~\mstar\ $> 10.2$, 2117 galaxies in our sample are classified as BD galaxies. This number is 453 for the CANDELS field, with BD galaxies identified in \citet{HC2015}. The ratio between these two numbers is roughly consistent with the ratio between our utilized area of COSMOS ($\approx 1.4$ deg$^2$) and the area of CANDELS ($\approx 0.25$ deg$^2$).

\section{PCOR analyses with different binning approaches} \label{a-bin}

\begin{figure*}
\begin{center}
\includegraphics[scale=0.34]{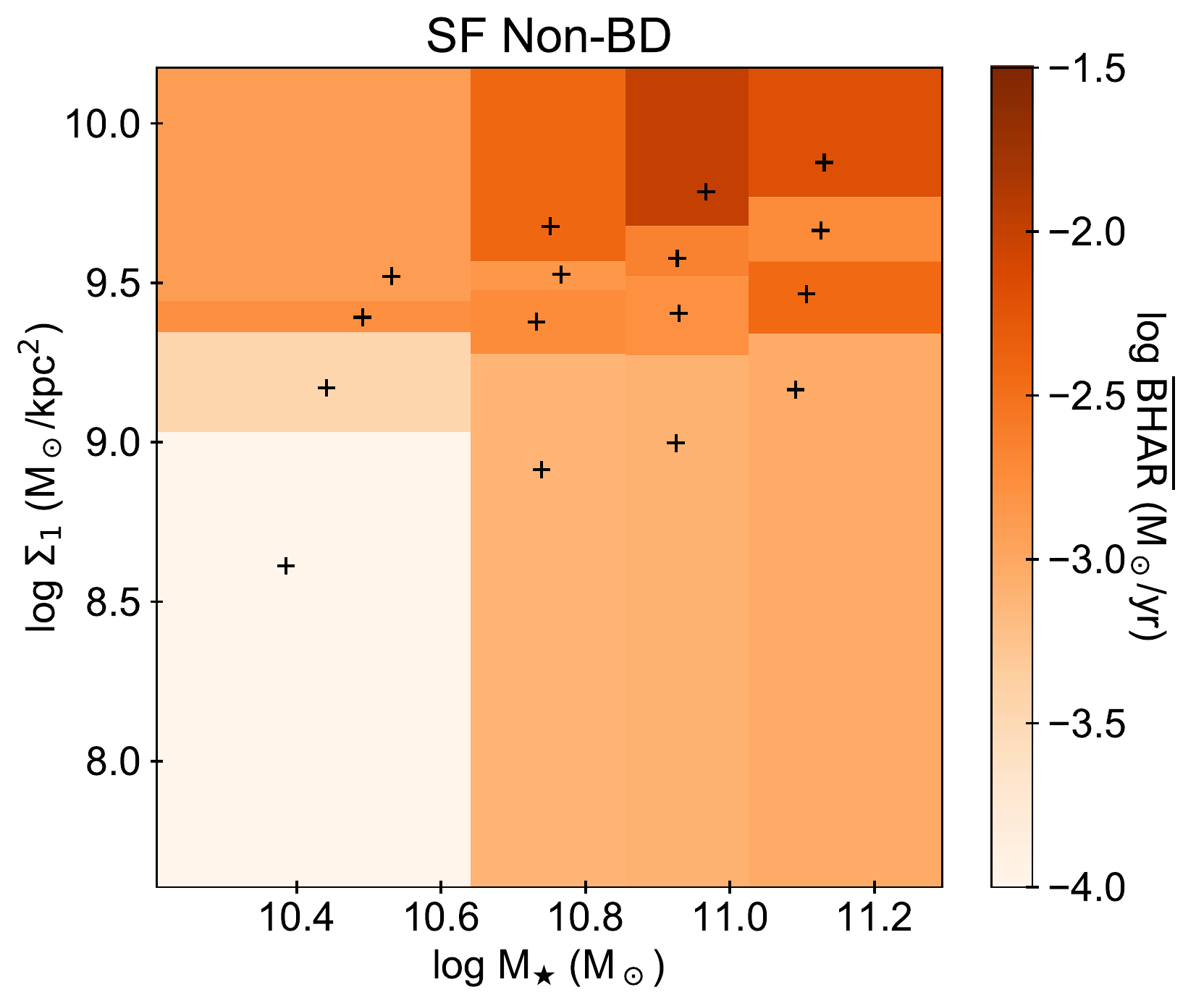}
\includegraphics[scale=0.34]{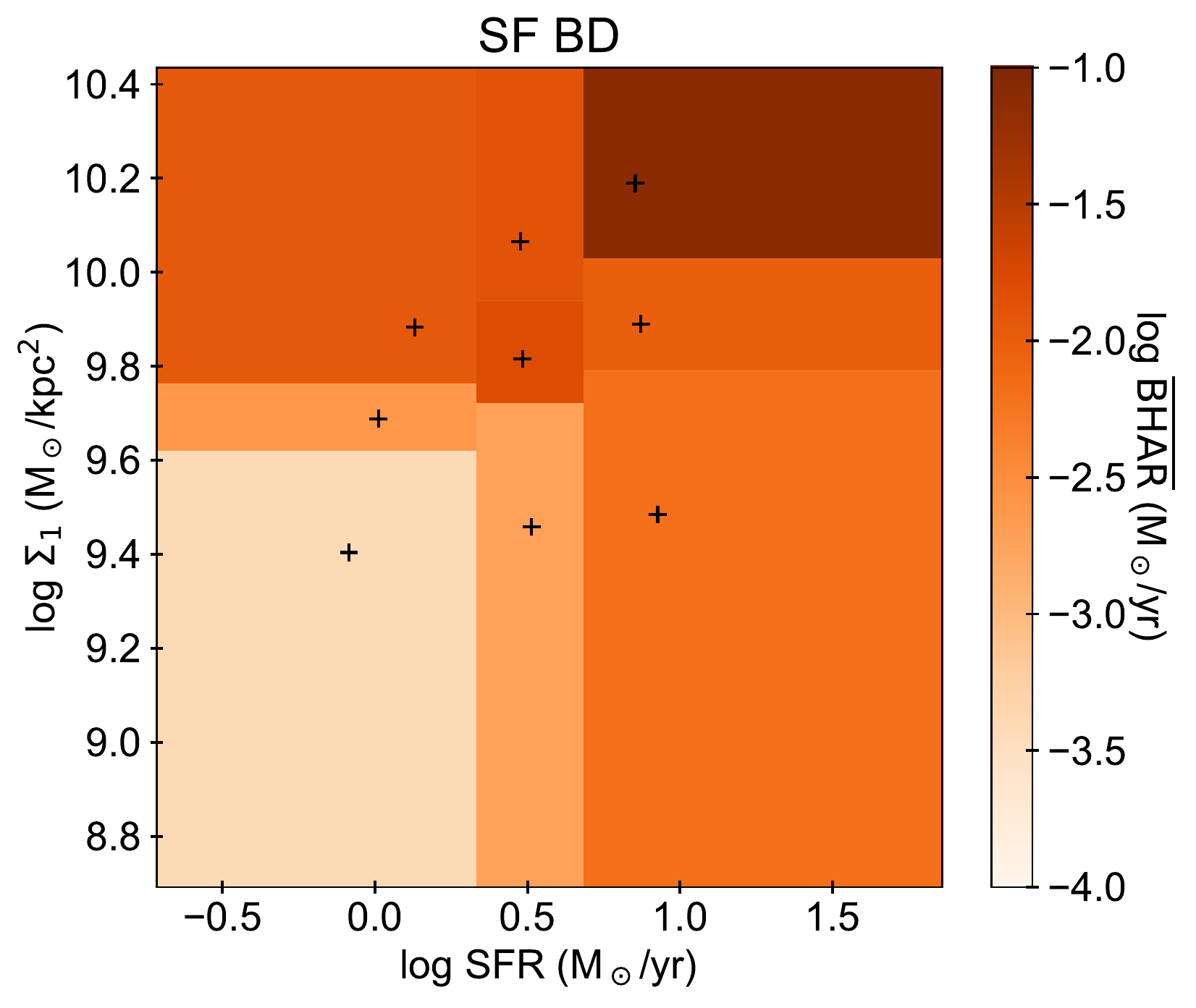}
\includegraphics[scale=0.34]{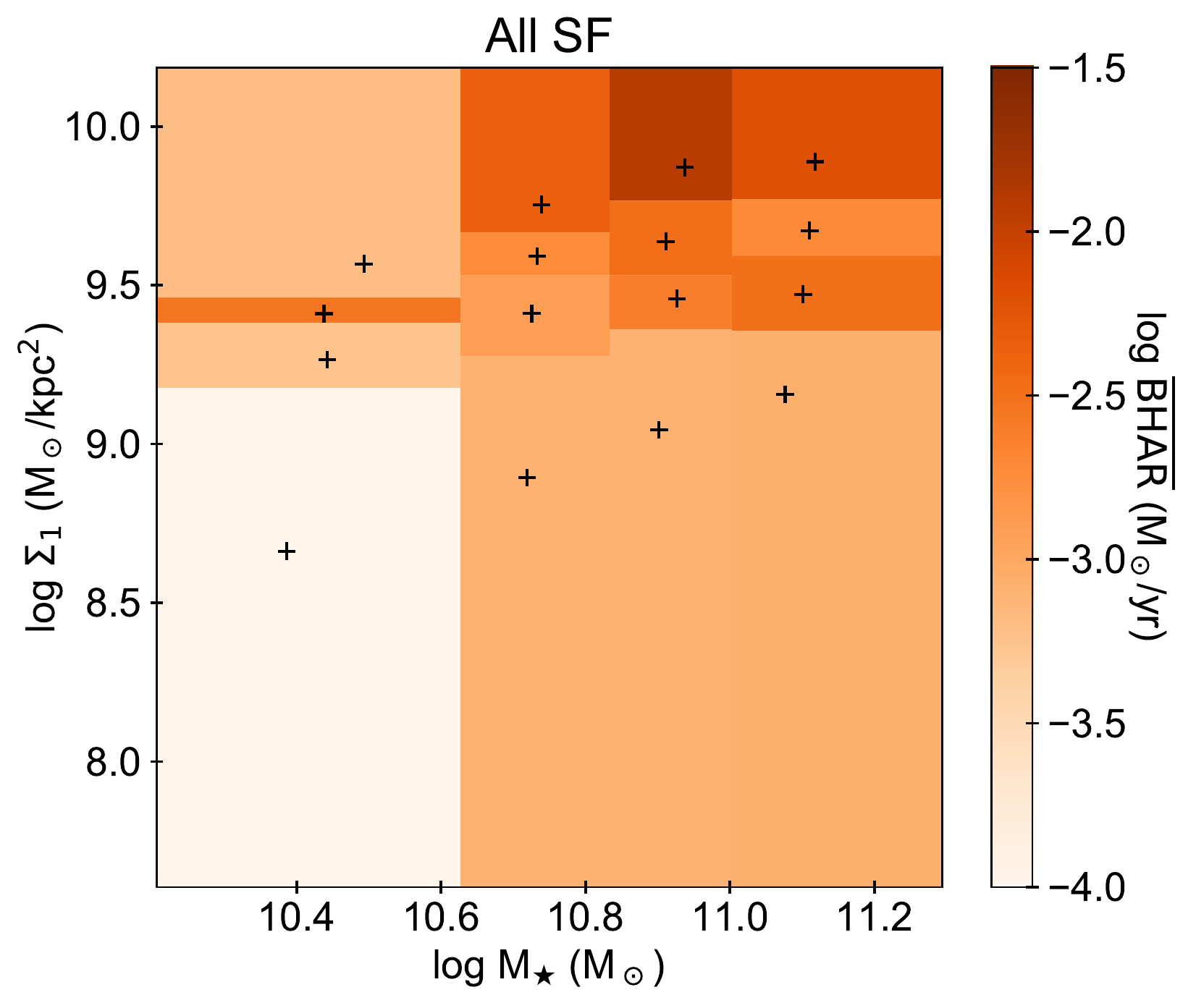}
\caption{\textit{Left panel:} Color-coded $\rm \overline{BHAR}$ in different bins of \mstar\ and \sigmaone\ for galaxies in the SF Non-BD sample. Each 2D bin contains $\approx$ 11 X-ray detected galaxies. The black plus sign indicates the median \mstar\ and \sigmaone\ of the sources in each bin. \textit{Middle panel:} Color-coded $\rm \overline{BHAR}$ in different bins of SFR and \sigmaone\ for galaxies in the SF BD sample. Each 2D bin contains $\approx$ 9 X-ray detected galaxies. The black plus sign indicates the median SFR and \sigmaone\ of the sources in each bin. \textit{Right panel:} Color-coded $\rm \overline{BHAR}$ in different bins of \mstar\ and \sigmaone\ for galaxies in the ALL SF sample. Each 2D bin contains $\approx$ 13 X-ray detected galaxies. The black plus sign indicates the median \mstar\ and \sigmaone\ of the sources in each bin.
}
\label{2dbin_xray}
\end{center}
\end{figure*}

We verified that the PCOR analysis results in Section~\ref{s-ar} do not change qualitatively when the binning approach changes.
Our finding in Section~\ref{ss-sfnonbd}/Section~\ref{ss-allsf} that the \bhar-\sigmaone\ relation is more fundamental than the \bhar-\mstar\ relation for the SF Non-BD/ALL SF sample holds true when we use $5 \times 5$ bins or $6 \times 6$ bins.
Also, if we bin objects so that each 2D bin has the same number of \xray\  detected galaxies (see Figure~\ref{2dbin_xray}), our results do not change qualitatively (see Table~\ref{xray-pcor}).
Similarly, our finding in Section~\ref{ss-sfbd} that the \bhar-\sigmaone\ relation exists when controlling for SFR for the SF BD sample holds true when we use $4 \times 4$ bins; this result does not change qualitatively when we bin objects based on the number of X-ray detected galaxies (see the middle panel of Figure~\ref{2dbin_xray} and Table~\ref{xray-pcor}).

\begin{table}
\begin{center}
\caption{$p$-values (significances) of partial correlation analyses for the samples binned by the number of X-ray detected galaxies}
\label{xray-pcor}
\begin{tabular}{ccccc}
\hline\hline
\multicolumn{3}{c}{SF Non-BD} \\ \hline\
Relation &   Pearson & Spearman \\\hline
\bhar-\sigmaone\            & $\boldsymbol {9 \times 10^{-5}~(3.9\sigma)}$  & $\boldsymbol {2 \times 10^{-4 }~(3.7\sigma)}$ &  \\
\bhar-$M_\star$               &  $0.08~(1.8\sigma)$  & $0.23~(1.2\sigma)$ \\
 \hline\hline
 \multicolumn{3}{c}{SF BD} \\ \hline 
Relation &   Pearson & Spearman \\\hline
\bhar-\sigmaone\            & $\boldsymbol {2\times 10^{-3}~(3.1\sigma)}$  & $7\times 10^{-3}~(2.7\sigma)$ &  \\
\bhar-SFR              &  $0.04~(2.1 \sigma)$  & $0.73~(0.4\sigma)$ \\
\hline\hline
\multicolumn{3}{c}{ALL SF} \\ \hline 
Relation &   Pearson & Spearman \\\hline
\bhar-\sigmaone\            & $\boldsymbol {2 \times 10^{-4}~(3.8\sigma)}$  & $5 \times 10^{-3}~(2.8\sigma)$ &  \\
\bhar-$M_\star$               &  $0.10~(1.6\sigma)$  & $0.22~(1.2\sigma)$ \\
 \hline\hline
\end{tabular}                                         
\end{center}
\end{table}

\section{The \bhar-SFR relation among BD galaxies in general} \label{a-allbd}

\begin{figure}
\begin{center}
\includegraphics[scale=0.58]{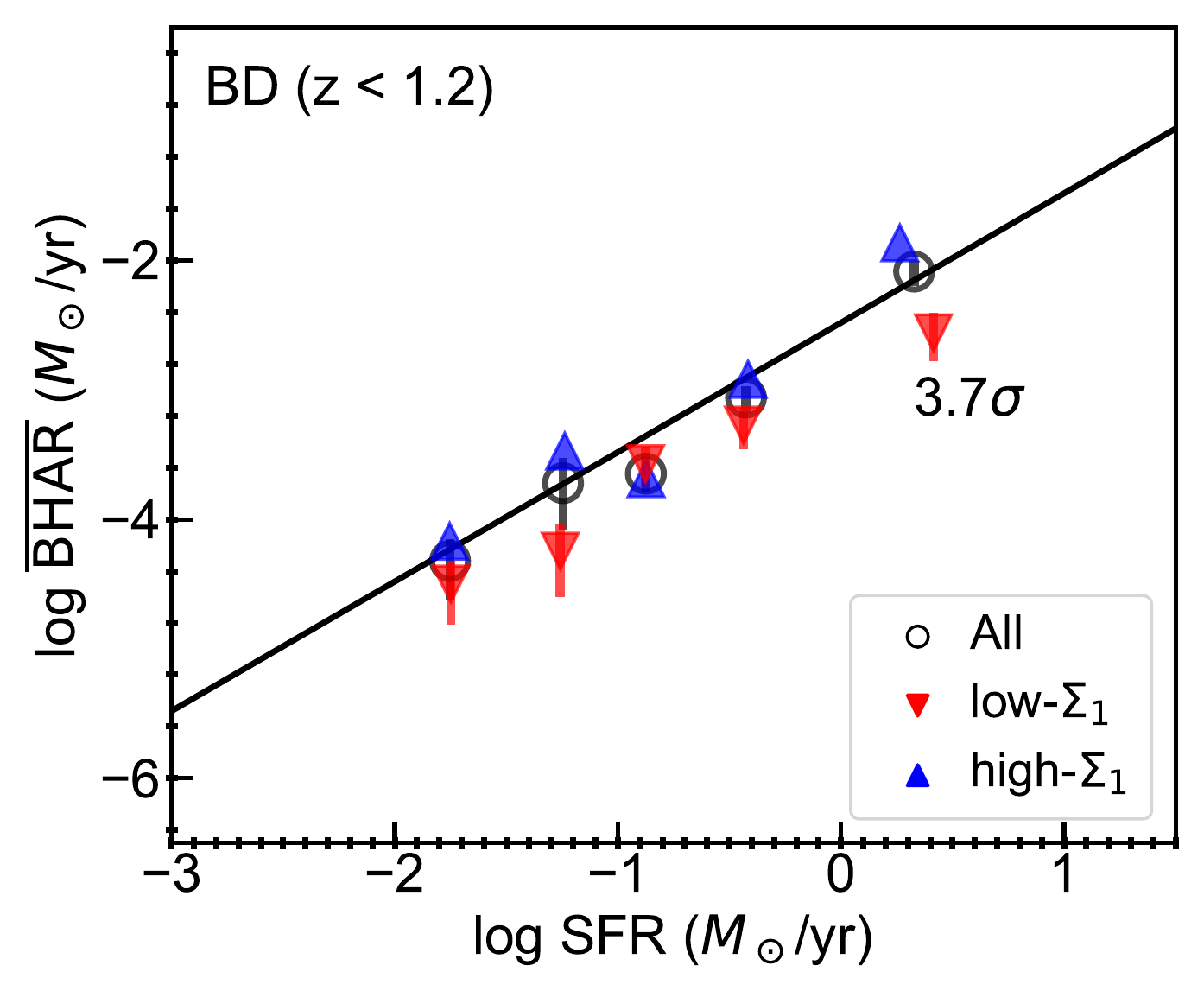}
\caption{\bhar\ vs. SFR for BD galaxies with \hbox{log \mstar\ $>$ 10} at $z < 1.2$ in the COSMOS field divided into SFR bins with $\approx 1000$ sources per bin. Each SFR bin (black circles) is further divided into two subsamples with \sigmaone\ above (blue upward-pointing triangles) and below (red downward-pointing triangles) the median \sigmaone\ of the bin, respectively. The black solid line represents the best-fit \bhar-SFR relation in \citet{Yang2019}. We can see that the general \bhar-SFR trend is close to that obtained in \citet{Yang2019}.}
\label{allbd}
\end{center}
\end{figure}

Though in Section~\ref{ss-sfbd} we found that among SF BD galaxies, the \bhar-\sigmaone\ relation is significant when controlling for SFR while the \bhar-SFR relation is not significant when controlling for \sigmaone, we note that the \bhar-SFR relation is still the dominant relation among BD galaxies in general (i.e. including quiescent BD galaxies; see Figure~\ref{allbd}), consistent with the findings in \citet{Yang2019}.
In Figure~\ref{allbd}, we show that the \bhar-SFR trend for all BD galaxies with \hbox{log \mstar\ $>$ 10} at $z < 1.2$ in the COSMOS field is close to the \bhar-SFR relation obtained in \citet{Yang2019} utilizing $z = 0.5$--3 galaxies in the CANDELS field;\footnote{The \citet{Yang2019} relation is well-constrained from log SFR $\approx 1.5$ to log SFR $\approx -2$, probing log \bhar\ from $\approx -1$ to $\approx -4.5$, so that it could be applied to the parameter space probed in this work.} we also show that the difference in \sigmaone\ at a given SFR value does not associate with a significant difference in \bhar\ except for the highest SFR bin (where a $\approx 3.7\sigma$ difference in \bhar\ is associated with \sigmaone).

As discussed in \citet{Ni2019} and Section~\ref{ss-monster}, the \bhar-SFR relation among BD galaxies and the \bhar-\sigmaone\ relation only among SF BD galaxies may reflect the same link between BH growth and the central $\sim$~kpc gas density of host galaxies. 
When \sigmagas\ (or $\Sigma_{\rm SFR}$)  is roughly uniform across the bulge, the SFR among BD galaxies could naturally serve as an indicator of the central $\sim$~kpc gas density. Due to the compact sizes (on $\sim$ kpc scale) of BD galaxies, even when the distribution of \sigmagas\ (or $\Sigma_{\rm SFR}$) among a BD galaxy is far from uniform, we still expect a significant fraction of SFR to be enclosed in the central $\sim$ kpc region. Thus, SFR could serve as an indicator of the central $\sim$~kpc gas density among BD galaxies, though this relation suffers from a considerable scatter which originates from the scatter in the fraction of gas/SFR enclosed in the central $\sim$~kpc region, similar to the the uncertainty associated with the SFR$_{\rm bulge}$-$\Sigma_{\rm SFR, 1~kpc}$ relation discussed in Footnote~\ref{fn-sfrbulge}.
Among SF BD galaxies, \sigmaone\ may serve as a better indicator of the central $\sim$~kpc gas density (though this indicator only works for SF galaxies).

\clearpage
\bibliography{cosmos.bib}
\bibliographystyle{mnras}

\bsp	
\label{lastpage}
\end{document}